\documentclass[10pt]{IEEEtran}
\usepackage[LGR,T1]{fontenc}
\usepackage[latin9]{inputenc}
\usepackage{color}
\usepackage{array}
\usepackage{cite} 
\usepackage{url}
\usepackage{multirow}
\usepackage{amsmath}
\usepackage{amsthm}
\usepackage{amssymb}
\usepackage{graphicx}
\usepackage[unicode=true,
 bookmarks=true,bookmarksnumbered=true,bookmarksopen=true,bookmarksopenlevel=3,
 breaklinks=false,pdfborder={0 0 0},pdfborderstyle={},backref=false,colorlinks=true]
 {hyperref}
\hypersetup{pdftitle={Renyi},
 pdfauthor={Lei Yu},
 pdfborderstyle={},pdfborderstyle={},pdfborderstyle={},pdfborderstyle={},pdfborderstyle={},pdfborderstyle={},pdfborderstyle={},pdfborderstyle={},pdfborderstyle={},pdfborderstyle={},pdfborderstyle={},pdfborderstyle={},pdfborderstyle={},pdfborderstyle={},pdfborderstyle={},pdfborderstyle={},pdfborderstyle={},pdfborderstyle={},pdfpagelayout=OneColumn,pdfnewwindow=true,pdfstartview=XYZ,plainpages=false,linkcolor=blue,urlcolor=blue,citecolor=red,anchorcolor=blue,linkcolor=blue,urlcolor=blue,citecolor=red,anchorcolor=blue}

\makeatletter

\providecommand{\tabularnewline}{\\}

\theoremstyle{plain}
\newtheorem{thm}{\protect\theoremname}
\theoremstyle{plain}
\newtheorem{lem}[]{\protect\lemmaname}
\theoremstyle{remark}
\newtheorem{rem}[]{\protect\remarkname}
\theoremstyle{plain}
\newtheorem{prop}[]{\protect\propositionname}
\theoremstyle{definition}
\newtheorem{defn}[]{\protect\definitionname}

\allowdisplaybreaks[1]
\newcounter{mytempeqncnt}

\newcommand{\Ebb}{\mathbb{E}}

\newcommand{\calK}{\mathcal{K}}

\newcommand{\calM}{\mathcal{M}}

\newcommand{\calP}{\mathcal{P}}

\newcommand{\calX}{\mathcal{X}}

\newcommand{\rme}{\mathrm{e}}

\newcommand{\bbE}{\mathbb{E}}

\newcommand{\bbN}{\mathbb{N}}

\newcommand{\bbR}{\mathbb{R}}

\DeclareMathAlphabet{\mathbsf}{OT1}{cmss}{bx}{n}
\DeclareMathAlphabet{\mathssf}{OT1}{cmss}{m}{sl}% slanted sans serif

\DeclareSymbolFont{bsfletters}{OT1}{cmss}{bx}{n}
\DeclareSymbolFont{ssfletters}{OT1}{cmss}{m}{n}
\DeclareMathSymbol{\bsfGamma}{0}{bsfletters}{'000}
\DeclareMathSymbol{\ssfGamma}{0}{ssfletters}{'000}
\DeclareMathSymbol{\bsfDelta}{0}{bsfletters}{'001}
\DeclareMathSymbol{\ssfDelta}{0}{ssfletters}{'001}
\DeclareMathSymbol{\bsfTheta}{0}{bsfletters}{'002}
\DeclareMathSymbol{\ssfTheta}{0}{ssfletters}{'002}
\DeclareMathSymbol{\bsfLambda}{0}{bsfletters}{'003}
\DeclareMathSymbol{\ssfLambda}{0}{ssfletters}{'003}
\DeclareMathSymbol{\bsfXi}{0}{bsfletters}{'004}
\DeclareMathSymbol{\ssfXi}{0}{ssfletters}{'004}
\DeclareMathSymbol{\bsfPi}{0}{bsfletters}{'005}
\DeclareMathSymbol{\ssfPi}{0}{ssfletters}{'005}
\DeclareMathSymbol{\bsfSigma}{0}{bsfletters}{'006}
\DeclareMathSymbol{\ssfSigma}{0}{ssfletters}{'006}
\DeclareMathSymbol{\bsfUpsilon}{0}{bsfletters}{'007}
\DeclareMathSymbol{\ssfUpsilon}{0}{ssfletters}{'007}
\DeclareMathSymbol{\bsfPhi}{0}{bsfletters}{'010}
\DeclareMathSymbol{\ssfPhi}{0}{ssfletters}{'010}
\DeclareMathSymbol{\bsfPsi}{0}{bsfletters}{'011}
\DeclareMathSymbol{\ssfPsi}{0}{ssfletters}{'011}
\DeclareMathSymbol{\bsfOmega}{0}{bsfletters}{'012}
\DeclareMathSymbol{\ssfOmega}{0}{ssfletters}{'012}

\newcommand{\dotleq}{\stackrel{.}{\leq}}

\DeclareMathOperator*{\argmin}{arg\,min}

\DeclareMathOperator{\var}{\mathsf{Var}}

\def\e{{\rm e}}

\def\dotle{\mathrel{\dot{\le}}}
\def\dotleq{\dotle}
\def\dotge{\mathrel{\dot{\ge}}}

\def\cP{{\cal P}}

\def\cT{{\cal T}}

\def\cX{{\cal X}}
\def\cY{{\cal Y}}

\def\rd{{\rm d}}

  \providecommand{\definitionname}{Definition}
  \providecommand{\lemmaname}{Lemma}
  \providecommand{\remarkname}{Remark}
\providecommand{\theoremname}{Theorem}

\providecommand{\definitionname}{Definition}
\providecommand{\lemmaname}{Lemma}
\providecommand{\propositionname}{Proposition}
\providecommand{\remarkname}{Remark}
\providecommand{\theoremname}{Theorem}

\providecommand{\definitionname}{Definition}
\providecommand{\lemmaname}{Lemma}
\providecommand{\propositionname}{Proposition}
\providecommand{\remarkname}{Remark}
\providecommand{\theoremname}{Theorem}

\providecommand{\definitionname}{Definition}
\providecommand{\lemmaname}{Lemma}
\providecommand{\propositionname}{Proposition}
\providecommand{\remarkname}{Remark}
\providecommand{\theoremname}{Theorem}

\makeatother

\providecommand{\definitionname}{Definition}
\providecommand{\lemmaname}{Lemma}
\providecommand{\propositionname}{Proposition}
\providecommand{\remarkname}{Remark}
\providecommand{\theoremname}{Theorem}

\begin{document}

\title{R\'enyi Resolvability and Its Applications to the Wiretap Channel }

%\author{Lei Yu and Vincent Y. F. Tan, \IEEEmembership{Senior Member,~IEEE}
%\thanks{The authors are with the Department of Electrical and Computer Engineering,
%National University of Singapore (Emails: \protect\protect\protect\protect\url{leiyu@nus.edu.sg},
%\protect\protect\protect\protect\url{vtan@nus.edu.sg}). V.~Y.~F.
%Tan is also with the Department of Mathematics, National University
%of Singapore. This paper was presented in part at the 2017 International
%Conference on Information Theoretic Security (ICITS) \cite{yu2017renyi}.} }

\author{Lei Yu and Vincent Y. F. Tan, \IEEEmembership{Senior Member,~IEEE}
\thanks{
%Manuscript received December 21, 2017; revised July 15, 2018; accepted July 15, 2018.
This work was  supported by a Singapore
National Research Foundation (NRF) National Cybersecurity R\&D Grant
(R-263-000-C74-281 and NRF2015NCR-NCR003-006). The first author was
also supported by a National Natural Science Foundation of China (NSFC)
under Grant (61631017).
%This work was supported
%by the Singapore National Research Foundation (NRF) National Cybersecurity
%R\&D Grant under Grants R-263-000-C74-281 and NRF2015NCR-NCR003-006.
This paper was presented in part at the 2017 International
Conference on Information Theoretic Security (ICITS) \cite{yu2017renyi}.}
\thanks{
L.~Yu is  with the Department of Electrical and Computer Engineering,
National University of Singapore (NUS), Singapore 117583 (e-mail: leiyu@nus.edu.sg).
V.~Y.~F.~Tan is  with the  Department of Electrical and Computer Engineering and the Department of Mathematics, NUS, Singapore 119076 (e-mail: vtan@nus.edu.sg).}
\thanks{
Communicated by  M. Bloch, Associate Editor for Shannon Theory. }
\thanks{Copyright
(c) 2018 IEEE. Personal use of this material is permitted. However,
permission to use this material for any other purposes must be obtained
from the IEEE by sending a request to pubs-permissions@ieee.org.}}

\maketitle
\begin{abstract}
The conventional channel resolvability problem refers to the determination
of the minimum rate required for an input process so that the output
distribution approximates a target distribution in either the total
variation distance or the relative entropy. In contrast to previous
works, in this paper, we use the (normalized or unnormalized) R\'enyi
divergence (with the R\'enyi parameter in $[0,2]\cup\{\infty\}$) to
measure the level of approximation. We also provide asymptotic expressions
for normalized R\'enyi divergence when the R\'enyi parameter is larger
than or equal to $1$ as well as (lower and upper) bounds for the
case when the same parameter is smaller than $1$. We characterize
the R\'enyi resolvability, which is defined as the minimum rate required
to ensure that the R\'enyi divergence vanishes asymptotically. The R\'enyi
resolvabilities are the same for both the normalized and unnormalized
divergence cases. In addition, when the R\'enyi parameter smaller than~$1$,
consistent with the traditional case where the R\'enyi parameter is
equal to~$1$, the R\'enyi resolvability equals the minimum mutual
information over all input distributions that induce the target output
distribution. When the R\'enyi parameter is larger than $1$ the R\'enyi
resolvability is, in general, larger than the mutual information.
The optimal R\'enyi divergence is proven to vanish at least exponentially
fast for both of these two cases, as long as the code rate is larger
than the R\'enyi resolvability. The optimal exponential rate of decay
for i.i.d.\ random codes is also characterized exactly. We apply
these results to the wiretap channel, and completely characterize
the optimal tradeoff between the rates of the secret and non-secret
messages when the leakage measure is given by the (unnormalized) R\'enyi
divergence. This tradeoff differs from the conventional setting when
the leakage is measured by the traditional mutual information. 
\end{abstract}


\section{\label{sec:Introduction}Introduction}

How much information is required to simulate a random process through
a given channel so that it mimics a target output distribution? This
is the so-called {\em channel resolvability problem}, studied by
Han and Verd\'u \cite{Han}. In \cite{Han}, the total variation (TV)
distance and the normalized relative entropy (Kullback-Leibler divergence)
were used to measure the level of approximation. The resolvability
problem with the \emph{unnormalized} relative entropy was studied
by Hayashi \cite{Hayashi06,Hayashi11}. In \cite{Han,Hayashi06,Hayashi11}
it was shown that in the memoryless case the minimum rates of randomness
needed for simulating a channel output under the TV, normalized relative
entropy, or unnormalized relative entropy measures are the same, and
are all equal to the minimum mutual information over all input distributions
that induce the target output distribution. Recently, Liu, Cuff, and
Verd\'u \cite{Liu} extended the theory of resolvability by using $E_{\gamma}$
metric with $\gamma\geq1$ to measure the level of approximation.
The $E_{\gamma}$ metric reduces to the TV distance when $\gamma=1$,
but it is weaker than the TV distance when $\gamma>1$. Hence, the
$E_{\gamma}$ metric generalizes the TV distance by {\em weakening}
the measure. In contrast, we generalize the channel resolvability
problem by {\em strengthening} the relative entropy measure and
considering a continuum of secrecy measures indexed by the R\'enyi parameter.
 {Furthermore, random variable simulation problems
under R\'enyi divergence measures of all orders in $[0,\infty]$, including
the source resolvability problem (the resolvability problem with the
identity channel), were studied by the present authors recently in \cite{yu2018simulation}.
The exact channel resolvability problem was studied by the present authors in \cite{yu2018asymptotic},
in which the output distribution is required to be exactly equal to
the target distribution, and meanwhile, the input process is allowed
to be an ``asymptotic function'' (i.e., not restricted to be a function) of a uniform  random variable (or the input process is
allowed to be compressed by variable-length codes, not restricted to   fixed-length codes).}

While the term ``channel resolvability'' was coined by Han and Verd\'u in \cite{Han}, the problem of approximating a given product measure
was first studied by Wyner \cite{Wyner}. In \cite{Wyner} Wyner investigated
the minimum rate of common randomness to simulate two correlated sources
in a distributed fashion such that the distance (e.g., TV distance
or relative entropy) between the code-induced distribution and the
target source distribution vanishes asymptotically; this rate was
coined the common information rate between the two sources. For the
achievability part, both channel resolvability and common information
problems rely on the so-called soft-covering lemmas~\cite{Cuff}.
The channel resolvability and common information problems have several
interesting applications\textemdash including secrecy, channel synthesis,
and source coding. For example, in \cite{Bloch} it was used to study
the performance of a wiretap channel system under different secrecy
measures. In~\cite{Han14} it was used to study the reliability and
secrecy exponents of a wiretap channel with cost constraints. In \cite{Parizi}
it was used to study the exact secrecy exponents of random code ensembles
for the wiretap channel. In \cite{hou2014effective}, Hou and Kramer
used ideas from the channel resolvability problem to study the \emph{effective
secrecy capacity} (the stealth-secrecy capacity) of wiretap channels.
This work is contrasted to the present work in greater detail in Section
\ref{sec:wiretap}.   {Furthermore, the perfectly stealthy
(or covert) communication problem, in which the distribution of the
signal overheard by the eavesdropper is required to be exactly equal to the target distribution, was studied by the present authors in \cite{yu2018asymptotic}.
The exact common information problem was studied in \cite{Kumar,li2017distributed,yu2018on},
in which the code-induced distribution is required to be exactly equal
to the target source distribution, and meanwhile, the common randomness
is allowed to be compressed by variable-length codes, not restricted to  fixed-length codes.}

In contrast to the aforementioned works, we use the (normalized or
unnormalized) R\'enyi divergence to measure the level of approximation
between the simulated and target output distributions. As expounded
by Iwamoto and Shikata~\cite{iwamoto}, we can quantify equivocation
using R\'enyi measures, thus obtaining a continuum of fundamental limits
of information leakage under the effect of various hash functions.
These fundamental limits are indexed by the R\'enyi parameter. Our work
is also partly motivated by Shikata~\cite{Shikata} who quantified
lengths of secret keys in terms of R\'enyi entropies of general orders
and Bai {\em et al.}~\cite{Bai2015} who showed that the R\'enyi
divergence is particularly suited for simplifying some security proofs.
Furthermore, it is worth noting that it is quite natural to use various
divergences to measure the discrepancy between two distributions.
Wyner \cite{Wyner} and Yu and Tan \cite{yu2018wyner,yu2018corrections,yu2018on}
respectively used the KL divergence and the R\'enyi divergence to measure
the level of approximation in the distributed source synthesis problem;
Hayashi~\cite{Hayashi06,Hayashi11} used the KL divergence to study
the channel resolvability problem, and showed the optimal decay exponents
of the KL divergence and the total variation are upper bounded by
an expression involving the R\'enyi divergence. In probability theory,
Barron \cite{barron1986entropy} and Bobkov, Chistyakov and G\"otze
\cite{bobkov2016r} respectively used the KL divergence and the R\'enyi
divergence to study the central limit theorem, i.e., they used them
to measure the discrepancy between the induced distribution of sum
of i.i.d.\ random variables and the normal distribution with the
same mean and variance. Furthermore, special instances of R\'enyi entropies
and divergences\textemdash including the KL divergence, the R\'enyi
divergence, the collision entropy (the R\'enyi entropy of order $2$),
and min-entropy (the R\'enyi entropy of order $\infty$)\textemdash were
used to study various information-theoretic problems (including security,
cryptography, and quantum information) in several works in the recent
literature~\cite{Bloch,hou2014effective,beigi2014quantum,dodis2013overcoming,Hayashi17,Tan,chou2017coding,Hayashi};
and these give some operational meanings of the R\'enyi divergence.
For example, in \cite{chou2017coding}, the normalized R\'enyi entropy
of order 2 was used to express an achievable rate for the secret communication
over the wiretap channel with non-uniform sources. In \cite{Hayashi},
the R\'enyi divergence was used to express an achievable exponent for
secure multiplex coding with the leakage measured by mutual information.

\subsection{Main Contributions}

Our main contributions are as follows: 
\begin{enumerate}
\item We provide finite length and asymptotic expressions for the R\'enyi
divergence between the simulated and target output distributions.
We distinguish between the case when the R\'enyi parameter is at least
$1$\textemdash in which case we have a tight expression\textemdash and
the case when the same parameter is smaller than $1$\textemdash in
which case we only have bounds (which are tight in some regime). 
\item We characterize the R\'enyi resolvability, which is defined as the minimum
rate needed to guarantee that the (normalized or unnormalized) R\'enyi
divergence vanishes asymptotically. Interestingly, these two R\'enyi
resolvabilities are the same regardless of whether we employ the normalized
or unnormalized R\'enyi divergences. The R\'enyi resolvability when the
R\'enyi parameter is at most $1$ is just equal to the minimum mutual
information over all input distribution that induce target output
distribution. This is similar to the traditional case \cite{Han,Hayashi11,Hayashi06}.
In contrast if the R\'enyi parameter is greater than $1$, the R\'enyi
resolvability is, in general, larger than the minimum mutual information. 
\item We prove that the optimal R\'enyi divergence between the simulated and
target output distributions vanishes (at least) exponentially fast
as long as the code rate is larger than the R\'enyi resolvability (cf.\ previous
point). We also exactly characterize the optimal (ensemble tight)
exponential decay rate for the ensemble of i.i.d.\ random codes.
These results are generalizations of the work by Parizi, Telatar and
Merhav~\cite{Parizi} in which the optimal exponent (leading to an
ensemble tight secrecy exponent for the wiretap channel) for the relative
entropy was studied. See Remark \ref{rem:parizi} for further comparisons
and contrasts to~\cite{Parizi}. 
\item As a concrete application of the above mathematical results, we consider
the wiretap channel and completely characterize the optimal tradeoff
between the rates of the secret and non-secret messages when the leakage
is measured by the unnormalized R\'enyi divergence. Note that different
from Csisz\'ar and K\"orner's work (with secrecy measured by the mutual
information) \cite{Csiszar78}, the optimal rates tradeoff provided
by us are achieved by a single-layered code. Hence, it has a different
expression from the one given in~\cite{Csiszar78}. See Remark~\ref{rmk:layer}
for a detailed discussion. 
\end{enumerate}
It is also worth noting that our work is partly motivated by the work
of Hayashi and Tan \cite{Hayashi17,Tan}. In their work, the R\'enyi
divergence was used to measure the level of approximation of a distribution
induced by a \emph{hash function}, typically used for source compression;
in our work, it is used to measure the level of approximation of an
input process that is sent through a {\em channel}. Hence our work
can be considered as a counterpart of theirs, just as the \emph{channel
coding} is a counterpart of the \emph{source hashing.}

\subsection{Notation }

\label{sec:notation}

In this paper, we use $P_{X}(x)$ to denote the probability distribution
of a random variable $X$, which is also shortly denoted as $P(x)$
(when the random variable $X$ is clear from the context). We also
use $P_{X}$, $\widetilde{P}_{X}$, and $Q_{X}$ to denote various
probability distributions with alphabet $\mathcal{X}$. All alphabets
considered in the sequel are finite. The set of probability distributions
on $\mathcal{X}$ is denoted as $\mathcal{P}\left(\mathcal{X}\right)$,
and the set of conditional probability distributions on $\mathcal{Y}$
given a variable in $\mathcal{X}$ is denoted as $\mathcal{P}\left(\mathcal{Y}|\mathcal{X}\right):=\left\{ P_{Y|X}:P_{Y|X}\left(\cdot|x\right)\in\mathcal{P}\left(\mathcal{Y}\right),x\in\mathcal{X}\right\} $.
Given $P_{X}$ and $P_{Y|X}$, we write $[P_{Y|X}\circ P_{X}](y):=\sum_{x}P_{Y|X}(y|x)P_{X}(x)$.

We use $T_{x^{n}}\left(x\right):=\frac{1}{n}\sum_{i=1}^{n}1\left\{ x_{i}=x\right\} $
to denote the type (empirical distribution) of a sequence $x^{n}$,
$T_{X}$ and $V_{Y|X}$ to respectively denote a type of sequences
in $\mathcal{X}^{n}$ and a conditional type of sequences in $\mathcal{Y}^{n}$
(given a sequence $x^{n}\in\calX^{n}$). For a type $T_{X}$, the
type class (set of sequences having the same type $T_{X}$) is denoted
by $\mathcal{T}_{T_{X}}$. For a conditional type $V_{Y|X}$ and a
sequence $x^{n}$, the \emph{$V$-shell of $x^{n}$} (the set of $y^{n}$
sequences having the same conditional type $V_{Y|X}$ given $x^{n}$)
is denoted by $\mathcal{T}_{V_{Y|X}}\left(x^{n}\right)$. The set
of types of sequences in $\mathcal{X}^{n}$ is denoted as $\mathcal{P}^{\left(n\right)}\left(\mathcal{X}\right):=\left\{ T_{x^{n}}:x^{n}\in\mathcal{X}^{n}\right\} $.
The set of conditional types of sequences in $\mathcal{Y}^{n}$ given
a sequence in $\mathcal{X}^{n}$ with the type $T_{X}$ is denoted
as $\mathcal{P}^{\left(n\right)}\left(\mathcal{Y}|T_{X}\right):=\{V_{Y|X}\in\mathcal{P}\left(\mathcal{Y}|\mathcal{X}\right):V_{Y|X}\times T_{X}\in\mathcal{P}^{\left(n\right)}\left(\mathcal{X}\times\mathcal{Y}\right)\}$.
For brevity, sometimes we use $T\left(x,y\right)$ to denote the joint
distributions $T\left(x\right)V\left(y|x\right)$ or $T\left(y\right)V\left(x|y\right)$.

The $\epsilon$-typical set relative to $Q_{X}$ is denoted as $\mathcal{T}_{\epsilon}^{n}\left(Q_{X}\right):=\left\{ x^{n}\in\mathcal{X}^{n}:\left|T_{x^{n}}\left(x\right)-Q_{X}\left(x\right)\right|\leq\epsilon Q_{X}\left(x\right),\forall x\in\mathcal{X}\right\} $.
The conditionally $\epsilon$-typical set relative to $Q_{XY}$ is
denoted as $\mathcal{T}_{\epsilon}^{n}\left(Q_{XY}|x^{n}\right):=\left\{ y^{n}\in\mathcal{X}^{n}:\left(x^{n},y^{n}\right)\in\mathcal{T}_{\epsilon}^{n}\left(Q_{XY}\right)\right\} $.
For brevity, we sometimes write $\mathcal{T}_{\epsilon}^{n}\left(Q_{X}\right)$
and $\mathcal{T}_{\epsilon}^{n}\left(Q_{XY}|x^{n}\right)$ as $\mathcal{T}_{\epsilon}^{n}$
and $\mathcal{T}_{\epsilon}^{n}\left(x^{n}\right)$ respectively.
Other notation generally follows the book by Csisz\'ar and K\"orner~\cite{Csi97}.

The total variation distance between two probability mass functions
$P$ and $Q$ with a common alphabet $\calX$ is defined by 
\begin{equation}
|P-Q|:=\frac{1}{2}\sum_{x\in\calX}|P(x)-Q(x)|.
\end{equation}
By the definition of $\epsilon$-typical set, we have that for any
$x^{n}\in\mathcal{T}_{\epsilon}^{n}\left(Q_{X}\right)$, $\left|T_{x^{n}}-Q_{X}\right|\leq\frac{\epsilon}{2}$.

Fix distributions $P_{X},Q_{X}\in\calP(\calX)$. Then the {\em relative
entropy} and the {\em R\'enyi divergence of order $1+s\in(0,1)\cup(1,\infty)$}
are respectively defined as 
\begin{align}
D(P_{X}\|Q_{X}) & :=\sum_{x\in\calX}P_{X}(x)\log\frac{P_{X}(x)}{Q_{X}(x)},\quad\mbox{and}\\*
D_{1+s}(P_{X}\|Q_{X}) & :=\frac{1}{s}\log\sum_{x\in\calX}P_{X}(x)^{1+s}Q_{X}(x)^{-s},
\end{align}
and the conditional versions are respectively defined as 
\begin{align}
D(P_{Y|X}\|Q_{Y|X}|P_{X}) & :=D(P_{X}P_{Y|X}\|P_{X}Q_{Y|X})\\*
D_{1+s}(P_{Y|X}\|Q_{Y|X}|P_{X}) & :=D_{1+s}(P_{X}P_{Y|X}\|P_{X}Q_{Y|X}),
\end{align}
where throughout, $\log$ is to the natural base $\rme$ and $s\geq-1$.The
R\'enyi divergences of order $0,1,$ and $\infty$ are respectively
defined as 
\begin{align}
D_{0}(P_{X}\|Q_{X}) & :=\lim_{s\downarrow-1}D_{1+s}(P_{X}\|Q_{X})\\
 & =-\log\{Q_{X}(P_{X}>0)\};\\
D_{1}(P_{X}\|Q_{X}) & :=\lim_{s\to0}D_{1+s}(P_{X}\|Q_{X})\\
 & =D(P_{X}\|Q_{X});\\
D_{\infty}(P_{X}\|Q_{X}) & :=\lim_{s\to\infty}D_{1+s}(P_{X}\|Q_{X})\\
 & =\log\sup_{x}\frac{P_{X}(x)}{Q_{X}(x)}.
\end{align}
 Hence a special case of the R\'enyi divergence is the usual relative
entropy. 

Finally, we write $f(n)\dotle g(n)$ if $\limsup_{n\to\infty}\frac{1}{n}\log\frac{f(n)}{g(n)}\le0$.
In addition, $f(n)\doteq g(n)$ means $f(n)\dotle g(n)$ and $g(n)\dotle f(n)$.
We use $o(1),\delta_{n},\delta_{n}',\delta_{n}''$ to denote generic
sequences tending to zero as $n\rightarrow\infty$. For $a\in\bbR$,
$[a]^{+}:=\max\{a,0\}$ denotes positive clipping.

\subsection{Problem Formulation}

\label{subsec:problem} 
\begin{figure}[t]
\centering \includegraphics[width=0.7\columnwidth]{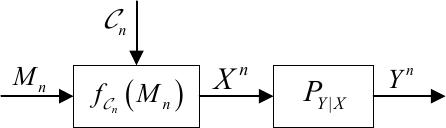}\caption{Channel resolvability problem: $\mathcal{C}_{n}$ is independent of
the message $M_{n}\in\calM_{n}$, and $f_{\mathcal{C}_{n}}$ is a
random function (induced by $\mathcal{C}_{n}$). }
\label{fig:Resolvability} 
\end{figure}

We consider the channel resolvability problem
illustrated in Fig.~\ref{fig:Resolvability}. Given a channel $P_{Y|X}$
and a target distribution $Q_{Y}$, we wish to minimize the alphabet
size of a random variable $M_{n}$ that is uniformly distributed over\footnote{For simplicity, we assume that $\e^{nR}$ and similar expressions
(such as $\e^{R}$) are integers.} $\calM_{n}:=\{1,\ldots,\e^{nR}\}$ ($R$ is a positive number known
as the {\em rate}), such that given common randomness $\mathcal{C}_{n}$,
the output distribution 
\begin{equation}
P_{Y^{n}|\mathcal{C}_{n}}\left(y^{n}|c_{n}\right):=\frac{1}{|{\cal M}_{n}|}\sum_{m\in{\cal M}_{n}}\prod_{i=1}^{n}P_{Y|X}\left(y_{i}|f_{c_{n},i}\left(m\right)\right)\label{eq:-113}
\end{equation}
forms a good approximation to the product distribution $Q_{Y}^{n}$.
Here $\mathcal{C}_{n}$ is a random variable independent of the random
variable $M_{n}$. If we set $\mathcal{C}_{n}=\left\{ X^{n}\left(m\right)\right\} _{m\in\calM_{n}}$
with $X^{n}(m)\sim P_{X^{n}}$ for all $m\in\calM_{n}$, and set $f_{\mathcal{C}_{n}}(m)=X^{n}(m)$,
then the random mapping is known as a {\em conventional random code}.
If the input distribution is i.i.d., i.e., $P_{X^{n}}=P_{X}^{n}$,
then it is known as an {\em i.i.d.\ random code}. In contrast
to previous works on the channel resolvability problem~\cite{Han},
here we employ the R\'enyi divergence 
\begin{equation}
D_{1+s}(P_{Y^{n}\mathcal{C}_{n}}\|Q_{Y}^{n}P_{\mathcal{C}_{n}})
\end{equation}
to measure the discrepancy between $P_{Y^{n}}$ and $Q_{Y}^{n}$.

Observe that 
\begin{align}
 & \e^{sD_{1+s}(P_{Y^{n}\mathcal{C}_{n}}\|Q_{Y}^{n}P_{\mathcal{C}_{n}})}\nonumber \\
 & =\mathbb{E}_{\mathcal{C}_{n}}\biggl[\sum_{y^{n}}\sum_{m}P(m)P(y^{n}|f_{\mathcal{C}_{n}}(m))\nonumber \\
 & \qquad\times\bigg(\frac{\sum_{m}P(m)P(y^{n}|f_{\mathcal{C}_{n}}(m))}{Q(y^{n})}\bigg)^{s}\biggr].
\end{align}
Hence to guarantee that $D_{1+s}(P_{Y^{n}\mathcal{C}_{n}}\|Q_{Y}^{n}P_{\mathcal{C}_{n}})$
is finite for $s\geq0$, we assume $P_{Y|X=x}\ll Q_{Y}$ for all $x\in\mathcal{X}$;
otherwise, we can remove all the values $x$ such that $P_{Y|X=x}\not\ll Q_{Y}$
from $\mathcal{X}$. However, it is worth noting that we do not need
to do so for $-1\leq s<0$, since $D_{1+s}(P_{Y^{n}\mathcal{C}_{n}}\|Q_{Y}^{n}P_{\mathcal{C}_{n}})$
is always finite regardless of whether $P_{Y|X=x}\ll Q_{Y}$ for all
$x\in\mathcal{X}$ or $P_{Y|X=x}\not\ll Q_{Y}$ for some $x\in\mathcal{X}$.
Furthermore, for simplicity, for the case $s=-1$ we assume $P_{Y|X=x}\gg Q_{Y}$
for some $x\in\mathcal{X}$.\footnote{Note that this condition is missing  in the conference version \cite{yu2017renyi}.}
Hence $D_{0}(P_{Y^{n}\mathcal{C}_{n}}\|Q_{Y}^{n}P_{\mathcal{C}_{n}})=0$
if the channel input is fixed to $x^{n}$.

Traditionally, the code $\mathcal{C}_{n}$ is deterministic and so
the measure $D_{1+s}(P_{Y^{n}}\|Q_{Y}^{n})$ is analyzed. However,
in our setting, especially in Section \ref{subsec:Exponential-Behavior},
we are interested in questions concerning the ensemble performance
of random codes $\mathcal{C}_{n}$. Hence, we analyze the discrepancy
measure $D_{1+s}(P_{Y^{n}\mathcal{C}_{n}}\|Q_{Y}^{n}P_{\mathcal{C}_{n}})$,
which represents the conditional R\'enyi divergence between the simulated
and target distributions given the random code $\mathcal{C}_{n}$.
Besides, we are also interested in another related discrepancy measure\footnote{Here we would like to thank Prof.\ Masahito Hayashi for inspiring
us to consider the measure $\mathbb{E}_{\mathcal{C}_{n}}\left[D_{1+s}(P_{Y^{n}|\mathcal{C}_{n}}\|Q_{Y}^{n})\right]$. }  $\mathbb{E}_{\mathcal{C}_{n}}\left[D_{1+s}(P_{Y^{n}|\mathcal{C}_{n}}\|Q_{Y}^{n})\right]$.
Since for $s=0$, 
\begin{align}
\mathbb{E}_{\mathcal{C}_{n}}\left[D_{1+s}(P_{Y^{n}|\mathcal{C}_{n}}\|Q_{Y}^{n})\right] & =D_{1+s}(P_{Y^{n}\mathcal{C}_{n}}\|Q_{Y}^{n}P_{\mathcal{C}_{n}})\\
 & =D(P_{Y^{n}\mathcal{C}_{n}}\|Q_{Y}^{n}P_{\mathcal{C}_{n}}),
\end{align}
%both $\mathbb{E}_{\mathcal{C}_{n}}\left[D_{1+s}(P_{Y^{n}|\mathcal{C}_{n}}\|Q_{Y}^{n})\right]$
these measures are consistent with the one used in %and $D_{1+s}(P_{Y^{n}\mathcal{C}_{n}}\|Q_{Y}^{n}P_{\mathcal{C}_{n}})$
%are equal to $D(P_{Y^{n}\mathcal{C}_{n}}\|Q_{Y}^{n}P_{\mathcal{C}_{n}})$
%which was used in 
Parizi et al.'s paper \cite{Parizi} (which is $D(P_{Y^{n}\mathcal{C}_{n}}\|Q_{Y}^{n}P_{\mathcal{C}_{n}})$).
Furthermore, the three measures above satisfy the relationship 
\begin{align}
\inf_{f}D_{1+s}(P_{Y^{n}}\|Q_{Y}^{n}) & \leq\inf_{f_{\mathcal{C}_{n}}}\mathbb{E}_{\mathcal{C}_{n}}\left[D_{1+s}(P_{Y^{n}|\mathcal{C}_{n}}\|Q_{Y}^{n})\right]\\
 & \leq\inf_{f_{\mathcal{C}_{n}}}D_{1+s}(P_{Y^{n}\mathcal{C}_{n}}\|Q_{Y}^{n}P_{\mathcal{C}_{n}}).
\end{align}
In fact, for the achievability parts, we bound $\inf_{f_{\mathcal{C}_{n}}}D_{1+s}(P_{Y^{n}\mathcal{C}_{n}}\|Q_{Y}^{n}P_{\mathcal{C}_{n}})$
from above and for the converse parts, we bound $\inf_{f}D_{1+s}(P_{Y^{n}}\|Q_{Y}^{n})$
from below. This implies, by the chain of inequalities above, that
our results in this paper hold for all these three measures. 

\section{Main Results}

% for R\'enyi Resolvability Problem

\subsection{One-Shot/Finite Blocklength Bounds }

\label{sec:one-shot} We first consider the one-shot (i.e., blocklength
$n$ equal to 1) or finite blocklength version of the problem. For
this case, we provide several bounds in the following two lemmas,
the proofs of which are given in Appendix \ref{sec:OneShot}. These
one-shot bounds will be used to derive asymptotic results in the next
subsection. We believe that similar techniques used to obtain these
bounds can be employed to derive second-order results, just as in~\cite{Hayashi17,Tan}. 
\begin{lem}[One-Shot Bounds for Direct Part]
\label{lem:oneshotach} Consider a random mapping $f_{\mathcal{C}}:\calM=\{1,\ldots,\e^{R}\}\rightarrow\calX$.
We set $\mathcal{C}=\left\{ X\left(m\right)\right\} _{m\in\calM}$
with $X\left(m\right),m\in\calM$ drawn independently for different
$m$'s and according to a same distribution $P_{X}$, and set $f_{\mathcal{C}}\left(m\right)=X\left(m\right)$.
This forms a random code. For this random code, we have for $s\in[0,1]$
and any distribution $Q_{Y}$, 
\begin{align}
 & \e^{sD_{1+s}(P_{Y\mathcal{C}}\|Q_{Y}P_{\mathcal{C}})}\nonumber \\
 & \leq\e^{sD_{1+s}\left(P_{XY}\|P_{X}Q_{Y}\right)-sR}+\e^{sD_{1+s}(P_{Y}\|Q_{Y})}\label{eq:-123}\\
 & \leq2\e^{s\Gamma_{1+s}\left(P_{X},P_{Y|X},Q_{Y},R\right)},
\end{align}
where 
\begin{align}
 & \Gamma_{1+s}\left(P_{X},P_{Y|X},Q_{Y},R\right)\nonumber \\
 & :=\max\left\{ D_{1+s}\left(P_{XY}\|P_{X}Q_{Y}\right)-R,D_{1+s}(P_{Y}\|Q_{Y})\right\} .\label{eq:-97}
\end{align}
In the other direction with $s\in[0,1)$, we have for any distribution
$Q_{Y}$, 
\begin{align}
 & \e^{-sD_{1-s}(P_{Y\mathcal{C}}\|Q_{Y}P_{\mathcal{C}})}\nonumber \\
 & \geq2^{-s}\biggl[\e^{sR}\sum_{x,y}P\left(x\right)P^{1-s}\left(y|x\right)Q^{s}\left(y\right)1\left\{ \frac{P\left(y|x\right)}{P\left(y\right)}\geq\e^{R}\right\} \nonumber \\
 & \qquad+\sum_{x,y}P\left(x\right)P\left(y|x\right)P^{-s}\left(y\right)Q^{s}\left(y\right)1\left\{ \frac{P\left(y|x\right)}{P\left(y\right)}<\e^{R}\right\} \biggr].
\end{align}
\end{lem}
\begin{rem}
A similar result to \eqref{eq:-123} was shown by Hayashi and Matsumoto
\cite[Thm.~14]{Hayashi}, but their result is a special case of ours
with the setting $Q_{Y}=P_{Y}$. 
\end{rem}
\begin{rem}
Since in the proof we require Lemma \ref{lem:norm} (see \eqref{eq:-150})
and the fact that $x\mapsto x^{s}$ is a concave function (see \eqref{eq:-42-2}),
the proof does not apply to the case in which the R\'enyi divergence
is of order $>2$. 
\end{rem}
\begin{lem}[One-Shot Bounds for Converse Part]
\label{lem:oneshotcon} For any deterministic mapping $f:\calM=\{1,\ldots,\e^{R}\}\rightarrow\calX$
and any $s\in[0,\infty]$, we have for any distribution $Q_{Y}$,
{} 
\begin{align}
\e^{sD_{1+s}(P_{Y}\|Q_{Y})} & \geq\e^{s\Gamma_{1+s}\left(P_{X},P_{Y|X},Q_{Y},R\right)}\label{eq:-146}
\end{align}
where 
\begin{align}
P\left(x\right) & :=\sum_{m}P\left(m\right)1\left\{ f\left(m\right)=x\right\} \label{eq:-37}\\
P\left(y\right) & :=\sum_{x}P\left(x\right)P\left(y|x\right)\label{eq:-38}
\end{align}
respectively denote the distributions of $X$ and $Y$ induced by
the mapping $f$, and $\Gamma_{1+s}\left(P_{X},P_{Y|X},Q_{Y},R\right)$
is given by \eqref{eq:-97}. In the other direction with $s\in[0,1)$,
we have for any distribution $Q_{Y}$, {} 
\begin{align}
 & \e^{-sD_{1-s}(P_{Y}\|Q_{Y})}\nonumber \\
 & \leq\e^{sR}\sum_{x,y}P\left(x\right)P^{1-s}\left(y|x\right)Q^{s}\left(y\right)1\left\{ \frac{P\left(y|x\right)}{P\left(y\right)}\geq\frac{\e^{R}}{2}\right\} \nonumber \\
 & \qquad+\sum_{x,y}P\left(x\right)P\left(y|x\right)P^{-s}\left(y\right)Q^{s}\left(y\right)1\left\{ \frac{P\left(y|x\right)}{P\left(y\right)}<\frac{\e^{R}}{2}\right\} \label{eq:-147}
\end{align}
where $P_{X}$ and $P_{Y}$ are given in \eqref{eq:-37}-\eqref{eq:-38}.
\end{lem}
\begin{rem}
Note that the direct and converse parts for the $1+s$ case only differ
by a factor of $2$. Similarly, the direct and converse parts for
the $1-s$ case differ by a factor of $2^{-s}$ and $R$ is replaced
by $R-\log2$. 
\end{rem}
\begin{rem}
For any random mapping $f_{\mathcal{C}}:\calM=\{1,\ldots,\e^{R}\}\rightarrow\calX$,
observe that for $s\in[0,\infty]$, 
\begin{align}
\e^{sD_{1+s}(P_{Y\mathcal{C}}\|Q_{Y}\times P_{\mathcal{C}})} & =\bbE_{\mathcal{C}}\sum_{y}P^{1+s}\left(y\right)Q^{s}\left(y\right)\\
 & \geq\min_{c}\sum_{y}P^{1+s}\left(y|c\right)Q^{s}\left(y\right)\\
 & =\min_{c}\e^{sD_{1+s}(P_{Y|\mathcal{C}=c}\|Q_{Y})},
\end{align}
and for $s\in[0,1)$, 
\begin{align}
\e^{-sD_{1-s}(P_{Y\mathcal{C}}\|Q_{Y}\times P_{\mathcal{C}})} & \leq\max_{c}\e^{-sD_{1-s}(P_{Y|\mathcal{C}=c}\|Q_{Y})}.
\end{align}
Therefore, the one-shot bounds in \eqref{eq:-146} and \eqref{eq:-147}
still hold for any random codes.  
\end{rem}
\begin{rem}
\label{rem:By-checking-our}By checking our proofs, it can be seen
that Lemmas \ref{lem:oneshotach} and \ref{lem:oneshotcon} hold not
only for channels with finite (input and output) alphabets, but also
for channels with countably infinite or continuous alphabets (e.g.,
Gaussian channels).
\end{rem}

\subsection{Asymptotic Expressions }

We now consider the asymptotics of the R\'enyi divergence as the blocklength
$n$ tends to infinity. The one-shot bounds can be used to prove the
following theorem, in which the asymptotics of the R\'enyi divergences
are characterized by multi-letter expressions. The proof of this theorem
is provided in Appendix \ref{sec:Proof-of-Theorem-asym}. 
\begin{prop}[Multi-letter Characterization]
\label{thm:multiletter} For any $s\in[0,1]$, we have 
\begin{align}
\frac{1}{n}\inf_{f_{\mathcal{C}_{n}}}D_{1+s}(P_{Y^{n}\mathcal{C}_{n}}\|Q_{Y}^{n}P_{\mathcal{C}_{n}}) & =\Gamma_{1+s}^{\left(n\right)}\left(P_{Y|X},Q_{Y},R\right)+o\left(1\right),\label{eq:multiletter}
\end{align}
where 
\begin{align}
 & \Gamma_{1+s}^{\left(n\right)}\left(P_{Y|X},Q_{Y},R\right):=\inf_{P_{X^{n}}}\max\biggl\{\nonumber \\
 & \;\frac{1}{n}D_{1+s}\left(P_{X^{n}Y^{n}}\|P_{X^{n}}Q_{Y}^{n}\right)-R,\frac{1}{n}D_{1+s}(P_{Y^{n}}\|Q_{Y}^{n})\biggr\}.\label{eq:Gamma}
\end{align}
Furthermore, for any $s\in(0,1)$, and any fixed positive integer
$k$, we have 
\begin{align}
 & \Gamma_{1-s}^{\left(n\right)}\left(P_{Y|X},Q_{Y},R\right)+o\left(1\right)\nonumber \\
 & \leq\frac{1}{n}\inf_{f_{\mathcal{C}_{n}}}D_{1-s}(P_{Y^{n}\mathcal{C}_{n}}\|Q_{Y}^{n}P_{\mathcal{C}_{n}})\label{eq:-2a}\\
 & \leq\Gamma_{1-s}^{\left(k\right)}\left(P_{Y|X},Q_{Y},R\right)+o\left(1\right),\label{eq:-2}
\end{align}
where in \eqref{eq:-2}, $o\left(1\right)$ is a term depending on
both $n$ and $k$, and vanishing as $n\to\infty$ for any fixed $k$,
and 
\begin{align}
 & \Gamma_{1-s}^{\left(n\right)}\left(P_{Y|X},Q_{Y},R\right):=\inf_{P_{X^{n}}}\max_{t\in\left[0,s\right]}\biggl\{-\frac{t}{s}R\nonumber \\
 & -\frac{1}{ns}\log\sum_{x^{n},y^{n}}P\left(x^{n},y^{n}\right)P^{-t}\left(y^{n}|x^{n}\right)P^{t-s}\left(y^{n}\right)Q^{s}\left(y^{n}\right)\biggr\}.\label{eq:Gamma2}
\end{align}
The infima in \eqref{eq:multiletter} and \eqref{eq:-2} are achieved
by a sequence of random codes described in Lemma \ref{lem:oneshotach}. 
\end{prop}
\begin{rem}
\label{rem:More-explicitly,-the}The converse part in \eqref{eq:multiletter}
also holds for $s\in(1,\infty]$. That is, for any $s\in(1,\infty]$,
\begin{align}
\frac{1}{n}\inf_{f_{\mathcal{C}_{n}}}D_{1+s}(P_{Y^{n}\mathcal{C}_{n}}\|Q_{Y}^{n}P_{\mathcal{C}_{n}}) & \geq\Gamma_{1+s}^{\left(n\right)}\left(P_{Y|X},Q_{Y},R\right)+o\left(1\right).\label{eqn:C1s-4-2-1}
\end{align}
\end{rem}
\begin{rem}
Note that in \eqref{eq:-2a} and~\eqref{eq:-2}, the lower bound
and the upper bounds differ only in the parameter of $\Gamma_{1-s}^{\left(\cdot\right)}$. 
\end{rem}
\begin{rem}
Proposition \ref{thm:multiletter} holds even when the alphabets are
not necessarily discrete. 
\end{rem}
\begin{rem}
From the definition of $\Gamma_{1+s}^{\left(n\right)}\left(P_{Y|X},Q_{Y},R\right)$,
we have 
\begin{align}
 & \Gamma_{1+s}^{\left(n\right)}\left(P_{Y|X},Q_{Y},R\right)=\inf_{P_{X^{n}}}\max_{t\in\left[0,s\right]}\biggl\{-\frac{t}{s}R\nonumber \\
 & +\frac{1}{ns}\log\sum_{x^{n},y^{n}}P\left(x^{n},y^{n}\right)P^{t}\left(y^{n}|x^{n}\right)P^{s-t}\left(y^{n}\right)Q^{-s}\left(y^{n}\right)\biggr\}.\label{eq:-57}
\end{align}
Therefore, the notations $\Gamma_{1+s}^{\left(n\right)}$ and $\Gamma_{1-s}^{\left(n\right)}$
are consistent in the sense that if we set $s$ to be $-s$ in $\Gamma_{1+s}^{\left(n\right)}$,
we obtain $\Gamma_{1-s}^{\left(n\right)}$. That is to say, $\Gamma_{1+s}^{\left(n\right)}$
for $s\in[0,\infty]$ and $\Gamma_{1-s}^{\left(n\right)}$ for $s\in(0,1)$
can be unified as in \eqref{eq:-57} for $s\in(-1,\infty]$. 
\end{rem}
Next, the asymptotics of the R\'enyi divergence is characterized by
single-letter expressions. We have an exact/tight result when the
R\'enyi parameter $\in[1,2]$ and upper and lower bounds when the R\'enyi
parameter $\in(0,1)$. This result is proved in Appendix \ref{sec:Proof-of-Theorem-asym-1}. 
\begin{thm}[Asymptotics of R\'enyi Divergence]
\label{thm:singleletter} For any $s\in[0,1]$, we have 
\begin{align}
 & \lim_{n\to\infty}\frac{1}{n}\inf_{f_{\mathcal{C}_{n}}}D_{1+s}(P_{Y^{n}\mathcal{C}_{n}}\|Q_{Y}^{n}P_{\mathcal{C}_{n}})\nonumber \\
 & =\min_{\widetilde{P}_{X}}\max\biggl\{\sum_{x}\widetilde{P}_{X}\left(x\right)D_{1+s}\left(P_{Y|X}\left(\cdot|x\right)\|Q_{Y}\right)-R,\nonumber \\
 & \qquad\qquad\qquad\max_{\widetilde{P}_{Y|X}}\eta_{1+s}\left(P_{Y|X},Q_{Y},\widetilde{P}_{X},\widetilde{P}_{Y|X}\right)\biggr\},\label{eq:-13}
\end{align}
where 
\begin{align}
 & \eta_{1+s}\left(P_{Y|X},Q_{Y},\widetilde{P}_{X},\widetilde{P}_{Y|X}\right)\nonumber \\
 & :=\left(-\frac{1}{s}-1\right)D\left(\widetilde{P}_{Y|X}\|P_{Y|X}|\widetilde{P}_{X}\right)+D\left(\widetilde{P}_{Y}\|Q_{Y}\right).
\end{align}
For any $s\in(0,1)$, we have 
\begin{align}
 & \Gamma_{1-s}^{\mathsf{LB}}\left(P_{Y|X},Q_{Y},R\right)\nonumber \\
 & \leq\liminf_{n\to\infty}\frac{1}{n}\inf_{f_{\mathcal{C}_{n}}}D_{1-s}(P_{Y^{n}\mathcal{C}_{n}}\|Q_{Y}^{n}P_{\mathcal{C}_{n}})\\
 & \leq\limsup_{n\to\infty}\frac{1}{n}\inf_{f_{\mathcal{C}_{n}}}D_{1-s}(P_{Y^{n}\mathcal{C}_{n}}\|Q_{Y}^{n}P_{\mathcal{C}_{n}})\\
 & \leq\Gamma_{1-s}^{\mathsf{UB}}\left(P_{Y|X},Q_{Y},R\right),\label{eq:-5}
\end{align}
where 
\begin{align}
 & \Gamma_{1-s}^{\mathsf{LB}}\left(P_{Y|X},Q_{Y},R\right)\nonumber \\
 & :=\min_{\widetilde{P}_{X},\widetilde{P}_{Y|X}}\max\biggl\{\left(\frac{1}{s}\!-\!1\right)D\left(\widetilde{P}_{Y|X}\|P_{Y|X}|\widetilde{P}_{X}\right)\nonumber \\
 & \qquad\qquad+\!D\left(\widetilde{P}_{Y|X}\|Q_{Y}|\widetilde{P}_{X}\right)\!-\!R,\nonumber \\
 & \qquad\left(\frac{1}{s}-1\right)D\left(\widetilde{P}_{Y|X}\|P_{Y|X}|\widetilde{P}_{X}\right)+D\left(\widetilde{P}_{Y}\|Q_{Y}\right)\biggr\},\label{eq:}\\
 & \Gamma_{1-s}^{\mathsf{UB}}\left(P_{Y|X},Q_{Y},R\right)\nonumber \\
 & :=\min_{\widetilde{P}_{X},\widetilde{P}_{Y|X}}\max\biggl\{\left(\frac{1}{s}\!-\!1\right)D\left(\widetilde{P}_{Y|X}\|P_{Y|X}|\widetilde{P}_{X}\right)\nonumber \\
 & \qquad\qquad+\!D\left(\widetilde{P}_{Y|X}\|Q_{Y}|\widetilde{P}_{X}\right)\!-\!R,\nonumber \\
 & \qquad\frac{1}{s}D\left(\widetilde{P}_{Y|X}\|P_{Y|X}|\widetilde{P}_{X}\right)+D\left(\widetilde{P}_{Y}\|Q_{Y}\right)\nonumber \\
 & \qquad\qquad-\min_{\widehat{P}_{Y|X}:\widehat{P}_{Y|X}\circ\widetilde{P}_{X}=\widetilde{P}_{Y|X}\circ\widetilde{P}_{X}}D\left(\widehat{P}_{Y|X}\|P_{Y|X}|\widetilde{P}_{X}\right)\biggr\}.\label{eq:-1}
\end{align}
We also have 
\begin{align}
\lim_{n\to\infty}\frac{1}{n}\inf_{f_{\mathcal{C}_{n}}}D_{0}(P_{Y^{n}\mathcal{C}_{n}}\|Q_{Y^{n}}P_{\mathcal{C}_{n}}) & =0.
\end{align}
Furthermore, the infima in \eqref{eq:-13} and $\Gamma_{1-s}^{\mathsf{UB}}\left(P_{Y|X},Q_{Y},R\right)$
are achieved by a sequence of constant composition codes. 
\end{thm}
\begin{rem}
\label{rem:Similar-to-Remark}Similar to Remark \ref{rem:More-explicitly,-the},
the converse part in \eqref{eq:-13} also holds for $s\in(1,\infty]$.
That is, for any $s\in(1,\infty]$, 
\begin{align}
 & \liminf_{n\to\infty}\frac{1}{n}\inf_{f_{\mathcal{C}_{n}}}D_{1+s}(P_{Y^{n}\mathcal{C}_{n}}\|Q_{Y}^{n}P_{\mathcal{C}_{n}})\nonumber \\
 & \geq\min_{\widetilde{P}_{X}}\max\biggl\{\sum_{x}\widetilde{P}_{X}\left(x\right)D_{1+s}\left(P_{Y|X}\left(\cdot|x\right)\|Q_{Y}\right)-R,\nonumber \\
 & \qquad\qquad\qquad\max_{\widetilde{P}_{Y|X}}\eta_{1+s}\left(P_{Y|X},Q_{Y},\widetilde{P}_{X},\widetilde{P}_{Y|X}\right)\biggr\}.
\end{align}
\end{rem}
\begin{rem}
The expression in \eqref{eq:-13} for $s\in[0,1]$ and $\Gamma_{1-s}^{\mathsf{LB}}\left(P_{Y|X},Q_{Y},R\right)$
or $\Gamma_{1-s}^{\mathsf{UB}}\left(P_{Y|X},Q_{Y},R\right)$ for $s\in(-1,0)$
may appear to be inconsistent; however, this is not true. It can be
easily shown that 
\begin{align}
 & \sum_{x}\widetilde{P}_{X}\left(x\right)D_{1+s}\left(P_{Y|X}\left(\cdot|x\right)\|Q_{Y}\right)\nonumber \\
 & =\max_{\widetilde{P}_{Y|X}}\biggl\{\left(-\frac{1}{s}-1\right)D\left(\widetilde{P}_{Y|X}\|P_{Y|X}|\widetilde{P}_{X}\right)\nonumber \\
 & \qquad\qquad+D\left(\widetilde{P}_{Y|X}\|Q_{Y}|\widetilde{P}_{X}\right)\biggr\}.
\end{align}
Hence we can rewrite \eqref{eq:-13} as 
\begin{align}
 & \lim_{n\to\infty}\frac{1}{n}\inf_{f_{\mathcal{C}_{n}}}D_{1+s}(P_{Y^{n}\mathcal{C}_{n}}\|Q_{Y^{n}}P_{\mathcal{C}_{n}})\nonumber \\
 & =\min_{\widetilde{P}_{X}}\max_{\widetilde{P}_{Y|X}}\max\biggl\{\left(-\!\frac{1}{s}\!-\!1\right)D\left(\widetilde{P}_{Y|X}\|P_{Y|X}|\widetilde{P}_{X}\right)\!\nonumber \\
 & \qquad+\!D\left(\widetilde{P}_{Y|X}\|Q_{Y}|\widetilde{P}_{X}\right)\!-\!R,\nonumber \\
 & \quad\left(-\frac{1}{s}-1\right)D\left(\widetilde{P}_{Y|X}\|P_{Y|X}|\widetilde{P}_{X}\right)+D\left(\widetilde{P}_{Y}\|Q_{Y}\right)\biggr\}.
\end{align}
In other words, the expression in~\eqref{eq:-13} for $s\in[0,1]$
is consistent with $\Gamma_{1-s}^{\mathsf{LB}}\left(P_{Y|X},Q_{Y},R\right)$
for $s\in(-1,0)$. 
\end{rem}
Note that $\Gamma_{1-s}^{\mathsf{UB}}\left(P_{Y|X},Q_{Y},R\right)$
and $\Gamma_{1-s}^{\mathsf{LB}}\left(P_{Y|X},Q_{Y},R\right)$ differ
only in the second term in the maximization. Moreover, when $R$ is
large enough, they are both equal to zero; see Theorem \ref{thm:Resolvability}
in the next subsection.

\begin{figure}[t]
\centering \includegraphics[width=1\columnwidth]{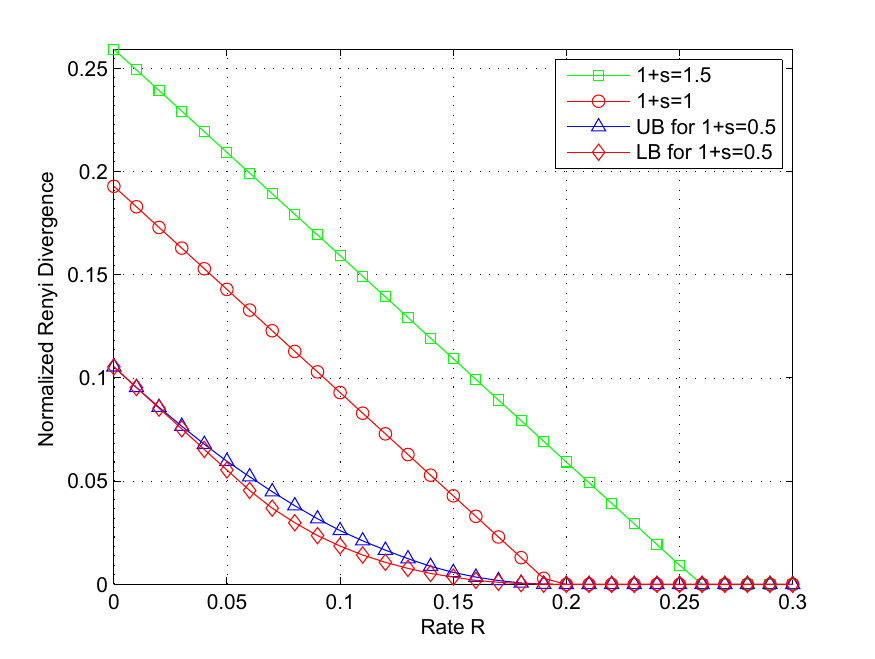}\caption{Illustration of the R\'enyi divergence measure $\frac{1}{n}\inf_{f_{\mathcal{C}_{n}}}D_{1+s}(P_{Y^{n}\mathcal{C}_{n}}\|Q_{Y}^{n}P_{\mathcal{C}_{n}})$
for $s\in[0,1]$ in \eqref{eq:-13} and the upper $\Gamma_{1-s}^{\mathsf{UB}}\left(P_{Y|X},Q_{Y},R\right)$
and lower bounds $\Gamma_{1-s}^{\mathsf{LB}}\left(P_{Y|X},Q_{Y},R\right)$
for $s\in(-1,0)$ in \eqref{eq:} and \eqref{eq:-1}, for the BSC
$Y=X\oplus V,V\sim\mathsf{Bern}\left(0.2\right)$ and the target distribution
$Q_{Y}=\mathsf{Bern}\left(0.5\right)$. }
\label{fig:Resolvability-1} 
\end{figure}

We numerically calculate the asymptotics of the normalized R\'enyi divergence
for binary symmetric channel (BSC) $Y=X\oplus V,V\sim\mathsf{Bern}\left(0.2\right)$
and $Q_{Y}=\mathsf{Bern}\left(0.5\right)$, and display the result
in Fig.~\ref{fig:Resolvability-1}. From this figure, we observe
that the normalized R\'enyi divergence decays as $R$ increases, and
finally vanishes for large enough $R$. Moreover, the rate at which
the normalized R\'enyi divergence transitions from a positive quantity
to zero increases in $s$ for the R\'enyi parameter $1+s\in[1,2]$,
and remains the same when $1+s\in(0,1]$. A rigorous statement of
this point will be provided in the next subsection.

\subsection{R\'enyi Resolvability}

\begin{figure}[t]
\centering \includegraphics[width=1\columnwidth]{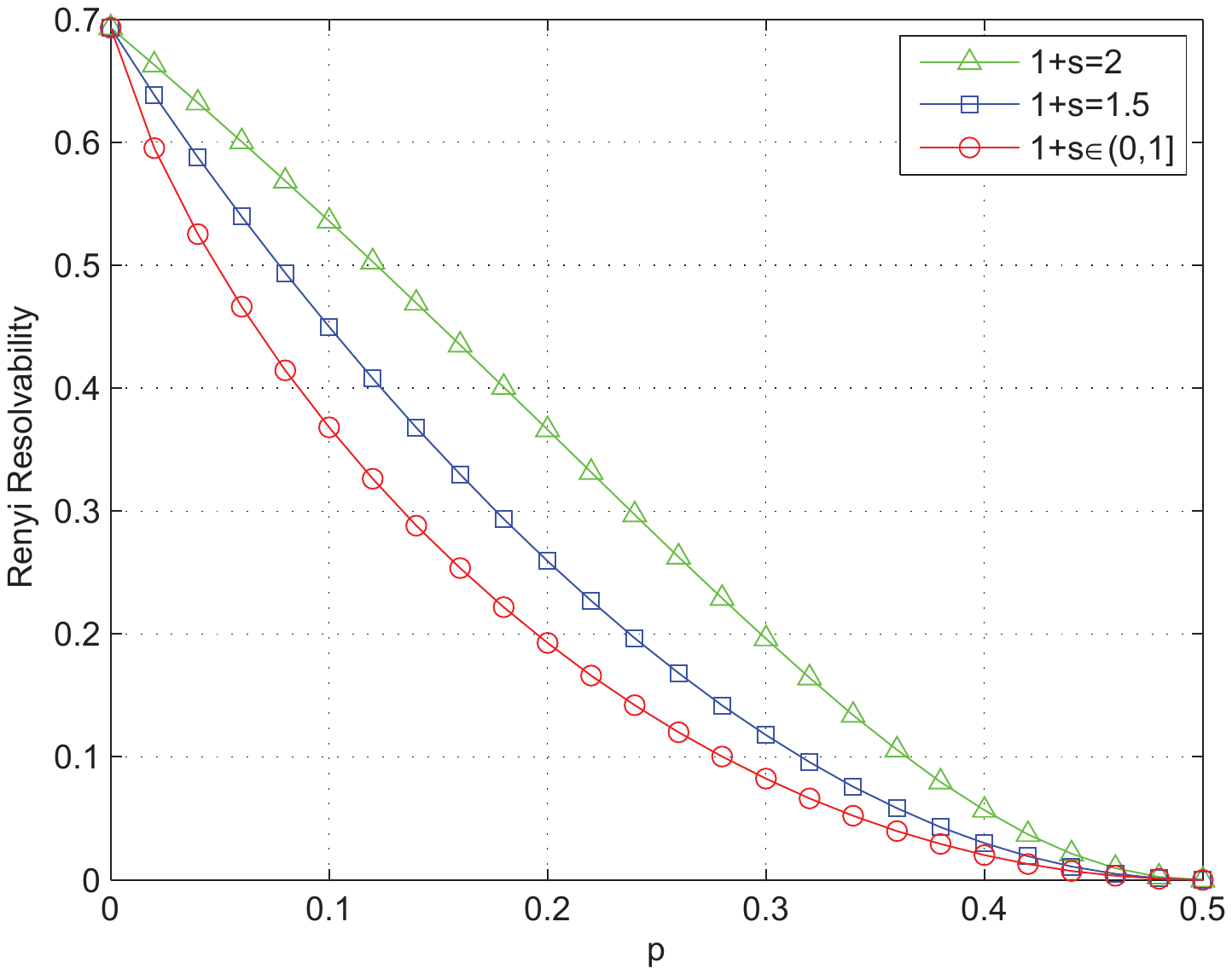}
\hfill{}\includegraphics[width=1\columnwidth]{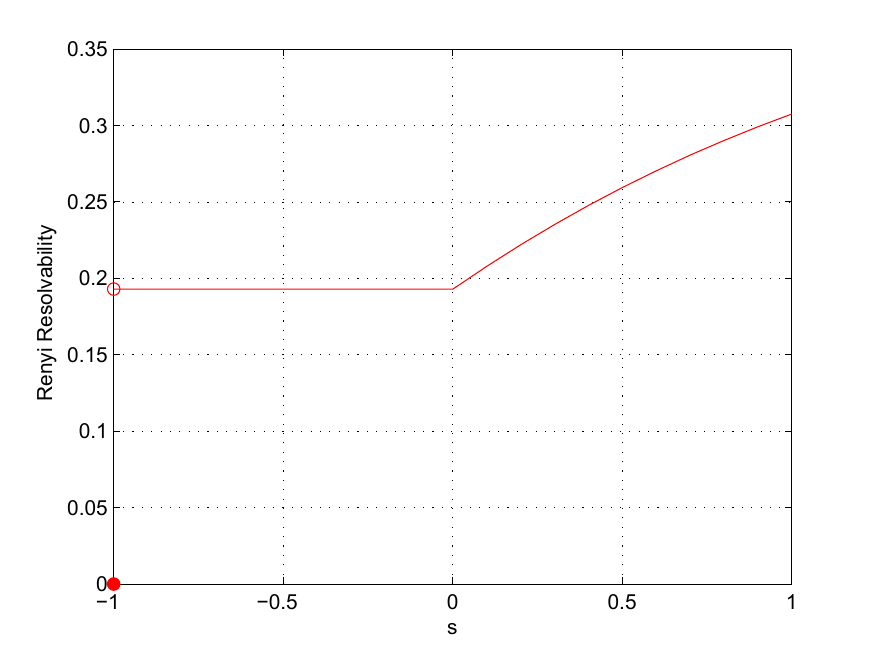}\caption{Illustration of the R\'enyi resolvability in \eqref{eq:-117} for the
BSC $Y=X\oplus V,V\sim\mathsf{Bern}\left(p\right)$ and $Q_{Y}=\mathsf{Bern}\left(0.5\right)$.
For the bottom subfigure, $p=0.2$.}
\label{fig:Resolvability-1-1} 
\end{figure}

We now compute the R\'enyi resolvability, which is defined as the minimum
rate $R$ of the input process $\{X^{n}(m):m\in\calM_{n}\}$ to ensure
that the unnormalized R\'enyi divergence $D_{1+s}(P_{Y^{n}\mathcal{C}_{n}}\|Q_{Y}^{n}P_{\mathcal{C}_{n}})$
or the normalized R\'enyi divergence $\frac{1}{n}D_{1+s}(P_{Y^{n}\mathcal{C}_{n}}\|Q_{Y}^{n}P_{\mathcal{C}_{n}})$
vanishes. We assume that 
\begin{equation}
\mathcal{P}\left(P_{Y|X},Q_{Y}\right):=\left\{ P_{X}:P_{Y|X}\circ P_{X}=Q_{Y}\right\} \ne\emptyset.\label{eq:-35}
\end{equation}
Otherwise, there does not exist a code such that $\frac{1}{n}D_{1+s}(P_{Y^{n}\mathcal{C}_{n}}\|Q_{Y}^{n}P_{\mathcal{C}_{n}})$
vanishes. By Theorem \ref{thm:singleletter} we easily obtain the
following result. The proof is provided in Appendix \ref{sec:Proof-of-Theorem-normalized}. 
\begin{thm}
\label{thm:Resolvability}(R\'enyi Resolvability) For $s\in[-1,1]\cup\{\infty\}$,
we have\footnote{We thank an anonymous reviewer for providing the achievability proof
for the case $s=\infty$ in Appendix \ref{subsec:Case}. Similar proof
ideas can be found in \cite{vellambi2016sufficient,vellambi2018new,yu2018on,yu2018exact}.} 
\begin{align}
 & \inf\left\{ R:\inf_{f_{\mathcal{C}_{n}}}D_{1+s}(P_{Y^{n}\mathcal{C}_{n}}\|Q_{Y}^{n}P_{\mathcal{C}_{n}})\rightarrow0\right\} \nonumber \\
 & =\inf\left\{ R:\frac{1}{n}\inf_{f_{\mathcal{C}_{n}}}D_{1+s}(P_{Y^{n}\mathcal{C}_{n}}\|Q_{Y}^{n}P_{\mathcal{C}_{n}})\rightarrow0\right\} \\
 & =R_{1+s}\left(P_{Y|X},Q_{Y}\right),\label{eq:-117}
\end{align}
where 
\begin{align}
 & R_{1+s}\left(P_{Y|X},Q_{Y}\right)\nonumber \\
 & :=\begin{cases}
\min\limits _{P_{X}\in\mathcal{P}\left(P_{Y|X},Q_{Y}\right)}\sum_{x}P_{X}\left(x\right)\\
\quad\times D_{1+s}\left(P_{Y|X}\left(\cdot|x\right)\|Q_{Y}\right), & s\in(0,1]\cup\{\infty\}\\
\min\limits _{P_{X}\in\mathcal{P}\left(P_{Y|X},Q_{Y}\right)}D(P_{Y|X}\|Q_{Y}|P_{X}), & s\in(-1,0]\\
0, & s=-1.
\end{cases}\label{eq:-18}
\end{align}
\end{thm}
\begin{rem}
Similar to Remarks \ref{rem:More-explicitly,-the} and \ref{rem:Similar-to-Remark},
the converse part in \eqref{eq:-117} also holds for any $s\in(1,\infty)$.
That is, for any $s\in(1,\infty)$, 
\begin{align}
 & \inf\left\{ R:\inf_{f_{\mathcal{C}_{n}}}D_{1+s}(P_{Y^{n}\mathcal{C}_{n}}\|Q_{Y}^{n}P_{\mathcal{C}_{n}})\rightarrow0\right\} \nonumber \\
 & \geq\inf\left\{ R:\frac{1}{n}\inf_{f_{\mathcal{C}_{n}}}D_{1+s}(P_{Y^{n}\mathcal{C}_{n}}\|Q_{Y}^{n}P_{\mathcal{C}_{n}})\rightarrow0\right\} \\
 & \geq\min\limits _{P_{X}\in\mathcal{P}\left(P_{Y|X},Q_{Y}\right)}\sum_{x}P_{X}\left(x\right)D_{1+s}\left(P_{Y|X}\left(\cdot|x\right)\|Q_{Y}\right).
\end{align}
\end{rem}
\begin{rem}
The R\'enyi resolvabilities for the normalized or unnormalized R\'enyi
divergence are the same. 
\end{rem}
\begin{rem}
\label{rmk:Rfunc} Note that for the case $s\in(-1,0]$, $R_{1+s}\left(P_{Y|X},Q_{Y}\right)$
can be also expressed as $\min_{P_{X}\in\mathcal{P}\left(P_{Y|X},Q_{Y}\right)}I(X;Y)$
where $(X,Y) \sim  P_{X}P_{Y|X}$, since $P_{X}\in\mathcal{P}\left(P_{Y|X},Q_{Y}\right)$. 
\end{rem}
\begin{rem}
Since $\mathcal{P}\left(P_{Y|X},Q_{Y}\right)$ is nonempty, $R_{1+s}\left(P_{Y|X},Q_{Y}\right)$
is finite. Hence it can be shown $\lim_{s\downarrow0}R_{1+s}\left(P_{Y|X},Q_{Y}\right)=R_{1}\left(P_{Y|X},Q_{Y}\right)$
(by using the continuity of R\'enyi divergence \cite{Erven}). Hence
$R_{1+s}\left(P_{Y|X},Q_{Y}\right)$ is continuous in $s$ for $s\in(-1,\infty]$.
See the bottom subfigure of Fig. \ref{fig:Resolvability-1-1}. 
\end{rem}
\begin{rem}
\label{rmk:unnormRenyi} This result for the case $s=0$ and the normalized
divergence (i.e., the normalized relative entropy) was first shown
by Wyner \cite{Wyner} for stationary memoryless channels, and was
extended to general channels by Han and Verd\'u \cite{Han}. Hence our
result for the normalized divergence is an extension of theirs to
the R\'enyi divergence $D_{1+s}$ of all orders $s\in[-1,1]\cup\{\infty\}$.
For the normalized divergence, our results for $s\in(0,1]\cup\{\infty\}$
and converse parts for $s\in(-1,0)$ are new. The case $s=0$ and
the unnormalized divergence (i.e., the unnormalized relative entropy
case) has been shown in other works, such as those by Hayashi~\cite{Hayashi06,Hayashi11},
which also imply the achievability result part for $s\in(-1,0)$ (since
the approximation measure $D_{1+s}$ for $s\in(-1,0)$ is weaker than
$D_{1}$). By Pinsker's inequality for the R\'enyi divergence \cite{Erven},
the resolvability result under the TV distance measure \cite{Han}
implies the converse for $s\in(-1,0]$. For the unnormalized divergence,
our results for $s\in(0,1]\cup\{\infty\}$ are new. All the results
above are summarized in Table \ref{tab:Summary-of-our}.

\begin{table*}
\centering \caption{\label{tab:Summary-of-our}Summary of results for normalized and unnormalized
R\'enyi resolvability   with different parameters. }
\begin{tabular}{|>{\centering}m{3cm}|>{\centering}p{3cm}|c|}
\hline 
\multicolumn{2}{|c|}{\textbf{Cases} } & \textbf{Results}\tabularnewline
\hline 
\hline 
\multirow{4}{3cm}{{Normalized Divergence } } & $s=0$  & Wyner \cite{Wyner} and Han-Verd\'u \cite{Han}\tabularnewline
\cline{2-3} 
 &  $s\in(-1,0)$  & Achievability: Wyner \cite{Wyner} and Han-Verd\'u \cite{Han};   Converse: Theorem \ref{thm:Resolvability}\tabularnewline
\cline{2-3} 
 & $s\in(0,1]\cup\{\infty\}$  & Theorem \ref{thm:Resolvability} \tabularnewline
\hline 
\multirow{3}{3cm}{{Unnormalized Divergence} } & $s=0$  & Hayashi~\cite{Hayashi06,Hayashi11} \tabularnewline
\cline{2-3} 
 & $s\in(-1,0)$  & Combining Pinsker's inequality \cite{Erven} and Han-Verd\'u \cite{Han}\tabularnewline
\cline{2-3} 
 & $s\in(0,1]\cup\{\infty\}$  & Theorem \ref{thm:Resolvability} \tabularnewline
\hline 
\end{tabular}
\end{table*}
\end{rem}
\begin{rem}
\label{rem:The-first-clause}The first clause in~\eqref{eq:-18}
is the minimization of an expectation of R\'enyi divergences $\sum_{x}P_{X}\left(x\right)D_{1+s}\left(P_{Y|X}\left(\cdot|x\right)\|Q_{Y}\right)$
but it is {\em not} (and in general smaller than) the conventional
conditional R\'enyi divergence $D_{1+s}\left(P_{XY}\|P_{X}Q_{Y}\right)$
(see Verd\'u \cite{Verdu} or Fong and Tan~\cite{Fong}). An optimal
i.i.d. code can achieve a rate equal to the minimization of conventional
conditional R\'enyi divergence $D_{1+s}(P_{XY}\|P_{X}Q_{Y})$ \cite[Thm.~14]{Hayashi},
while an optimal \emph{constant composition code} for the normalized
R\'enyi divergence or an optimal \emph{typical set code} (a code with
channel input distributed according to the target distribution $Q_{X}^{n}$
but truncated to an appropriate typical set) for both the unnormalized
and normalized R\'enyi divergences can achieve a better (smaller) rate
equal to the first clause in \eqref{eq:-18}. This shows that the
expectation of R\'enyi divergences also admits an operational interpretation
as the minimum rate needed to drive the R\'enyi divergence to zero when
its parameter is $\ge1$. Besides, a similar definition for the conditional
R\'enyi entropy can be found in \cite{cachin1997entropy,fehr2014conditional}.
In addition, observe that any constant composition code can be approximated
arbitrarily well by a typical set code by setting the typical set
parameter $\epsilon$ arbitrarily close to 0. But conversely, compared
to typical set codes, constant composition codes are easier to analyze.
This simplifies the proofs of our results significantly (e.g., that
of Theorem \ref{thm:singleletter}). Furthermore, for constant composition
codes, the codewords are each independently drawn from the uniform
distribution on a type class. It is worth noting that in Han and Verd\'u s
paper \cite[Example~1]{Han}, the extremal input process that results
in the worst (largest) resolvability is also the uniform distribution
on a type class.  
\end{rem}
The result in Theorem \ref{thm:Resolvability} for the BSC $Y=X\oplus V,V\sim\mathsf{Bern}\left(p\right)$
and $Q_{Y}=\mathsf{Bern}\left(0.5\right)$ is illustrated in Fig.~\ref{fig:Resolvability-1-1}.
For this case, 
\begin{align}
 & R_{1+s}\left(P_{Y|X},Q_{Y}\right)\nonumber \\
 & =\begin{cases}
\log\left(2\max\left\{ p,\overline{p}\right\} \right) & s=\infty\\
\frac{1}{s}\log\left(p^{1+s}2^{s}+\overline{p}^{1+s}2^{s}\right) & s\in(0,1]\\
1-H_{2}\left(p\right) & s=(-1,0]\\
0 & s=-1
\end{cases}.
\end{align}

\subsection{\label{subsec:Exponential-Behavior}Exponential Behavior }

\begin{figure}[t]
\centering \includegraphics[width=1\columnwidth]{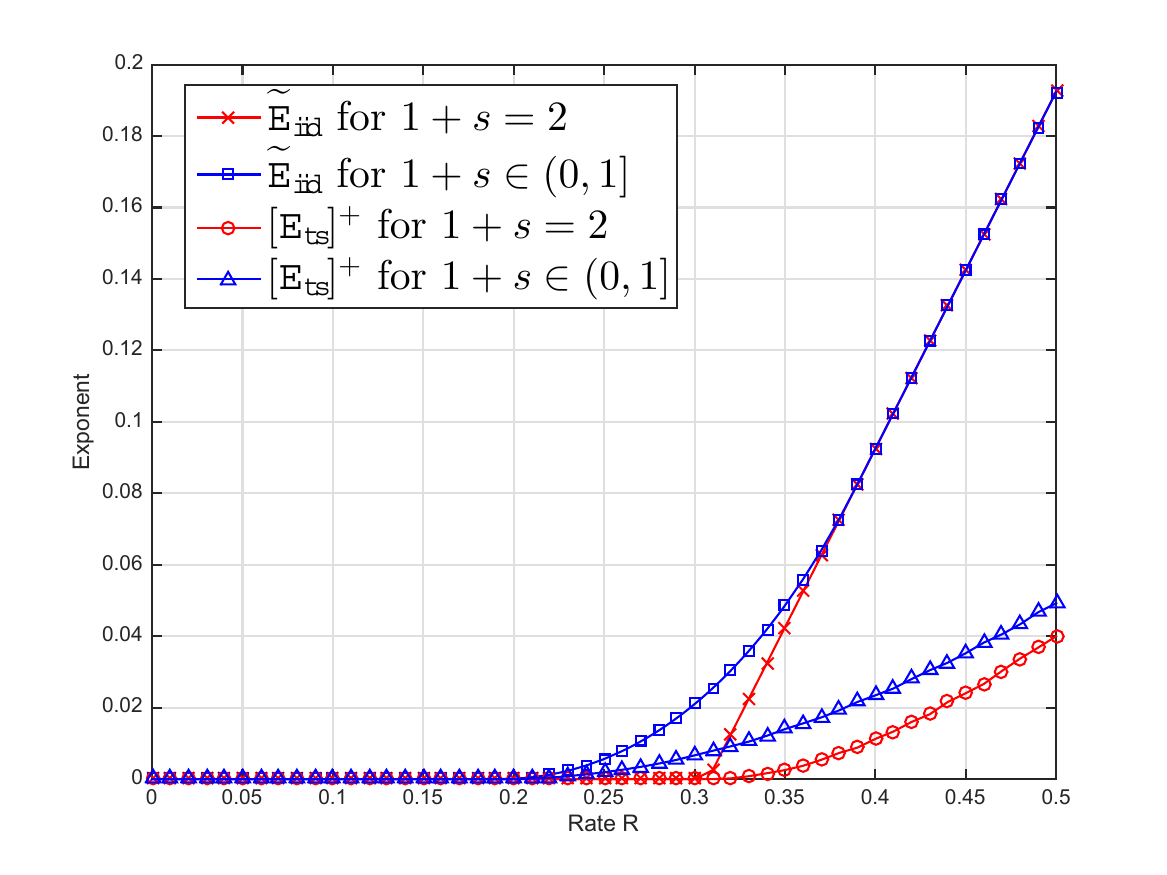}
\centering \includegraphics[width=1\columnwidth]{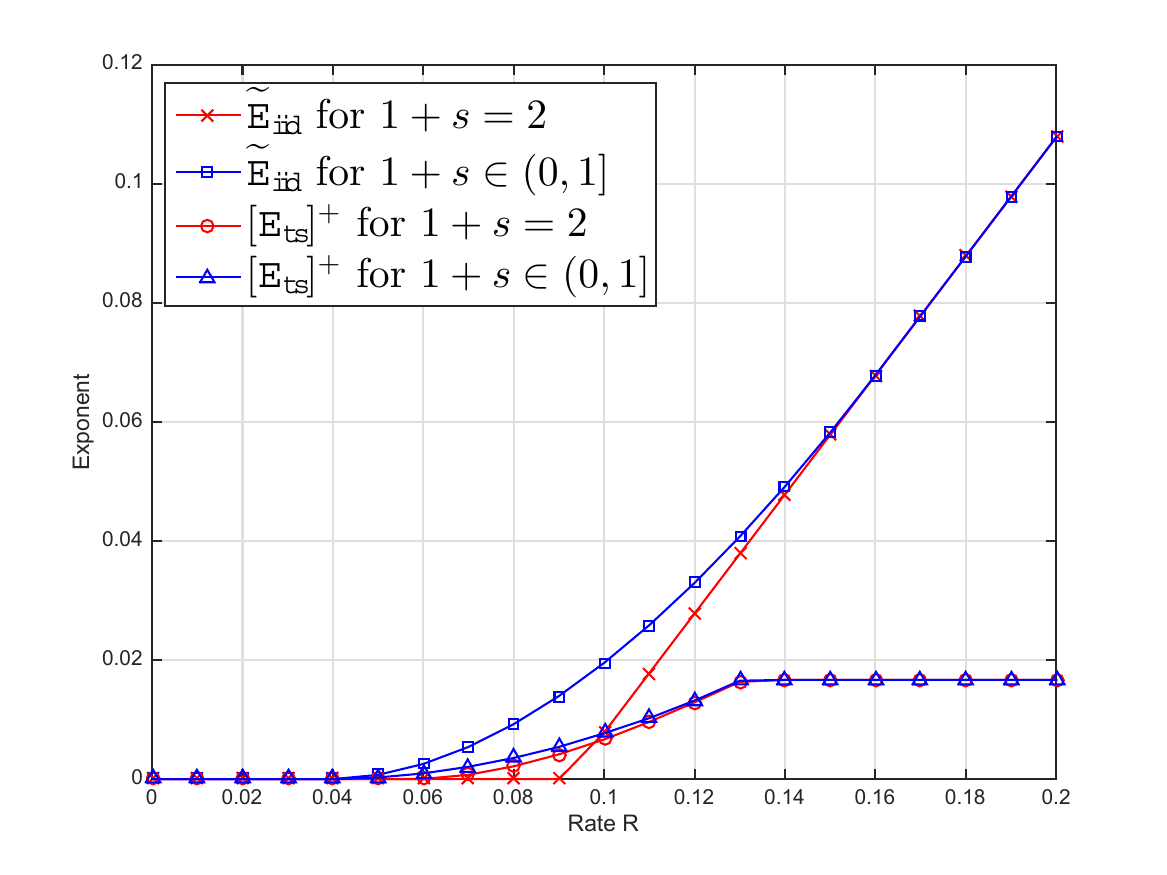}
\caption{Illustration of the exponent $\mathsf{\widetilde{E}_{iid}}\left(P_{X},P_{Y|X},Q_{Y}\right)$
in \eqref{eqn:Eiid2} for the ensemble of i.i.d.\ random codes and
the lower bound $[\mathsf{{E}_{ts}}\left(P_{X},P_{Y|X},Q_{Y}\right)]^{+}$
in \eqref{eqn:Ets} on the exponent for the ensemble of typical set
codes for R\'enyi parameter $1+s\in(0,2]$ for the BSC $Y=X\oplus V,V\sim\mathsf{Bern}\left(0.2\right)$.
For the top subfigure, $Q_{Y}=\mathsf{Bern}\left(0.5\right)$, and
for the bottom subfigure, $Q_{Y}=\mathsf{Bern}\left(0.23\right)$. }
\label{fig:exponent} 
\end{figure}

We now consider the exponent of $D_{1+s}(P_{Y^{n}\mathcal{C}_{n}}\|Q_{Y}^{n}P_{\mathcal{C}_{n}})$
when the codebook is generated in an i.i.d.\ fashion. In this case,
we can characterize the optimal exponent for this ensemble exactly.
The proof of the following theorem is provided in Appendix~\ref{sec:Proof-of-Theorem-expiid}. 
\begin{thm}[Exponential Behavior of i.i.d.\ Random Codes]
\label{thm:exponentforiid} Let $\mathcal{C}_{n}=\left\{ X^{n}\left(m\right)\right\} _{m\in\calM_{n}}$
with $X^{n}\left(m\right)\sim P_{X}^{n},m\in\calM_{n}$, and set $f_{\mathcal{C}_{n}}\left(m\right)=X^{n}\left(m\right)$,
where $P_{X}\in\mathcal{P}\left(P_{Y|X},Q_{Y}\right)$. For this i.i.d.\ code,
if the rate $R$ satisfies for $s\in(0,1]$, 
\begin{equation}
R>D_{1+s}\left(P_{XY}\|P_{X}\times Q_{Y}\right)
\end{equation}
and for $s\in(-1,0]$, 
\begin{equation}
R>D(P_{XY}\|P_{X}\times Q_{Y})=I(X;Y),
\end{equation}
then we have 
\begin{equation}
\lim_{n\to\infty}-\frac{1}{n}\log D_{1+s}(P_{Y^{n}\mathcal{C}_{n}}\|Q_{Y}^{n}P_{\mathcal{C}_{n}})=\mathsf{E_{iid}}\left(P_{X},P_{Y|X},Q_{Y}\right),\label{eq:-118}
\end{equation}
where 
\begin{align}
 & \mathsf{E_{iid}}\left(P_{X},P_{Y|X},Q_{Y}\right)\nonumber \\
 & :=\begin{cases}
\underset{t\in[s,1]}{\max}t\left(R-D_{1+t}\left(P_{XY}\|P_{X}\times Q_{Y}\right)\right) & s\in(0,1]\\
\underset{t\in[0,1]}{\max}t\left(R-D_{1+t}\left(P_{XY}\|P_{X}\times Q_{Y}\right)\right) & s\in(-1,0]
\end{cases}.\label{eqn:Eiid}
\end{align}
\end{thm}
\begin{rem}
By checking the proof, we can obtain that for any $s\in(1,\infty]$,
if $R>D_{1+s}\left(P_{XY}\|P_{X}\times Q_{Y}\right)$, then the i.i.d.\ code
above satisfies 
\begin{align}
 & \liminf_{n\to\infty}-\frac{1}{n}\log D_{1+s}(P_{Y^{n}\mathcal{C}_{n}}\|Q_{Y}^{n}P_{\mathcal{C}_{n}})\nonumber \\
 & \leq\sup_{t\ge1}t\left(R-D_{1+t}\left(P_{XY}\|P_{X}\times Q_{Y}\right)\right).
\end{align}
\end{rem}
\begin{rem}
Similar to Remark \ref{rem:By-checking-our}, by checking our proofs,
it can be seen that the achievability part in this theorem 
\begin{equation}
\lim_{n\to\infty}-\frac{1}{n}\log D_{1+s}(P_{Y^{n}\mathcal{C}_{n}}\|Q_{Y}^{n}P_{\mathcal{C}_{n}})\geq\mathsf{E_{iid}}\left(P_{X},P_{Y|X},Q_{Y}\right),\label{eq:-118-1}
\end{equation}
holds not only for channels with finite (input and output) alphabets,
but also for channels with countably infinite or continuous alphabets
(e.g., Gaussian channels). Hence Theorem \ref{thm:exponentforiid}
gives an exponential achievability result for channel resolvability
problems with countable or continuous alphabets. 
\end{rem}
\begin{rem}
\label{rmk:iid} Observe that the exponent of i.i.d.\ random codes
cannot be negative (see Lemma \ref{lem:oneshotach}) and the exponent
is non-decreasing in $R$. Hence for the i.i.d.\ code above with
any rate $R>0$, we have 
\begin{align}
 & \lim_{n\to\infty}-\frac{1}{n}\log D_{1+s}(P_{Y^{n}\mathcal{C}_{n}}\|Q_{Y}^{n}P_{\mathcal{C}_{n}})\nonumber \\
 & =\mathsf{\widetilde{E}_{iid}}\left(P_{X},P_{Y|X},Q_{Y}\right),
\end{align}
where 
\begin{align}
 & \mathsf{\widetilde{E}_{iid}}\left(P_{X},P_{Y|X},Q_{Y}\right)\nonumber \\
 & :=\begin{cases}
\left[\underset{t\in[s,1]}{\max}t\left(R-D_{1+t}\left(P_{XY}\|P_{X}\times Q_{Y}\right)\right)\right]^{+} & s\in(0,1]\\
\underset{t\in[0,1]}{\max}t\left(R-D_{1+t}\left(P_{XY}\|P_{X}\times Q_{Y}\right)\right) & s\in(-1,0]
\end{cases}.\label{eqn:Eiid2}
\end{align}
\end{rem}
The result for $s=0$ (relative entropy) can be found in Parizi, Telatar
and Merhav's paper \cite{Parizi}. The results for the other cases
$s\in[-1,1]\setminus\{0\}$ are new. The result of Theorem \ref{thm:exponentforiid}
for a BSC is illustrated in Fig.~\ref{fig:exponent}.

%\begin{figure}[t]
%\centering \includegraphics[width=0.55\columnwidth]{Exponent}\caption{Illustration of the exponent for the ensemble of i.i.d.\
%random codes for R\'enyi parameter $1+s\in(0,2]$ in \eqref{eq:-118}
%and \eqref{eqn:Eiid} for the BSC $Y=X\oplus V,V\sim\mathsf{Bern}\left(0.2\right)$
%and $P_{X}=Q_{Y}=\mathsf{Bern}\left(0.5\right)$. }
%\label{fig:Resolvability-1-1-1}
%\end{figure}

Furthermore, for general codes, we show that the R\'enyi divergence
decays at least exponentially fast, as long as the code rate is larger
than the R\'enyi resolvability given in the previous subsection. The
proof is provided in Appendix \ref{sec:Proof-of-Theorem-exp}. 
\begin{thm}[General Lower Bound on the R\'enyi Divergence Exponent]
\label{thm:exponent} Let $s\in[-1,1]$. If 
\begin{equation}
R>R_{1+s}\left(P_{Y|X},Q_{Y}\right),
\end{equation}
then we have 
\begin{align}
 & \liminf_{n\to\infty}-\frac{1}{n}\log\inf_{f_{\mathcal{C}_{n}}}\left(D_{1+s}(P_{Y^{n}\mathcal{C}_{n}}\|Q_{Y}^{n}P_{\mathcal{C}_{n}})\right)\nonumber \\
 & \geq\max\limits _{P_{X}\in\mathcal{P}\left(P_{Y|X},Q_{Y}\right)}\max\bigl\{\mathsf{\widetilde{E}_{iid}}\left(P_{X},P_{Y|X},Q_{Y}\right),\nonumber \\
 & \qquad\qquad\qquad\qquad\mathsf{E_{ts}}\left(P_{X},P_{Y|X},Q_{Y}\right)\bigr\},\label{eq:-84}
\end{align}
where $\mathsf{\widetilde{E}_{iid}}\left(P_{X},P_{Y|X},Q_{Y}\right)$
is defined in \eqref{eqn:Eiid2}, and\footnote{Here the subscript of $\mathsf{E_{ts}}$ refers to \emph{typical set}.
The achievability scheme for this exponent is one with channel input
following a truncated version of the target distribution $Q_{Y}^{n}$
to some typical set; hence we term this \emph{typical set code}.} 
\begin{align}
 & \mathsf{E_{ts}}\left(P_{X},P_{Y|X},Q_{Y}\right)\nonumber \\
 & :=\begin{cases}
\sup_{\epsilon\in(0,1]}\min\left\{ \frac{\epsilon^{2}P_{\mathsf{min}}}{3},\theta\left(s,\epsilon,P_{X}\right)\right\}  & s\in(0,1]\\
\sup_{\epsilon\in(0,1]}\min\left\{ \frac{\epsilon^{2}P_{\mathsf{min}}}{3},\theta\left(0,\epsilon,P_{X}\right)\right\}  & s\in(-1,0]
\end{cases}\label{eqn:Ets}
\end{align}
with $P_{\mathsf{min}}:=\min_{x}P_{X}\left(x\right)$, and 
\begin{align}
 & \theta\left(s,\epsilon,P_{X}\right)\nonumber \\
 & :=\sup_{t\in[s,1]}t\Bigl(R-\left(1+\epsilon\right)\sum_{x}P_{X}\left(x\right)D_{1+t}\left(P_{Y|X}\left(\cdot|x\right)\|Q_{Y}\right)\Bigr).
\end{align}
\end{thm}
\begin{rem}
From Theorem \ref{thm:exponent} and Remark \ref{rmk:iid}, it can
be easily observed that for $s\in(0,1]$, if $R_{1+s}\left(P_{Y|X},Q_{Y}\right)<R<\min_{P_{X}\in\mathcal{P}\left(P_{Y|X},Q_{Y}\right)}D_{1+s}\left(P_{XY}\|P_{X}\times Q_{Y}\right),$
then 
\begin{align}
 & \max\limits _{P_{X}\in\mathcal{P}\left(P_{Y|X},Q_{Y}\right)}\mathsf{E_{ts}}\left(P_{X},P_{Y|X},Q_{Y}\right)\nonumber \\
 & >\max\limits _{P_{X}\in\mathcal{P}\left(P_{Y|X},Q_{Y}\right)}\mathsf{\widetilde{E}_{iid}}\left(P_{X},P_{Y|X},Q_{Y}\right)\\
 & =0.
\end{align}
For this case, the optimal R\'enyi divergence of i.i.d.\ codes increases
almost linearly as $n\to\infty$. This can be observed from the one-shot
bounds (Lemmas \ref{lem:oneshotach} and \ref{lem:oneshotcon}). Hence
i.i.d.\ codes are, in general, not optimal in achieving the best
exponent for $s\in(0,1]$. This point is unsurprising given Remark
\ref{rem:The-first-clause}, since as stated in Remark \ref{rem:The-first-clause},
optimal i.i.d. codes are not optimal in achieving even the R\'enyi resolvability.
Hence, optimal i.i.d.\ codes are certainly not optimal in achieving
the R\'enyi divergence exponent. This point can be seen from Fig. \ref{fig:exponent}. 
\end{rem}
\begin{rem}
\label{rem:parizi} The optimal exponent of constant composition codes
\cite{Csi97} has been studied by Parizi, Telatar, and Merhav \cite{Parizi},
but different from our case, they consider the relative entropy between
the channel output and the corresponding expected version (over the
codebook) as the channel resolvability. Note that according to \cite[Equation (13)]{Parizi},
even in the $s=0$ case, this quantity is weaker than the R\'enyi divergence
considered by us. They obtained ensemble tight results for constant
composition and i.i.d.\ codes (and applied these results to the wiretap
channel) but we are only able to do the same for the simpler i.i.d.\ codes. 
\end{rem}

\section{Application to the Wiretap Channel }

\label{sec:wiretap}

We apply the preceding results to the wiretap channel \cite{Wyner75},
\cite{Csiszar78}. In \cite{hou2014effective}, Hou and Kramer proposed
a new security measure, termed {\em effective secrecy}, for wiretap
channels by exploiting the unnormalized KL divergence to quantify
not only (the wiretapper's) \emph{confusion} but also \emph{stealth}.
In this section, we generalize Hou and Kramer's result to a generalized
divergence measure\textemdash the R\'enyi divergence. We provide a complete
characterization of the secrecy capacity region under this new and
generalized leakage measure.

Consider a discrete memoryless wiretap channel $P_{YZ|X}$, and two
messages $\left(M_{0},M_{1}\right)$ that are uniformly distributed
over $\calM_{0}:=\{1,\ldots,\e^{nR_{0}}\}$ and $\calM_{1}:=\{1,\ldots,\e^{nR_{1}}\}$
respectively. A sender wants to transmit the pair $\left(M_{0},M_{1}\right)$
to a legitimate user, and, at the same time, ensure that $M_{1}$
as almost independent from the wiretapper's observation $Z^{n}$. 
\begin{defn}
An $\left(n,R_{0},R_{1}\right)$ secrecy code is defined by two stochastic
mappings $P_{X^{n}|M_{0}M_{1}}:\mathcal{M}_{0}\times\mathcal{M}_{1}\rightarrow\mathcal{X}^{n}$
and $P_{\widehat{M}_{0}\widehat{M}_{1}|Y^{n}}:\mathcal{Y}^{n}\rightarrow\mathcal{M}_{0}\times\mathcal{M}_{1}$. 
\end{defn}
Given a target distribution $Q_{Z}$, we wish to maximize the alphabet
size (or rate) of $M_{1}$ such that the distribution $P_{M_{1}Z^{n}}$
induced by the code is approximately equal to the target distribution
$P_{M_{1}}Q_{Z}^{n}$ and $M_{1}$ can be decoded correctly asymptotically. 
\begin{defn}
The tuple $(R_{0},R_{1})$ is {\em $(Q_{Z},1+s)$-achievable} if
there exists a sequence of $\left(n,R_{0},R_{1}\right)$ secrecy codes
with induced distribution $P$ such that 
\begin{enumerate}
\item Error constraint: 
\begin{equation}
\lim_{n\rightarrow\infty}\mathbb{P}\left(\left(M_{0},M_{1}\right)\neq(\widehat{M}_{0},\widehat{M}_{1})\right)=0;\label{eq:-30-1}
\end{equation}
\item Secrecy constraint (generalized effective secrecy): 
\begin{equation}
{\displaystyle \lim_{n\rightarrow\infty}D_{1+s}(P_{M_{1}Z^{n}}\|P_{M_{1}}Q_{Z}^{n})=0}.\label{eq:-31}
\end{equation}
\end{enumerate}
\end{defn}
It is worth noting that \eqref{eq:-31} is a generalized version of
the notion of effective secrecy considered in \cite{hou2014effective}.
Here we assume $Q_{Z}$ satisfies $\mathcal{P}\left(P_{Z|X},Q_{Z}\right)\neq\emptyset$
($\mathcal{P}\left(P_{Z|X},Q_{Z}\right)$ is defined in \eqref{eq:-35});
otherwise, \eqref{eq:-31} cannot be satisfied by any secrecy code. 
\begin{defn}
The {\em $(Q_{Z},1+s)$-admissible region} is defined as 
\begin{equation}
\mathcal{R}_{1+s}(Q_{Z}):=\textrm{Closure}\left\{ \textrm{\ensuremath{Q_{Z}}-achievable }(R_{0},R_{1})\right\} .
\end{equation}
\end{defn}
It is worth noting that our secrecy metric (even when $s=0$) is stronger
than the unnormalized relative entropy $D(P_{M_{1}Z^{n}}\|P_{M_{1}}P_{Z^{n}})$
(or $I\left(M_{1};Z^{n}\right)$) considered in Bloch and Laneman~\cite{Bloch},
since 
\begin{align}
D(P_{M_{1}Z^{n}}\|P_{M_{1}}Q_{Z}^{n}) & =I\left(M_{1};Z^{n}\right)+D(P_{Z^{n}}\|Q_{Z}^{n})\label{eq:effective}\\
 & \geq I\left(M_{1};Z^{n}\right).\label{eq:effective-1}
\end{align}
For our secrecy metric, in addition to requiring that $M_{1}$ and
$Z^{n}$ are approximately independent, we also require that the wiretapper's
observation $Z^{n}$ is close to the product distribution $Q_{Z}^{n}$.
This is similar to Hou and Kramer's work \cite{hou2014effective}
and Csisz\'ar and Narayan's work~\cite[Eqn.~(6)]{CN04}, but we consider
a continuum of secrecy measures indexed by $s\in[-1,1]$.

The interpretation of our secrecy measure with $s=0$ can be found
in \cite{hou2014effective}, where the authors interpreted $I(M_{1};Z^{n})$
in~\eqref{eq:effective-1} as a measure of ``non-confusion'' and
$D(P_{Z^{n}}\|Q_{Z}^{n})$ in~\eqref{eq:effective} as a measure
of ``non-stealth''. Under this interpretation, we set $Q_{Z}^{n}$
to be the distribution of the signal that the wiretapper observes
if the sender is not sending useful information. Hence if the secrecy
constraint \eqref{eq:-31} is satisfied then we can say that {\em
useful} information is being transmitted stealthily.

\subsection{Main Result for Deterministic Encoders}

\label{subsec:DetEnc} 
\begin{figure}[t]
\centering \includegraphics[width=1\columnwidth]{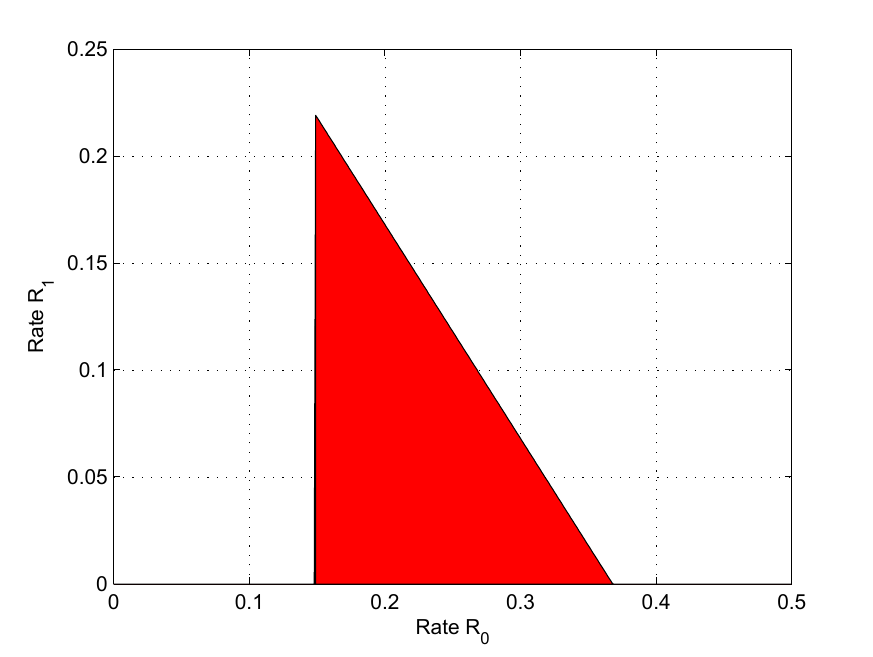}\caption{Illustration of the admissible region for case of using a deterministic
encoder and with R\'enyi parameter $1+s=2$ in \eqref{eq:-40} for the
binary wiretap channel. }
%binary
%channel $Y=X\oplus V_{1},V_{1}\sim\mathsf{Bern}\left(0.1\right)$,
%$Z=X\oplus V_{2},V_{2}\sim\mathsf{Bern}\left(0.3\right)$ and $Q_{Z}=\mathsf{Bern}\left(0.5\right)$. }
\label{fig:DetEnc} 
\end{figure}

Before solving the problem, in this subsection
we consider a simpler version of the problem\textemdash namely, a
system with a deterministic encoder. That is, the encoder is restricted
to a deterministic (non-stochastic) function $f:\mathcal{M}_{0}\times\mathcal{M}_{1}\rightarrow\mathcal{X}^{n}$
(denote the $(Q_{Z},1+s)$-admissible region for this case as $\mathcal{R}_{1+s}^{\mathrm{det}}(Q_{Z})$).
Using Theorem \ref{thm:Resolvability}, we obtain the following theorem.
The detailed proof is provided in Appendix~\ref{sec:Proof-of-Theorem-DetEnc}. 
\begin{thm}
\label{thm:DetEnc} For $s\in[-1,1]\cup\{\infty\}$, we have 
\begin{align}
 & \mathcal{R}_{1+s}^{\mathrm{det}}(Q_{Z})\nonumber \\
 & =\bigcup_{P_{X}\in\mathcal{P}\left(P_{Z|X},Q_{Z}\right)}\left\{ \begin{array}{c}
(R_{0},R_{1}):R_{0}+R_{1}\leq I\left(X;Y\right)\\
R_{0}\geq\widetilde{R}_{1+s}\left(P_{X},P_{Z|X},Q_{Z}\right)
\end{array}\right\} ,\label{eq:-40}
\end{align}
where $\widetilde{R}_{1+s}\left(P_{X},P_{Z|X},Q_{Z}\right)$ is defined
as 
\begin{align}
 & \widetilde{R}_{1+s}\left(P_{X},P_{Z|X},Q_{Z}\right)\nonumber \\
 & :=\begin{cases}
\sum_{x}P_{X}\left(x\right)D_{1+s}\left(P_{Z|X}\left(\cdot|x\right)\|Q_{Z}\right) & s\in(0,1]\cup\{\infty\}\\
D(P_{Z|X}\|Q_{Z}|P_{X}) & s\in(-1,0]\\
0 & s=-1
\end{cases}.\label{eq:-28}
\end{align}
\end{thm}
\begin{rem}
This theorem provides an expression for the admissible rate region
for the case with no extra randomness (i.e., the case with deterministic
encoders). Related works on determining the amount of randomness needed
to realize stochastic encoding include Watanabe and Oohama's work
\cite{watanabe2015optimal} and Chou, Vellambi, Bloch, and Kliewer's
work \cite{chou2017coding}. Hence studying the deterministic encoder
case is of independent interest. 
\end{rem}
\begin{rem}
Similar to the exponential behavior for the R\'enyi resolvability problem,
it is easy to see that if $(R_{0},R_{1})$ is an interior point of
$\mathcal{R}_{1+s}^{\mathrm{det}}(Q_{Z})$, then the R\'enyi divergence
for the wiretap channel with deterministic encoder also decays at
least exponetially fast. 
\end{rem}
The result of Theorem \ref{thm:DetEnc} for the binary wiretap channel
is illustrated in Fig. \ref{fig:DetEnc}. From the figure (or the
theorem), we observe that for the problem with deterministic encoder,
the achievability of a rate pair $(R_{0},R_{1})$ does not necessarily
imply the achievability of a rate pair $(R_{0}',R_{1}')$ such that
$R_{0}'\leq R_{0},R_{1}'\leq R_{1}$. This is because to meet the
resolvability constraint, a certain amount of local randomness (besides
the secret message $M_{1})$ at the sender is needed; this local randomness
only comes from the non-secret message $M_{0}$ (since the encoder
is a deterministic function of $M_{0},M_{1}$). Therefore, a rate
less than $R_{0}$ may not satisfy the resolvability constraint.

\subsection{Main Result for Stochastic Encoders}

\begin{figure}[t]
\centering \includegraphics[width=1\columnwidth]{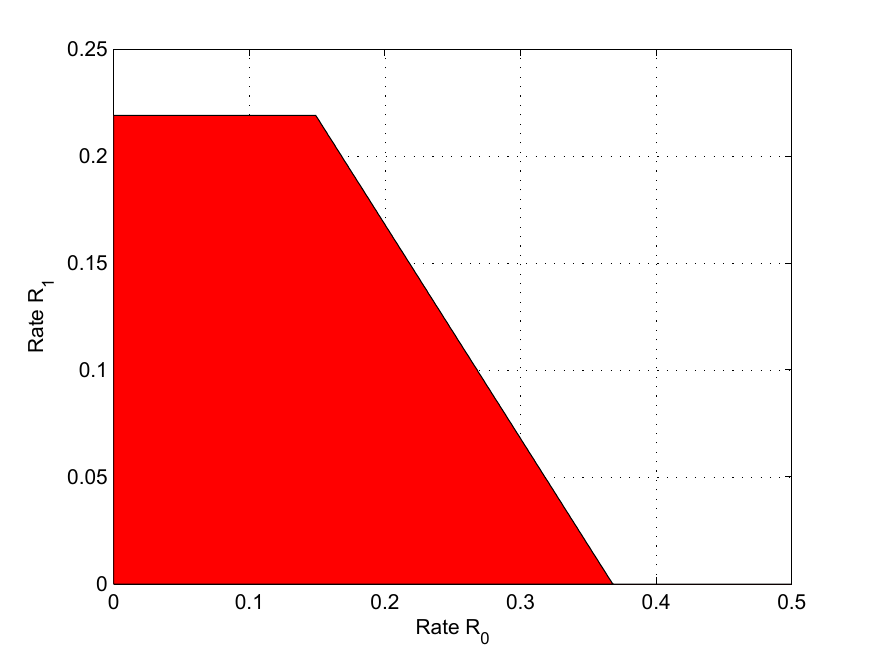}\caption{Illustration of the admissible region for case of using a stochastic
encoder and with R\'enyi parameter $1+s=2$ in \eqref{eq:-39} or \eqref{eq:-120}
for the binary wiretap channel. }
\label{fig:Resolvability-1-1-1-1-1} 
\end{figure}

If a stochastic encoder is allowed, we can add a virtual memoryless
channel $P_{X|W}^{n}$ between the deterministic encoder and the channel.
Then we have the following achievability result. 
\begin{prop}
\label{thm:StochasticEncoder0} For $s\in[-1,1]\cup\{\infty\}$, we
have 
\begin{align}
 & \mathcal{R}_{1+s}(Q_{Z})\nonumber \\
 & \supseteq\bigcup_{\substack{P_{XW}:\\
P_{W}\in\mathcal{P}\left(P_{Z|W},Q_{Z}\right)
}
}\left\{ \begin{array}{c}
(R_{0},R_{1}):R_{0}+R_{1}\leq I\left(W;Y\right),\\
R_{0}\geq\widetilde{R}_{1+s}\left(P_{W},P_{Z|W},Q_{Z}\right)
\end{array}\right\} ,
\end{align}
where $\widetilde{R}_{1+s}\left(P_{W},P_{Z|W},Q_{Z}\right)$ is given
by \eqref{eq:-28}. 
\end{prop}
However, adding a memoryless channel is not optimal in general. In
the following theorem, we completely characterize the admissible region,
and show that adding a channel with memory between the encoder and
channel is optimal. The proof of this theorem is given in Appendix
\ref{sec:Proof-of-Theorem-StocEnc}. 
\begin{thm}
\label{thm:StochasticEncoder} For $s\in[-1,1]\cup\{\infty\}$, we
have 
\begin{align}
 & \mathcal{R}_{1+s}(Q_{Z})\nonumber \\
 & =\bigcup_{\substack{\widetilde{P}_{XW}:\\
\widetilde{P}_{X}\in\mathcal{P}\left(P_{Z|X},Q_{Z}\right)
}
}\!\left\{ \begin{array}{c}
(R_{0},R_{1}):R_{0}+R_{1}\leq I_{\widetilde{P}}\left(W;Y\right)\\
R_{0}\geq\widetilde{R}'_{1+s}\left(\widetilde{P}_{W|X}\widetilde{P}_{X},P_{Z|X},Q_{Z}\right)
\end{array}\right\} \label{eq:-39}\\
 & =\bigcup_{\substack{\widetilde{P}_{XW}:\\
\widetilde{P}_{X}\in\mathcal{P}\left(P_{Z|X},Q_{Z}\right)
}
}\left\{ \begin{array}{l}
(R_{0},R_{1}):R_{0}+R_{1}\leq I_{\widetilde{P}}\left(W;Y\right)\\
R_{1}\leq I_{\widetilde{P}}\left(W;Y\right)\\
\;-\widetilde{R}'_{1+s}\left(\widetilde{P}_{W|X}\widetilde{P}_{X},P_{Z|X},Q_{Z}\right)
\end{array}\right\} ,\label{eq:-120}
\end{align}
where $\widetilde{R}'_{1+s}\left(\widetilde{P}_{W|X}\widetilde{P}_{X},P_{Z|X},Q_{Z}\right)$
is given by 
\begin{align}
 & \widetilde{R}'_{1+s}\left(\widetilde{P}_{W|X}\widetilde{P}_{X},P_{Z|X},Q_{Z}\right)\nonumber \\
 & :=\begin{cases}
\underset{\widetilde{P}_{Z|WX}}{\max}\Big\{-\frac{1+s}{s}D(\widetilde{P}_{Z|WX}\|P_{Z|X}|\widetilde{P}_{XW})\\
\qquad\qquad+D(\widetilde{P}_{Z|W}\|Q_{Z}|\widetilde{P}_{W})\Big\}, & s\in(0,1]\\
 & \quad\;\cup\{\infty\}\\
I_{\widetilde{P}}\left(W;Z\right), & s\in(-1,0]\\
0, & s=-1
\end{cases}.\label{eq:-6}
\end{align}
Here $I_{\widetilde{P}}\left(W;Y\right)$ in \eqref{eq:-39} and \eqref{eq:-120}
and $I_{\widetilde{P}}\left(W;Z\right)$ in \eqref{eq:-6} are the
mutual informations evaluated under the distribution $\widetilde{P}_{WX}P_{YZ|X}$.
Furthermore, the ranges of $W$ in \eqref{eq:-39} and \eqref{eq:-120}
may be assumed to satisfy $\left|\mathcal{W}\right|\leq\left|\mathcal{X}\right|+1$. 
\end{thm}
\begin{rem}
It is easy to show that if $R$ is an interior point of $\mathcal{R}_{1+s}(Q_{Z})$,
then the R\'enyi divergence for the wiretap channel problem with stochastic
encoder also decays at least exponentially fast. 
\end{rem}
\begin{rem}
\label{rmk:layer} We can define the \emph{effective secrecy capacity}
with the leakage measured by the R\'enyi divergence with parameter $1+s$
and with target output distribution $Q_{Z}$ as $C_{1+s}\left(Q_{Z}\right):=\max_{(R_{0},R_{1})\in\mathcal{R}_{1+s}(Q_{Z})}R_{1}$.
The special case with $s=0$ was defined by Hou and Kramer \cite{hou2014effective},
and they showed 
\begin{equation}
C_{1}\left(Q_{Z}\right)=\max_{\substack{\widetilde{P}_{XW}:\\
\widetilde{P}_{X}\in\mathcal{P}\left(P_{Z|X},Q_{Z}\right)
}
}\left\{ I_{\widetilde{P}}\left(W;Y\right)-I_{\widetilde{P}}\left(W;Z\right)\right\} .
\end{equation}
For the general case $s\in[-1,1]\cup\{\infty\}$, by Theorem \ref{thm:StochasticEncoder},
we have 
\begin{align}
 & C_{1+s}\left(Q_{Z}\right)=\nonumber \\
 & \max_{\substack{\widetilde{P}_{XW}:\\
\widetilde{P}_{X}\in\mathcal{P}\left(P_{Z|X},Q_{Z}\right)
}
}\left\{ I_{\widetilde{P}}\left(W;Y\right)-\widetilde{R}'_{1+s}(\widetilde{P}_{W|X}\widetilde{P}_{X},P_{Z|X},Q_{Z})\right\} ,
\end{align}
which has a similar form as the conventional secrecy capacity (with
secrecy measured by the normalized mutual information $\frac{1}{n}I\left(M;Z^{n}\right)$
or unnormalized mutual information $I\left(M;Z^{n}\right)$) given
in \cite{Csiszar78,Hayashi11,Hayashi06}, 
\begin{equation}
C_{\mathsf{MI}}=\max_{P_{WX}}\left\{ I\left(W;Y\right)-I\left(W;Z\right)\right\} .
\end{equation}
Note that $C_{\mathsf{MI}}\geq\max_{Q_{Z}}C_{1+s}(Q_{Z})$ for $s\in(0,1]$
and $C_{\mathsf{MI}}=\max_{Q_{Z}}C_{1+s}(Q_{Z})$ for $s\in(-1,0]$.
This is because our secrecy measure is stronger than the conventional
one. Furthermore, when considering the simultaneous transmission of
secret and non-secret messages, the optimal rate region \cite{xu2008broadcast}
\cite[Cor. 2]{Csiszar78}\footnote{Note that here we refer to Corollary 2 of \cite{Csiszar78}, in which
the common message rate is set to zero and the $R_{1}$ and $R_{e}$
there respectively correspond to the $R_{0}+R_{1}$ and $R_{1}$ of
this paper. Although the setting in Corollary 2 of \cite{Csiszar78}
does not implicitly indicate the secret and non-secret parts, it is
easy to show that if divide the total rate into these two parts, the
admissible region does not change.} is 
\begin{align}
\mathcal{R}_{\mathsf{MI}} & =\bigcup_{\substack{P_{U|W}P_{W|X}P_{X}:\\
I(U;Y)\leq I(U;Z)
}
}\left\{ \begin{array}{c}
(R_{0},R_{1}):R_{0}+R_{1}\leq I\left(W;Y\right),\\
R_{1}\leq I\left(W;Y|U\right)-I\left(W;Z|U\right)
\end{array}\right\} ,
\end{align}
which is different from the optimal region $\mathcal{R}_{1+s}$ given
by us. Obviously $\bigcup_{Q_{Z}}\mathcal{R}_{1+s}$ $(Q_{Z})\subseteq\mathcal{R}_{\mathsf{MI}}$.
Xu and Chen \cite{xu2008broadcast} and Csisz\'ar and K\"orner~\cite[Cor. 2]{Csiszar78}
derived the optimal region $\mathcal{R}_{\mathsf{MI}}$ by using a
two-layered code, but for our case, a single-layered code is sufficient
to achieve the optimality; a similar conclusion for the $s=0$ case
can be drawn from the results in \cite{kobayashi2013secure}. This
is because our secrecy measure requires that $M_{1}$ and $Z^{n}$
are approximately independent (similarly to the conventional setting)
but also requires the wiretapper's observation $Z^{n}$ to approximately
follow a target memoryless distribution $Q_{Z}^{n}$ (soft-covering
the space according to the target distribution). We provide an intuitive
interpretation for why a two-layered code is not necessary to achieve
the optimal region for our problem. For simplicity, we consider the
case with the R\'enyi parameter equal to $1$; If we apply a two-layered
code to our setting then to guarantee the soft-covering property (under
the TV distance measure, which is weaker than the R\'enyi divergence),
the non-secret message for each layer has to have rates that are appropriately
lower bounded as follows: $R_{0}^{(1)}>I\left(U;Z\right),$ $R_{0}^{(1)}+R_{0}^{(2)}>I\left(UW;Z\right)$
for some $P_{UW|X}$ and ${P}_{X}\in\mathcal{P}\left(P_{Z|X},Q_{Z}\right)$
\cite{Gohari}, where $R_{0}^{(1)}$ and $R_{0}^{(2)}$ respectively
denote the transmission rate of the non-secret message for the first
and second layer. On the other hand, the total rate is still constrained
by $I\left(W;Y\right)$, i.e., $R_{0}^{(1)}+R_{0}^{(2)}+R_{1}\leq I\left(W;Y\right)$.
Hence the achievable rate pair $(R_{0}^{(1)}+R_{0}^{(2)},R_{1})$
is still in $\mathcal{R}_{1}(P_{Z})$. Note that this is true even
for the TV distance. As a result, it must also be true for the stronger
distance measures such as relative entropy or R\'enyi divergence. 
\end{rem}
\begin{rem}
Both the coding scheme in this paper and that in \cite[Cor. 2]{Csiszar78}
require stochastic encoding to achieve the optimal rate regions. The
amount of randomness needed to realize the stochastic encoding for
the setting similar to that in~\cite[Cor. 2]{Csiszar78} was studied
in \cite{watanabe2015optimal}, and the case with only an asymptotically
vanishing rate of extra randomness available but with non-uniform
sources to be transmitted was studied in \cite{chou2017coding}. For
our setting, the admissible rate region for the case with no extra
randomness (i.e., the case with deterministic encoders) was provided
in Subsection \ref{subsec:DetEnc}. 
\end{rem}
\begin{rem}
\label{rmk:ssc} The {\em semantic-security capacity} $C_{\mathsf{SS}}$
(with the secrecy measure\footnote{This measure comes from \cite[Thm.~2]{Goldfeld}, but is different
from and stronger than the original one $\max_{P_{M}\in\mathcal{P}\left(\mathcal{M}\right)}I\left(M;Z^{n}\right)$,
also considered by Goldfeld, Cuff, and Permuter in~\cite{Goldfeld}.
However, both measures result in the same secrecy capacity~\cite{Goldfeld}.} $\max_{m_{1}}D\left(P_{Z^{n}|M_{1}=m_{1}}\|Q_{Z}^{n}\right)\rightarrow0$),
studied in~\cite{Goldfeld}, is proven to be equal to $C_{\mathsf{MI}}$.
Obviously, this secrecy measure is not weaker than the one considered
in this paper (when the R\'enyi divergence parameter is equal to $1$).
In fact, by a simple expurgation argument, it is easy to show that
the secrecy measure of $D\left(P_{Z^{n}|M_{1}}\|Q_{Z}^{n}|P_{M_{1}}\right)\rightarrow0$
implies semantic secrecy (see for example \cite[Appendix A]{Parizi}
or \cite[footnote on p.~6825]{Han14}. So these two measures are equivalent.
In \cite{Goldfeld} Goldfeld, Cuff, and Permuter focused only on the
secrecy capacity $C_{\mathsf{SS}}$, i.e., the maximum transmission
rate of the secret message without a constraint on non-secret message
required by the legitimate user. Here we consider a more general scenario:
the simultaneous transmission of the secret and non-secret messages.
By the above-mentioned expurgation argument, we can obtain a complete
characterization of the admissible region of $(R_{0},R_{1})$ under
the secrecy constraint $\max_{m_{1}}D_{1+s}\left(P_{Z^{n}|M_{1}=m_{1}}\|Q_{Z^{n}}\right)\rightarrow0,s\in[-1,1]\cup\{\infty\}$,
which turns out to be the same as $\mathcal{R}_{1+s}(Q_{Z})$. 
\end{rem}
The result of Theorem \ref{thm:StochasticEncoder} for the binary
wiretap channel $Y=X\oplus V_{1},V_{1}\sim\mathsf{Bern}\left(0.1\right)$
and $Z=X\oplus V_{2},V_{2}\sim\mathsf{Bern}\left(0.3\right)$ with
target distribution $Q_{Z}=\mathsf{Bern}\left(0.5\right)$ and $s=1$
is illustrated in Fig.~\ref{fig:Resolvability-1-1-1-1-1}. From the
figure, we observe that different from the deterministic encoder case,
for this case the achievability of a rate pair $(R_{0},R_{1})$ indeed
implies the achievability of a rate pair $(R_{0}',R_{1}')$ such that
$R_{0}'\leq R_{0},R_{1}'\leq R_{1}$.

\section{Conclusion and Future Work}

\label{sec:concl}In this paper, we studied a generalized version
of channel resolvability problem, in which the (normalized or unnormalized)
R\'enyi divergence is used to measure the level of approximation. We
also applied these results to the wiretap channel.

Our results generalize or extend several classical and recent results.
Our resolvability results extend those by Han and Verd\'u \cite{Han}
and by Hayashi \cite{Hayashi06,Hayashi11} as we consider R\'enyi divergences
with orders in $[0,2]\cup\{\infty\}$. Our results for the wiretap
channel generalize those by Hou and Kramer \cite{hou2014effective},
and extend those by Wyner \cite{Wyner75} and Csisz\'ar and K\"orner \cite{Csiszar78},
as we measure the effective secrecy (or the leakage) using the R\'enyi
divergence. As discussed in Remark \ref{rmk:ssc}, our result on the
wiretap channel is also related to the semantic-security capacity
studied by Golfeld, Cuff, and Permuter~\cite{Goldfeld}.

In the future, we plan to explore various closely related problems
to the one contained herein. 
\begin{enumerate}
\item {\em R\'enyi common information}: Wyner \cite{Wyner} defined the
common information between two sources is the minimum rate of commonness
needed to simulate these two source in a distributed fashion. In his
original work, the normalized relative entropy was used to measure
the level of approximation. We can generalize his problem by replacing
the relative entropy with the R\'enyi divergence, and define the minimum
rate for this case as \emph{R\'enyi common information}. In fact, a
complete characterization of the R\'enyi common information for order
$\in[0,1]$ and bounds for order $\in(1,2]\cup\{\infty\}$ were provided
by us in \cite{yu2018wyner,yu2018corrections,yu2018on}. Furthermore,
the equivalence between the R\'enyi common information with order $\infty$
and the exact common information was given in \cite{yu2018on}.
\item {\em Distributed channel synthesis under the R\'enyi divergence}:
The coordination problem or distributed channel synthesis problem
was studied by Cuff, Permuter, and Cover \cite{Cuff10,Cuff}. In this
problem, an observer (encoder) of a source sequence describes the
sequence to a distant random number generator (decoder) that produces
another sequence. What is the minimum rate of description needed to
achieve a joint distribution that is statistically indistinguishable,
under the TV distance, from the distribution induced by a given channel?
For this problem, Cuff \cite{Cuff} provided a complete characterization
of the minimum rate. We can enhance the level of coordination by replacing
the TV distance measure with the R\'enyi divergence. For this enhanced
version of the problem, we are interested in characterizing the corresponding
admissible rate region. A variant of the infinity-order case was used to study
exact channel synthesis by us in \cite{yu2018exact}.
\end{enumerate}

\appendices{ }

\section{Preliminaries for the Proofs}
\begin{lem}
\label{lem:typecovering} 
\begin{enumerate}
\item Assume $\mathcal{X}$ is a finite set. Then for any $P_{X}\in\mathcal{P}\left(\mathcal{X}\right)$,
one can find a sequence of types $T_{X}^{\left(n\right)}\in\mathcal{P}^{\left(n\right)}\left(\mathcal{X}\right),n\in\bbN$
such that $\big|P_{X}-T_{X}^{\left(n\right)}\big| \le\frac{\left|\mathcal{X}\right|}{2n}$
as $n\rightarrow\infty$. 
\item Assume $\mathcal{X},\mathcal{Y}$ are finite sets. Then for any sequence
of types $T_{X}^{\left(n\right)}\in\mathcal{P}^{\left(n\right)}\left(\mathcal{X}\right),n\in\bbN$
and any $P_{Y|X}\in\mathcal{P}\left(\mathcal{Y}|\mathcal{X}\right)$,
one can find a sequence of conditional types $V_{Y|X}^{(n)}\in\mathcal{P}^{\left(n\right)}\big(\mathcal{Y}|T_{X}^{\left(n\right)}\big),n\in\bbN$
such that $\big|T_{X}^{\left(n\right)}P_{Y|X}-T_{X}^{\left(n\right)}V_{Y|X}^{(n)}\big| \le \frac{\left|\mathcal{X}\right|\left|\mathcal{Y}\right|}{2n}$
as $n\rightarrow\infty$. 
\end{enumerate}
\end{lem}
Statement 1) is exactly~\cite[Lem.~2.1.2]{Dembo}. The proof of statement
2) follows similarly so its proof is omitted.

We also have the following property concerning the optimization over
the set of types and conditional types. To save space, the proof is
omitted. 
\begin{lem}
\label{lem:minequality} 
\begin{enumerate}
\item Assume $\mathcal{X}$ is a finite set. Then for any continuous (under
TV distance) function $f:\mathcal{P}\left(\mathcal{X}\right)\rightarrow\mathbb{R}$,
we have\footnote{Since $\mathcal{P}\left(\mathcal{X}\right)$ and $\mathcal{P}^{\left(n\right)}\left(\mathcal{X}\right)$
are compact (closed and bounded) and $f$ is continuous on $\mathcal{P}\left(\mathcal{X}\right)$,
the infima of $\inf_{P_{X}\in\mathcal{P}\left(\mathcal{X}\right)}f\left(P_{X}\right)$
and $\inf_{P_{X}\in\mathcal{P}^{\left(n\right)}\left(\mathcal{X}\right)}f\left(P_{X}\right)$
are actually minima. } 
\begin{equation}
\lim_{n\rightarrow\infty}\min_{P_{X}\in\mathcal{P}^{\left(n\right)}\left(\mathcal{X}\right)}f\left(P_{X}\right)=\min_{P_{X}\in\mathcal{P}\left(\mathcal{X}\right)}f\left(P_{X}\right).
\end{equation}
\item Assume $\mathcal{X},\mathcal{Y}$ are finite sets. Then for any continuous
function $f:\mathcal{P}\left(\mathcal{X}\times\mathcal{Y}\right)\rightarrow\mathbb{R}$
and any sequence of types $T_{X}^{\left(n\right)}\in\mathcal{P}^{\left(n\right)}\left(\mathcal{X}\right),n\in\bbN$,
we have 
\begin{align}
 & \min_{P_{Y|X}\in\mathcal{P}^{\left(n\right)}(\mathcal{Y}|T_{X}^{\left(n\right)})}f\big(T_{X}^{\left(n\right)}P_{Y|X}\big)\nonumber \\
 & =\min_{P_{Y|X}\in\mathcal{P}\left(\mathcal{Y}|\mathcal{X}\right)}f\big(T_{X}^{\left(n\right)}P_{Y|X}\big)+o\left(1\right).\label{eqn:state2}
\end{align}
\end{enumerate}
\end{lem}
\begin{rem}
We have 
\begin{align}
 & \lim_{n\rightarrow\infty}\min_{P_{Y|X}\in\mathcal{P}^{\left(n\right)}(\mathcal{Y}|T_{X}^{\left(n\right)})}f\big(T_{X}^{\left(n\right)}P_{Y|X}\big)\nonumber \\
 & =\lim_{n\rightarrow\infty}\min_{P_{Y|X}\in\mathcal{P}\left(\mathcal{Y}|\mathcal{X}\right)}f\big(T_{X}^{\left(n\right)}P_{Y|X}\big)
\end{align}
if either one of the limits above exists. 
\end{rem}
\begin{lem}
\label{lem:typeequality} For any joint type $T_{Y}V_{X|Y}\in\mathcal{P}^{\left(n\right)}\left(\mathcal{X\times Y}\right)$
and any distribution $P_{X^{n}}\in\mathcal{P}\left(\mathcal{X}^{n}\right)$
(not restricted to be i.i.d.), we have 
\begin{equation}
\sum_{y^{n}\in\mathcal{T}_{T_{Y}}}P_{X^{n}}\big(\mathcal{T}_{V_{X|Y}}(y^{n})\big)=\e^{nH\left(V_{Y|X}|T_{X}\right)+n\delta_{n}}P_{X^{n}}(\mathcal{T}_{T_{X}}),
\end{equation}
where $T_{X}V_{Y|X}=T_{Y}V_{X|Y}$. 
\end{lem}
The proof of Lemma \ref{lem:typeequality} follows from a straightforward
application of the method of types~\cite{Csi97} and so is omitted.
\iffalse 
\begin{IEEEproof}
For brevity, we write $P=P_{X^{n}}$. By the standard method of types~\cite{Csi97},
\begin{align}
 & \sum_{y^{n}\in\mathcal{T}_{T_{Y}}}P\Bigl(\mathcal{T}_{V_{X|Y}}(y^{n})\Bigr)=\sum_{y^{n}\in\mathcal{T}_{T_{Y}}}\sum_{x^{n}\in\mathcal{T}_{V_{X|Y}}\left(y^{n}\right)}P\left(x^{n}\right)\\
 & =\sum_{x^{n}\in\mathcal{T}_{T_{X}}}\sum_{y^{n}\in\mathcal{T}_{V_{Y|X}}\left(x^{n}\right)}P\left(x^{n}\right)=\sum_{x^{n}\in\mathcal{T}_{T_{X}}}\e^{nH\left(V_{Y|X}|T_{X}\right)+n\delta_{n}}P\left(x^{n}\right)\\
 & =\e^{nH\left(V_{Y|X}|T_{X}\right)+n\delta_{n}}P(\mathcal{T}_{T_{X}}).
\end{align}
\end{IEEEproof}
\fi 
\begin{lem}
\label{lem:norm} \cite[Problem 4.15(f)]{Gallager} Assume $\left\{ a_{i}\right\} $
are non-negative real numbers. Then for $p\geq1$, we have 
\begin{equation}
\sum_{i}a_{i}^{p}\leq\left(\sum_{i}a_{i}\right)^{p},
\end{equation}
and for $0<p\leq1$, we have 
\begin{equation}
\sum_{i}a_{i}^{p}\geq\left(\sum_{i}a_{i}\right)^{p}.\label{eqn:p_ineq}
\end{equation}
\end{lem}
%\section{\label{sec:min}Proof of Lemma \ref{lem:minequality}}

Note that $\left(\sum_{i}a_{i}^{p}\right)^{1/p}$ is a norm for $p\geq1$,
but not for $0<p<1$. 

\section{\label{sec:OneShot}Proofs of Lemmas \ref{lem:oneshotach} and \ref{lem:oneshotcon} }

\subsection{\label{subsec:Direct-Part-for-1}Direct Part for Case $1+s$ with
$s\in[0,1]$}

Observe that 
\begin{align}
 & \e^{sD_{1+s}(P_{Y\mathcal{C}}\|Q_{Y}\times P_{\mathcal{C}})}\nonumber \\
 & =\mathbb{E}_{\mathcal{C}}\sum_{y}P^{1+s}\left(y|\mathcal{C}\right)Q^{-s}\left(y\right)\\
 & =\mathbb{E}_{\mathcal{C}}\sum_{y}\sum_{m}P\left(m\right)P\left(y|f_{\mathcal{C}}\left(m\right)\right)\biggl(P\left(m\right)P\left(y|f_{\mathcal{C}}\left(m\right)\right)\nonumber \\
 & \qquad+\sum_{m'\neq m}P(m')P\left(y|f_{\mathcal{C}}\left(m'\right)\right)\biggr)^{s}Q^{-s}\left(y\right).\label{eq:-149}
\end{align}
Then using Lemma \ref{lem:norm}, we get 
\begin{align}
 & \e^{sD_{1+s}(P_{Y\mathcal{C}}\|Q_{Y}\times P_{\mathcal{C}})}\leq L_{1}+L_{2},\label{eq:-150}
\end{align}
where 
\begin{align}
 & L_{1}:=\sum_{y}\sum_{m}P^{1+s}\left(m\right)\bbE_{\mathcal{C}}\left[P^{1+s}\left(y|f_{\mathcal{C}}\left(m\right)\right)\right]Q^{-s}\left(y\right)\\
 & L_{2}:=\mathbb{E}_{\mathcal{C}}\sum_{y}\sum_{m}P\left(m\right)P\left(y|f_{\mathcal{C}}\left(m\right)\right)\nonumber \\
 & \qquad\times\left(\sum_{m'\neq m}P(m')P\left(y|f_{\mathcal{C}}\left(m'\right)\right)\right)^{s}Q^{-s}\left(y\right).
\end{align}

Furthermore, $L_{1}$ and $L_{2}$ can be respectively expressed and
upper bounded as follows.

{} 
\begin{align}
L_{1} & =\sum_{y}\sum_{m}P^{1+s}\left(m\right)\sum_{x}P\left(x\right)P^{1+s}\left(y|x\right)Q^{-s}\left(y\right)\\
 & =\e^{\log\sum_{x,y}P\left(x\right)P^{1+s}\left(y|x\right)Q^{-s}\left(y\right)-sR},\label{eq:-3-3-1}
\end{align}
and 
\begin{align}
L_{2} & =\sum_{y}\sum_{m}P\left(m\right)\bbE_{\mathcal{C}}\left[P\left(y|f_{\mathcal{C}}\left(m\right)\right)\right]\nonumber \\
 & \qquad\times\bbE_{\mathcal{C}}\left[\left(\sum_{m'\neq m}P(m')P\left(y|f_{\mathcal{C}}\left(m'\right)\right)\right)^{s}\right]Q^{-s}\left(y\right)\label{eq:-114-1}\\
 & \leq\sum_{y}\sum_{m}P\left(m\right)\bbE_{\mathcal{C}}\left[P\left(y|f_{\mathcal{C}}\left(m\right)\right)\right]\nonumber \\
 & \qquad\times\left(\sum_{m'\neq m}P(m')\bbE_{\mathcal{C}}\left[P\left(y|f_{\mathcal{C}}\left(m'\right)\right)\right]\right)^{s}Q^{-s}\left(y\right)\label{eq:-42-2}\\
 & =\sum_{y}\sum_{m}P\left(m\right)\sum_{x}P\left(x\right)P\left(y|x\right)\nonumber \\
 & \qquad\times\left(\sum_{m'\neq m}P(m')\sum_{x}P\left(x\right)P\left(y|x\right)\right)^{s}Q^{-s}\left(y\right)\\
 & \leq\sum_{y}P^{1+s}\left(y\right)Q^{-s}\left(y\right)\\
 & =\e^{sD_{1+s}(P_{Y}\|Q_{Y})}.\label{eq:-148}
\end{align}
where \eqref{eq:-114-1} follows since $f_{\mathcal{C}}\left(m\right)$
and $f_{\mathcal{C}}\left(m'\right)$ are independent for $m\neq m'$,
and \eqref{eq:-42-2} follows since $x\mapsto x^{s}$ is a concave
function.

Combining \eqref{eq:-149}, \eqref{eq:-3-3-1} and \eqref{eq:-148}
gives us 
\begin{align}
 & \e^{sD_{1+s}(P_{Y\mathcal{C}}\|Q_{Y}\times P_{\mathcal{C}})}\nonumber \\
 & \leq\e^{\log\sum_{x,y}P\left(x\right)P^{1+s}\left(y|x\right)Q^{-s}\left(y\right)-sR}+\e^{sD_{1+s}(P_{Y}\|Q_{Y})}\\
 & \leq2\max\{\e^{\log\sum_{x,y}P\left(x\right)P^{1+s}\left(y|x\right)Q^{-s}\left(y\right)-sR},\e^{sD_{1+s}(P_{Y}\|Q_{Y})}\}\\
 & =2\e^{s\Gamma_{1+s}\left(P_{Y|X},Q_{Y},R\right)}.
\end{align}

\subsection{Direct Part for Case $1-s$ with $s\in(0,1)$}

For the random code given in Lemma \ref{lem:oneshotach}, we have
\begin{align}
 & \e^{-sD_{1-s}(P_{Y\mathcal{C}}\|Q_{Y}\times P_{\mathcal{C}})}\nonumber \\
 & =\bbE_{\mathcal{C}}\sum_{y}P^{1-s}\left(y|\mathcal{C}\right)Q^{s}\left(y\right)\\
 & =\bbE_{\mathcal{C}}\sum_{y}\sum_{m}P\left(m\right)P\left(y|f_{\mathcal{C}}\left(m\right)\right)\nonumber \\
 & \qquad\times\left(\sum_{m}P\left(m\right)P\left(y|f_{\mathcal{C}}\left(m\right)\right)\right)^{-s}Q^{s}\left(y\right)\\
 & =\bbE_{\mathcal{C}}\sum_{y}\sum_{m}P\left(m\right)P\left(y|f_{\mathcal{C}}\left(m\right)\right)\Bigl(P\left(m\right)P\left(y|f_{\mathcal{C}}\left(m\right)\right)\nonumber \\
 & \qquad+\sum_{m'\neq m}P(m')P\left(y|f_{\mathcal{C}}\left(m'\right)\right)\Bigr)^{-s}Q^{s}\left(y\right)\\
 & \geq\sum_{y}\sum_{m}P\left(m\right)\bbE_{\mathcal{C}}\biggl[P\left(y|f_{\mathcal{C}}\left(m\right)\right)\biggl(P\left(m\right)P\left(y|f_{\mathcal{C}}\left(m\right)\right)\nonumber \\
 & \qquad+\sum_{m'\neq m}P(m')\bbE_{\mathcal{C}}\left[P\left(y|f_{\mathcal{C}}\left(m'\right)\right)\right]\biggr)^{-s}\biggr]Q^{s}\left(y\right)\label{eq:-44-1}\\
 & =\sum_{y}\sum_{m}P\left(m\right)\bbE_{\mathcal{C}}\biggl[P\left(y|f_{\mathcal{C}}\left(m\right)\right)\biggl(P\left(m\right)P\left(y|f_{\mathcal{C}}\left(m\right)\right)\nonumber \\
 & \qquad+\sum_{m'\neq m}P(m')\sum_{x}P\left(x\right)P\left(y|x\right)\biggr)^{-s}\biggr]Q^{s}\left(y\right)\\
 & \geq\sum_{y}\sum_{m}P\left(m\right)\bbE_{\mathcal{C}}\biggl[P\left(y|f_{\mathcal{C}}\left(m\right)\right)\biggl(P\left(m\right)P\left(y|f_{\mathcal{C}}\left(m\right)\right)\nonumber \\
 & \qquad+P\left(y\right)\biggr)^{-s}\biggr]Q^{s}\left(y\right)\\
 & \geq\sum_{y}\sum_{m}P\left(m\right)\bbE_{\mathcal{C}}\biggl[P\left(y|f_{\mathcal{C}}\left(m\right)\right)\nonumber \\
 & \qquad\times\left(2\max\left\{ P\left(m\right)P\left(y|f_{\mathcal{C}}\left(m\right)\right),P\left(y\right)\right\} \right)^{-s}\biggr]Q^{s}\left(y\right)\\
 & =2^{-s}\sum_{y}\sum_{m}P\left(m\right)\bbE_{\mathcal{C}}\Biggl[P\left(y|f_{\mathcal{C}}\left(m\right)\right)\biggl(P\left(m\right)^{-s}\nonumber \\
 & \qquad\times P\left(y|f_{\mathcal{C}}\left(m\right)\right)^{-s}1\left\{ P\left(m\right)P\left(y|f_{\mathcal{C}}\left(m\right)\right)\geq P\left(y\right)\right\} \nonumber \\
 & \qquad+P^{-s}\left(y\right)1\left\{ P\left(m\right)P\left(y|f_{\mathcal{C}}\left(m\right)\right)<P\left(y\right)\right\} \biggr)\Biggr]Q^{s}\left(y\right)\\
 & =2^{-s}\biggl(\sum_{y}\sum_{m}\bbE_{\mathcal{C}}\biggl[\left(P\left(m\right)P\left(y|f_{\mathcal{C}}\left(m\right)\right)\right)^{1-s}\nonumber \\
 & \qquad\times Q^{s}\left(y\right)1\left\{ P\left(m\right)P\left(y|f_{\mathcal{C}}\left(m\right)\right)\geq P\left(y\right)\right\} \biggr]\nonumber \\
 & \qquad+\sum_{y}\sum_{m}P\left(m\right)\bbE_{\mathcal{C}}\biggl[P\left(y|f_{\mathcal{C}}\left(m\right)\right)P^{-s}\left(y\right)\nonumber \\
 & \qquad\times Q^{s}\left(y\right)1\left\{ P\left(m\right)P\left(y|f_{\mathcal{C}}\left(m\right)\right)<P\left(y\right)\right\} \biggr]\biggr)\\
 & =2^{-s}\biggl(\sum_{m,x,y}\e^{-\left(1-s\right)R}P\left(x\right)P^{1-s}\left(y|x\right)\nonumber \\
 & \qquad\times Q^{s}\left(y\right)1\left\{ \frac{P\left(y|x\right)}{P\left(y\right)}\geq\e^{R}\right\} \nonumber \\
 & \qquad+\sum_{m,x,y}P\left(m\right)P\left(x\right)P\left(y|x\right)P^{-s}\left(y\right)\nonumber \\
 & \qquad\times Q^{s}\left(y\right)1\left\{ \frac{P\left(y\right)}{P\left(y|x\right)}>\e^{-R}\right\} \biggr)\label{eq:-136}\\
 & =2^{-s}\biggl(\e^{sR}\sum_{x,y}P\left(x\right)P^{1-s}\left(y|x\right)Q^{s}\left(y\right)1\left\{ \frac{P\left(y|x\right)}{P\left(y\right)}\geq\e^{R}\right\} \nonumber \\
 & \qquad+\sum_{x,y}P\left(x,y\right)P^{-s}\left(y\right)Q^{s}\left(y\right)1\left\{ \frac{P\left(y\right)}{P\left(y|x\right)}>\e^{-R}\right\} \biggr),
\end{align}
where \eqref{eq:-44-1} follows from that $x\mapsto x^{-s}$ is a
convex function and $f_{\mathcal{C}}\left(m\right)$ and $f_{\mathcal{C}}\left(m'\right)$
are independent for $m\neq m'$, and \eqref{eq:-136} follows since
by the construction of the code, $\mathbb{P}\left(f_{\mathcal{C}}\left(m\right)=x\right)=P_{X}(x)$.

\subsection{Converse Part for Case $1+s$ with $s\in(0,\infty]$}

Observe that 
\begin{align}
 & \e^{sD_{1+s}(P_{Y}\|Q_{Y})}\nonumber \\
 & =\sum_{y}P^{1+s}\left(y\right)Q^{-s}\left(y\right)\\
 & =\sum_{y}\sum_{m}P\left(m\right)P\left(y|f\left(m\right)\right)\biggl(P\left(m\right)P\left(y|f\left(m\right)\right)\nonumber \\
 & \qquad+\sum_{m'\neq m}P(m')P\left(y|f\left(m'\right)\right)\biggr)^{s}Q^{-s}\left(y\right)\\
 & \geq\sum_{y}\sum_{m}P\left(m\right)P\left(y|f\left(m\right)\right)\left(P\left(m\right)P\left(y|f\left(m\right)\right)\right)^{s}Q^{-s}\left(y\right)\\
 & =\e^{-sR}\sum_{y}\sum_{m}P\left(m\right)P^{1+s}\left(y|f\left(m\right)\right)Q^{-s}\left(y\right)\\
 & =\e^{-sR}\sum_{y,x,m}P\left(m\right)P^{1+s}\left(y|x\right)Q^{-s}\left(y\right)1\left\{ f\left(m\right)=x\right\} \\
 & =\e^{-sR}\sum_{x,y}P\left(x\right)P^{1+s}\left(y|x\right)Q^{-s}\left(y\right)\\
 & =\e^{\log\sum_{x,y}P\left(x\right)P^{1+s}\left(y|x\right)Q^{-s}\left(y\right)-sR},\label{eq:-68}
\end{align}
where $P\left(x\right):=\sum_{m}P\left(m\right)1\left\{ f\left(m\right)=x\right\} $
and $P\left(y\right):=\sum_{x}P\left(x\right)P\left(y|x\right)$ respectively
denote the distributions of $X$ and $Y$ induced by the mapping $f$. 

On the other hand, 
\begin{align}
 & \e^{sD_{1+s}(P_{Y}\|Q_{Y})}\nonumber \\
 & =\sum_{y}\left(\sum_{x}P\left(x\right)P\left(y|x\right)\right)^{1+s}Q^{-s}\left(y\right).\label{eq:-135}
\end{align}
Putting \eqref{eq:-68} and \eqref{eq:-135} together yields the desired
result.

\subsection{Converse Part for Case $1-s$ with $s\in[0,1)$}

Observe that 
\begin{align}
 & \e^{-sD_{1-s}(P_{Y}\|Q_{Y})}\nonumber \\
 & =\sum_{y}P^{1-s}\left(y\right)Q^{s}\left(y\right)\\
 & =\sum_{y}\left(\sum_{m}P\left(m\right)P\left(y|f\left(m\right)\right)\right)^{1-s}Q^{s}\left(y\right)\\
 & =\sum_{y}\left(\sum_{m}P\left(m\right)P\left(y|f\left(m\right)\right)\right)^{1-s}Q^{s}\left(y\right)\nonumber \\*
 & \quad\times\biggl(1\Bigl\{ P\left(m\right)P\left(y|f\left(m\right)\right)\geq\sum_{m'\neq m}P(m')P\left(y|f\left(m'\right)\right)\Bigr\}\nonumber \\*
 & \qquad+1\Bigl\{ P\left(m\right)P\left(y|f\left(m\right)\right)<\sum_{m'\neq m}P(m')P\left(y|f\left(m'\right)\right)\Bigr\}\biggr)\\
 & \leq\sum_{y}\sum_{m}\left(P\left(m\right)P\left(y|f\left(m\right)\right)\right)^{1-s}Q^{s}\left(y\right)\nonumber \\
 & \qquad\times1\left\{ P\left(m\right)P\left(y|f\left(m\right)\right)\geq\sum_{m'\neq m}P(m')P\left(y|f\left(m'\right)\right)\right\} \nonumber \\
 & \quad+\sum_{y}\left(\sum_{m}P\left(m\right)P\left(y|f\left(m\right)\right)\right)^{1-s}Q^{s}\left(y\right)\nonumber \\
 & \qquad\times1\left\{ P\left(m\right)P\left(y|f\left(m\right)\right)<\sum_{m'\neq m}P(m')P\left(y|f\left(m'\right)\right)\right\} \\
 & =\e^{sR}\sum_{y}\sum_{m}P\left(m\right)P^{1-s}\left(y|f\left(m\right)\right)Q^{s}\left(y\right)\nonumber \\
 & \qquad\times1\left\{ 2P\left(m\right)P\left(y|f\left(m\right)\right)\geq P\left(y\right)\right\} \nonumber \\
 & \quad+\sum_{y}\left(\sum_{m}P\left(m\right)P\left(y|f\left(m\right)\right)\right)^{1-s}Q^{s}\left(y\right)\nonumber \\
 & \qquad\times1\left\{ 2P\left(m\right)P\left(y|f\left(m\right)\right)<P\left(y\right)\right\} \\
 & =\e^{sR}\sum_{y}\sum_{m}P\left(m\right)P^{1-s}\left(y|f\left(m\right)\right)Q^{s}\left(y\right)\nonumber \\
 & \qquad\times1\left\{ \frac{P\left(y|f\left(m\right)\right)}{P\left(y\right)}\geq\frac{\e^{R}}{2}\right\} \nonumber \\
 & \quad+\sum_{y}\sum_{m}P\left(m\right)P\left(y|f\left(m\right)\right)P\left(y\right)^{-s}Q^{s}\left(y\right)\nonumber \\
 & \qquad\times1\left\{ \frac{P\left(y|f\left(m\right)\right)}{P\left(y\right)}<\frac{\e^{R}}{2}\right\} \\
 & \leq\e^{sR}\sum_{x,y}P\left(x\right)P^{1-s}\left(y|x\right)Q^{s}\left(y\right)1\left\{ \frac{P\left(y|x\right)}{P\left(y\right)}\geq\frac{\e^{R}}{2}\right\} \nonumber \\
 & \quad+\sum_{x,y}P\left(x\right)P\left(y|x\right)P^{-s}\left(y\right)Q^{s}\left(y\right)1\left\{ \frac{P\left(y|x\right)}{P\left(y\right)}<\frac{\e^{R}}{2}\right\} ,
\end{align}
where $P\left(x\right):=\sum_{m}P\left(m\right)1\left\{ f\left(m\right)=x\right\} $
and $P\left(y\right):=\sum_{x}P\left(x\right)P\left(y|x\right)$ respectively
denote the distributions of $X$ and $Y$ induced by the mapping $f$.

\section{\label{sec:Proof-of-Theorem-asym}Proof of Proposition \ref{thm:multiletter} }

For the $n$-letter version of the problem, $\calM_{n}=\{1,\ldots,\e^{nR}\}$,
and the channel $P_{Y|X}^{n}$, used $n$ times, can be considered
as a superletter channel. Hence the one-shot bounds given in Lemmas
\ref{lem:oneshotach} and \ref{lem:oneshotcon} can be used to prove
Proposition \ref{thm:multiletter}.

\subsection{\label{subsec:Direct-Part-for}Direct Part for Case $1+s$ with $s\in[0,1]$}

By Lemma \ref{lem:oneshotach}, we have 
\begin{align}
 & \frac{1}{n}D_{1+s}(P_{Y^{n}\mathcal{C}_{n}}\|Q_{Y}^{n}\times P_{\mathcal{C}_{n}})\nonumber \\
 & \leq\frac{1}{n}\Gamma_{1+s}\left(P_{X^{n}},P_{Y^{n}|X^{n}},Q_{Y}^{n},nR\right)+\frac{1}{ns}\log2\\
 & =\max\bigl\{\frac{1}{n}D_{1+s}\left(P_{X^{n}Y^{n}}\|P_{X^{n}}\times Q_{Y}^{n}\right)-R,\nonumber \\
 & \qquad\frac{1}{n}D_{1+s}(P_{Y^{n}}\|Q_{Y}^{n})\bigr\}+o(1).
\end{align}

Since $P_{X^{n}}$ is arbitrary, we have 
\begin{align}
 & \frac{1}{n}\inf_{f_{\mathcal{C}_{n}}}D_{1+s}(P_{Y^{n}\mathcal{C}_{n}}\|Q_{Y}^{n}\times P_{\mathcal{C}_{n}})\nonumber \\
 & \leq\inf_{P_{X^{n}}}\max\bigl\{\frac{1}{n}D_{1+s}\left(P_{X^{n}Y^{n}}\|P_{X^{n}}\times Q_{Y}^{n}\right)-R,\nonumber \\
 & \qquad\frac{1}{n}D_{1+s}(P_{Y^{n}}\|Q_{Y}^{n})\bigr\}+o(1)\\
 & =\Gamma_{1+s}^{\left(n\right)}\left(P_{Y|X},Q_{Y},R\right)+o(1).
\end{align}

\subsection{Converse Part for Case $1+s$ with $s\in(0,\infty]$}

By Lemma \ref{lem:oneshotcon}, we have 
\begin{align}
 & \frac{1}{n}D_{1+s}(P_{Y^{n}\mathcal{C}_{n}}\|Q_{Y}^{n}\times P_{\mathcal{C}_{n}})\nonumber \\
 & \geq\frac{1}{n}\Gamma_{1+s}\left(P_{X^{n}},P_{Y^{n}|X^{n}},Q_{Y}^{n},nR\right)\\
 & =\max\bigl\{\frac{1}{n}D_{1+s}\left(P_{X^{n}Y^{n}}\|P_{X^{n}}\times Q_{Y}^{n}\right)-R,\nonumber \\
 & \qquad\frac{1}{n}D_{1+s}(P_{Y^{n}}\|Q_{Y}^{n})\bigr\}\\
 & \geq\inf_{P_{X^{n}}}\max\bigl\{\frac{1}{n}D_{1+s}\left(P_{X^{n}Y^{n}}\|P_{X^{n}}\times Q_{Y}^{n}\right)-R,\nonumber \\
 & \qquad\frac{1}{n}D_{1+s}(P_{Y^{n}}\|Q_{Y}^{n})\bigr\}\\
 & =\Gamma_{1+s}^{\left(n\right)}\left(P_{Y|X},Q_{Y},R\right).
\end{align}

\subsection{Direct Part for Case $1-s$ with $s\in(0,1)$}

Choose $P_{X^{n}}=P_{X}^{n}$ for some $P_{X}$. By Lemma \ref{lem:oneshotach},
we have 
\begin{align}
 & \e^{-sD_{1-s}(P_{Y^{n}\mathcal{C}_{n}}\|Q_{Y}^{n}\times P_{\mathcal{C}_{n}})}\nonumber \\
\geq & 2^{-s}\biggl[\e^{nsR}\sum_{x^{n},y^{n}}P\left(x^{n}\right)P^{1-s}\left(y^{n}|x^{n}\right)Q^{s}\left(y^{n}\right)\nonumber \\
 & \qquad\times1\left\{ \frac{P\left(y^{n}|x^{n}\right)}{P\left(y^{n}\right)}\geq\e^{nR}\right\} \nonumber \\
 & +\sum_{x^{n},y^{n}}P\left(x^{n}\right)P\left(y^{n}|x^{n}\right)P^{-s}\left(y^{n}\right)Q^{s}\left(y^{n}\right)\nonumber \\
 & \qquad\times1\left\{ \frac{P\left(y^{n}|x^{n}\right)}{P\left(y^{n}\right)}<\e^{nR}\right\} \biggr]\\
= & 2^{-s}\biggl[\e^{nsR}\Phi_{1}^{n}\sum_{x^{n},y^{n}}\frac{P\left(x^{n}\right)P^{1-s}\left(y^{n}|x^{n}\right)Q^{s}\left(y^{n}\right)}{\Phi_{1}^{n}}\nonumber \\
 & \qquad\times1\left\{ \frac{P\left(y^{n}|x^{n}\right)}{P\left(y^{n}\right)}\geq\e^{nR}\right\} \nonumber \\
 & +\Phi_{2}^{n}\sum_{x^{n},y^{n}}\frac{P\left(x^{n}\right)P\left(y^{n}|x^{n}\right)P^{-s}\left(y^{n}\right)Q^{s}\left(y^{n}\right)}{\Phi_{2}^{n}}\nonumber \\
 & \qquad\times1\left\{ \frac{P\left(y^{n}|x^{n}\right)}{P\left(y^{n}\right)}<\e^{nR}\right\} \biggr],\label{eq:-5-1-1}
\end{align}
where 
\begin{align}
\Phi_{1} & :=\left(\sum_{x^{n},y^{n}}P\left(x^{n}\right)P^{1-s}\left(y^{n}|x^{n}\right)Q^{s}\left(y^{n}\right)\right)^{1/n}\label{eq:-137}\\
 & =\e^{-sD_{1-s}\left(P_{XY}\|P_{X}\times Q_{Y}\right)},\\
\Phi_{2} & :=\left(\sum_{x^{n},y^{n}}P\left(x^{n}\right)P\left(y^{n}|x^{n}\right)P^{-s}\left(y^{n}\right)Q^{s}\left(y^{n}\right)\right)^{1/n}\label{eq:-138}\\
 & =\e^{-sD_{1-s}(P_{Y}\|Q_{Y})}.
\end{align}
According to large deviation theory~\cite{Dembo} (Cram\'er's theorem),
we have 
\begin{align}
 & \lim_{n\rightarrow\infty}-\frac{1}{n}\log\sum_{x^{n},y^{n}}\frac{P\left(x^{n}\right)P^{1-s}\left(y^{n}|x^{n}\right)Q^{s}\left(y^{n}\right)}{\Phi_{1}^{n}}\nonumber \\
 & \qquad\times1\left\{ \frac{P\left(y^{n}|x^{n}\right)}{P\left(y^{n}\right)}\geq\e^{nR}\right\} \nonumber \\
 & =\max_{t\geq0}\left(-\log\sum_{x,y}\frac{P\left(x\right)P^{1-s}\left(y|x\right)Q^{s}\left(y\right)}{\Phi_{1}}\left(\frac{P\left(y|x\right)}{P\left(y\right)\e^{R}}\right)^{t}\right)\\
 & =\max_{t\geq0}\left(tR-\log\sum_{x,y}P\left(x\right)P^{1-\left(s-t\right)}\left(y|x\right)P^{-t}\left(y\right)Q^{s}\left(y\right)\right)\nonumber \\
 & \qquad+\log\Phi_{1},
\end{align}
and 
\begin{align}
 & \lim_{n\rightarrow\infty}-\frac{1}{n}\log\sum_{x^{n},y^{n}}\frac{P\left(x^{n}\right)P\left(y^{n}|x^{n}\right)P^{-s}\left(y^{n}\right)Q^{s}\left(y^{n}\right)}{\Phi_{2}^{n}}\nonumber \\
 & \qquad\times1\left\{ \frac{P\left(y^{n}|x^{n}\right)}{P\left(y^{n}\right)}<\e^{nR}\right\} \nonumber \\
 & =\max_{t\geq0}\left(-\log\sum_{x,y}\frac{P\left(x,y\right)P^{-s}\left(y\right)Q^{s}\left(y\right)}{\Phi_{2}}\left(\frac{\e^{R}P\left(y\right)}{P\left(y|x\right)}\right)^{t}\right)\\
 & =\max_{t\geq0}\left(-tR-\log\sum_{x,y}P\left(x\right)P^{1-t}\left(y|x\right)P^{t-s}\left(y\right)Q^{s}\left(y\right)\right)\nonumber \\
 & \qquad+\log\Phi_{2}.
\end{align}
Substituting these into \eqref{eq:-5-1-1}, we have 
\begin{align}
 & \e^{-sD_{1-s}(P_{Y^{n}\mathcal{C}_{n}}\|Q_{Y}^{n}\times P_{\mathcal{C}_{n}})}\nonumber \\
 & \geq2^{-s}\{\e^{-n\max_{t\geq0}\tau\left(R,s,s-t\right)-n\delta_{n}}+\e^{-n\max_{t\geq0}\tau\left(R,s,t\right)-n\delta_{n}'}\},
\end{align}
where 
\begin{align}
 & \tau\left(R,s,t\right)\nonumber \\
 & :=-tR-\log\sum_{x,y}P\left(x\right)P^{1-t}\left(y|x\right)P^{-\left(s-t\right)}\left(y\right)Q^{s}\left(y\right).
\end{align}
That is, 
\begin{align}
 & \frac{1}{n}D_{1-s}(P_{Y^{n}\mathcal{C}_{n}}\|Q_{Y}^{n}\times P_{\mathcal{C}_{n}})\nonumber \\
 & \leq\frac{1}{s}\min\left\{ \max_{t\geq0}\tau\left(R,s,s-t\right),\max_{t\geq0}\tau\left(R,s,t\right)\right\} +\delta_{n}+\delta_{n}'\\
 & =\frac{1}{s}\min\left\{ \max_{t\leq s}\tau\left(R,s,t\right),\max_{t\geq0}\tau\left(R,s,t\right)\right\} +\delta_{n}+\delta_{n}',
\end{align}
where 
\begin{align}
 & \tau\left(R,s,t\right)\nonumber \\
 & :=-tR-\log\sum_{x,y}P\left(x\right)P^{1-t}\left(y|x\right)P^{-\left(s-t\right)}\left(y\right)Q^{s}\left(y\right).\label{eq:-124}
\end{align}

We claim that given $R$ and $s$, $\tau\left(R,s,t\right)$ is concave
in $t$; see Lemma \ref{lem:convex} below. This implies that 
\begin{equation}
\min\left\{ \max_{t\leq s}\tau\left(R,s,t\right),\max_{t\geq0}\tau\left(R,s,t\right)\right\} =\max_{t\in\left[0,s\right]}\tau\left(R,s,t\right).\label{eq:-8}
\end{equation}
Hence we have 
\begin{equation}
\limsup_{n\rightarrow\infty}\frac{1}{n}D_{1-s}(P_{Y^{n}\mathcal{C}_{n}}\|Q_{Y}^{n}\times P_{\mathcal{C}_{n}})\leq\frac{1}{s}\max_{t\in\left[0,s\right]}\tau\left(R,s,t\right).
\end{equation}
Moreover, $P_{X}$ is arbitrary, hence 
\begin{align}
 & \inf_{P_{X^{n}}}\frac{1}{n}D_{1-s}(P_{Y^{n}\mathcal{C}_{n}}\|Q_{Y}^{n}\times P_{\mathcal{C}_{n}})\nonumber \\
 & \leq\frac{1}{s}\min_{P_{X}}\max_{t\in\left[0,s\right]}\tau\left(R,s,t\right)+\delta_{n}+\delta_{n}'\\
 & =\Gamma_{1-s}^{\left(1\right)}\left(P_{Y|X},Q_{Y},R\right)+\delta_{n}+\delta_{n}'.\label{eq:-41}
\end{align}

Note that $\Gamma_{1-s}^{\left(1\right)}\left(P_{Y|X},Q_{Y},R\right)$
is a single-letter version of $\Gamma_{1-s}^{\left(n\right)}\left(P_{Y|X},Q_{Y},R\right)$.
To achieve the desired result, we set $P_{X^{mk}}=P_{X^{k}}^{m}$
%$P_{X^{mk}}=\prod_{i=1}^{m}P_{X^{k}}$ \red{should this be $P_{X}^{mk}$?}
for some fixed $k$. Consider $X^{k}$ as a super-letter, then applying
the derivations above, we have as $m\rightarrow\infty$, 
\begin{align}
 & \inf_{P_{X^{mk}}}\frac{1}{mk}D_{1-s}(P_{Y^{mk}\mathcal{C}_{mk}}\|Q_{Y}^{mk}\times P_{\mathcal{C}_{mk}})\nonumber \\
 & \leq\Gamma_{1-s}^{\left(k\right)}\left(P_{Y|X},Q_{Y},R\right)+\delta_{mk}+\delta_{mk}',
\end{align}
where $\delta_{mk},\delta_{mk}'\to0$ as $m\to\infty$ for fixed $k$.
When $n$ is not a multiple of $k$, we consider $X^{k},Y^{k}$ as
super-letters, and then apply the code to the first $m:=\left\lfloor \frac{n}{k}\right\rfloor $
super-letters. Then we have 
\begin{align}
 & \inf_{P_{X^{n}}}\frac{1}{n}D_{1-s}(P_{Y^{n}\mathcal{C}_{n}}\|Q_{Y}^{n}\times P_{\mathcal{C}_{n}})\nonumber \\
 & \leq\inf_{P_{X^{mk}}}\frac{1}{n}D_{1-s}(P_{Y^{mk}\mathcal{C}_{mk}}\|Q_{Y}^{mk}\times P_{\mathcal{C}_{mk}})\nonumber \\
 & \qquad+\inf_{P_{X^{l}}}\frac{1}{n}D_{1-s}(P_{Y^{l}\mathcal{C}_{l}}\|Q_{Y}^{l}\times P_{\mathcal{C}_{l}})\\
 & \leq\inf_{P_{X^{mk}}}\frac{1}{mk}D_{1-s}(P_{Y^{mk}\mathcal{C}_{mk}}\|Q_{Y}^{mk}\times P_{\mathcal{C}_{mk}})\nonumber \\
 & \qquad+\frac{1}{m}\inf_{P_{X^{l}}}\frac{1}{l}D_{1-s}(P_{Y^{l}\mathcal{C}_{l}}\|Q_{Y}^{l}\times P_{\mathcal{C}_{l}}),
\end{align}
where $l:=n-mk<k$. Observe that 
\begin{align}
 & \inf_{P_{X^{l}}}\frac{1}{l}D_{1-s}(P_{Y^{l}\mathcal{C}_{l}}\|Q_{Y}^{l}\times P_{\mathcal{C}_{l}})\nonumber \\
 & \le\inf_{P_{X}^{l}}\frac{1}{l}D_{1-s}(P_{Y^{l}\mathcal{C}_{l}}\|Q_{Y}^{l}\times P_{\mathcal{C}_{l}})\\
 & =\inf_{P_{X}}D_{1-s}(P_{Y\mathcal{C}_{1}}\|Q_{Y}\times P_{\mathcal{C}_{1}}),
\end{align}
and the RHS of the inequality above is finite (as assumed in Section
\ref{subsec:problem}). Hence the LHS of the inequality above is also
finite. Hence for fixed $k$, we have 
\begin{align}
 & \inf_{P_{X^{n}}}\frac{1}{n}D_{1-s}(P_{Y^{n}\mathcal{C}_{n}}\|Q_{Y}^{n}\times P_{\mathcal{C}_{n}})\nonumber \\
 & \leq\inf_{P_{X^{mk}}}\frac{1}{mk}D_{1-s}(P_{Y^{mk}\mathcal{C}_{mk}}\|Q_{Y}^{mk}\times P_{\mathcal{C}_{mk}})+\delta_{m}''\\
 & \leq\Gamma_{1-s}^{\left(k\right)}\left(P_{Y|X},Q_{Y},R\right)+\delta_{mk}+\delta_{mk}'+\delta_{m}''\\
 & =\Gamma_{1-s}^{\left(k\right)}\left(P_{Y|X},Q_{Y},R\right)+o\left(1\right),
\end{align}
where $o(1)$ is a term tending to zero as $m\to\infty$ or $n\to\infty$
since $k$ is fixed. Since $k$ is arbitrary, we obtain the desired
result. 
\begin{lem}
\label{lem:convex} Given $R$ and $s$, $\tau\left(R,s,t\right)$
is concave in $t$. 
\end{lem}
\begin{IEEEproof}
Define $f(x,y):=P\left(x\right)P\left(y|x\right)P^{-s}\left(y\right)Q^{s}\left(y\right)$
and $g(x,y):=\frac{P\left(y\right)}{P\left(y|x\right)}$. Then 
\begin{align}
 & -\log\sum_{x,y}P\left(x\right)P^{1-t}\left(y|x\right)P^{-\left(s-t\right)}\left(y\right)Q^{s}\left(y\right)\nonumber \\
 & =-\log\sum_{x,y}f(x,y)g^{t}(x,y).
\end{align}
Assume $t=\lambda t_{1}+\left(1-\lambda\right)t_{2}$ for $\lambda\in[0,1]$,
then 
\begin{align}
 & \sum_{x,y}f(x,y)g^{t}(x,y)\nonumber \\
 & =\sum_{x,y}f(x,y)g^{\lambda t_{1}+\left(1-\lambda\right)t_{2}}(x,y)\\
 & =\sum_{x,y}\left(f(x,y)g^{t_{1}}(x,y)\right)^{\lambda}\left(f(x,y)g^{t_{2}}(x,y)\right)^{1-\lambda}\\
 & \leq\left(\sum_{x,y}f(x,y)g^{t_{1}}(x,y)\right)^{\lambda}\left(\sum_{x,y}f(x,y)g^{t_{2}}(x,y)\right)^{1-\lambda},\label{eq:-42}
\end{align}
where \eqref{eq:-42} follows from H\"older's inequality.

Hence 
\begin{align}
 & -\log\sum_{x,y}f(x,y)g^{t}(x,y)\nonumber \\
 & \geq-\lambda\log\sum_{x,y}f(x,y)g^{t_{1}}(x,y)\nonumber \\
 & \qquad-\left(1-\lambda\right)\log\sum_{x,y}f(x,y)g^{t_{2}}(x,y).
\end{align}
That is, $\tau\left(R,s,t\right)$ is concave in $t$. 
\end{IEEEproof}

\subsection{Converse Part for Case $1-s$ with $s\in[0,1)$}

By Lemma \ref{lem:oneshotcon}, we have for some $P_{X^{n}}$, 
\begin{align}
 & \e^{-sD_{1-s}(P_{Y^{n}\mathcal{C}_{n}}\|Q_{Y}^{n}\times P_{\mathcal{C}_{n}})}\nonumber \\
 & \leq\e^{nsR}\sum_{x^{n},y^{n}}P\left(x^{n}\right)P^{1-s}\left(y^{n}|x^{n}\right)Q^{s}\left(y^{n}\right)\nonumber \\
 & \quad\times1\left\{ \frac{P\left(y^{n}|x^{n}\right)}{P\left(y^{n}\right)}\geq\frac{\e^{nR}}{2}\right\} \nonumber \\
 & \quad+\sum_{x^{n},y^{n}}P\left(x^{n}\right)P\left(y^{n}|x^{n}\right)P^{-s}\left(y^{n}\right)Q^{s}\left(y^{n}\right)\nonumber \\
 & \quad\times1\left\{ \frac{P\left(y^{n}|x^{n}\right)}{P\left(y^{n}\right)}<\frac{\e^{nR}}{2}\right\} .\label{eq:-3-1}
\end{align}

Denote $R':=R-\frac{1}{n}\log2$. From Markov's inequality, we have
\begin{align}
 & -\frac{1}{n}\log\sum_{x^{n},y^{n}}\frac{P\left(x^{n}\right)P^{1-s}\left(y^{n}|x^{n}\right)Q^{s}\left(y^{n}\right)}{\Phi_{1}^{n}}\nonumber \\
 & \quad\times1\left\{ \frac{P\left(y^{n}|x^{n}\right)}{P\left(y^{n}\right)}\geq\frac{\e^{nR}}{2}\right\} \nonumber \\
 & \geq\max_{t\geq0}\left(t\left(R-\frac{1}{n}\log2\right)-\kappa_{1}\right)+\log\Phi_{1}\\
 & =\max_{t\geq0}\left(tR'-\kappa_{1}\right)+\log\Phi_{1},
\end{align}
and 
\begin{align}
 & -\frac{1}{n}\log\sum_{x^{n},y^{n}}\frac{P\left(x^{n}\right)P\left(y^{n}|x^{n}\right)P^{-s}\left(y^{n}\right)Q^{s}\left(y^{n}\right)}{\Phi_{2}^{n}}\nonumber \\
 & \quad\times1\left\{ \frac{P\left(y^{n}|x^{n}\right)}{P\left(y^{n}\right)}<\frac{\e^{nR}}{2}\right\} \nonumber \\
 & \geq\max_{t\geq0}\left(-tR'-\kappa_{2}\right)+\log\Phi_{2},
\end{align}
where 
\begin{align}
 & \Phi_{1}:=\left(\sum_{x^{n},y^{n}}P\left(x^{n}\right)P^{1-s}\left(y^{n}|x^{n}\right)Q^{s}\left(y^{n}\right)\right)^{1/n},\label{eq:-137-1}\\
 & \Phi_{2}:=\left(\sum_{x^{n},y^{n}}P\left(x^{n}\right)P\left(y^{n}|x^{n}\right)P^{-s}\left(y^{n}\right)Q^{s}\left(y^{n}\right)\right)^{1/n},\label{eq:-138-1}
\end{align}
and 
\begin{align}
 & \kappa_{1}\nonumber \\
 & :=\frac{1}{n}\log\sum_{x^{n},y^{n}}P\left(x^{n}\right)P^{1-s}\left(y^{n}|x^{n}\right)Q^{s}\left(y^{n}\right)\left(\frac{P\left(y^{n}|x^{n}\right)}{P\left(y^{n}\right)}\right)^{t}\\
 & =\frac{1}{n}\log\sum_{x^{n},y^{n}}P\left(x^{n}\right)P^{1-\left(s-t\right)}\left(y^{n}|x^{n}\right)P^{-t}\left(y^{n}\right)Q^{s}\left(y^{n}\right),\\
 & \kappa_{2}\nonumber \\
 & :=\frac{1}{n}\log\sum_{x^{n},y^{n}}P\left(x^{n},y^{n}\right)P^{-s}\left(y^{n}\right)Q^{s}\left(y^{n}\right)\left(\frac{P\left(y^{n}|x^{n}\right)}{P\left(y^{n}\right)}\right)^{-t}\\
 & =\frac{1}{n}\log\sum_{x^{n},y^{n}}P\left(x^{n}\right)P^{1-t}\left(y^{n}|x^{n}\right)P^{t-s}\left(y^{n}\right)Q^{s}\left(y^{n}\right).
\end{align}

Substituting these into \eqref{eq:-3-1}, we have 
\begin{align}
 & \e^{-sD_{1-s}(P_{Y^{n}\mathcal{C}_{n}}\|Q_{Y}^{n}\times P_{\mathcal{C}_{n}})}\nonumber \\
 & \leq\e^{n\left(sR-\max_{t\geq0}\left(tR'-\kappa_{1}\right)\right)}+\e^{-n\max_{t\geq0}\left(-tR'-\kappa_{2}\right)}\\
 & \leq2\max\{\e^{n\left(sR-\max_{t\geq0}\left(tR'-\kappa_{1}\right)\right)},\e^{-n\max_{t\geq0}\left(-tR'-\kappa_{2}\right)}\}\\
 & =2\e^{-ns\min\left\{ \frac{1}{s}\max_{t\geq0}\left(tR'-\kappa_{1}\right)-R,\frac{1}{s}\max_{t\geq0}\left(-tR'-\kappa_{2}\right)\right\} }.
\end{align}
That is, 
\begin{align}
 & \frac{1}{n}D_{1-s}(P_{Y^{n}\mathcal{C}_{n}}\|Q_{Y}^{n}\times P_{\mathcal{C}_{n}})\nonumber \\
 & \geq\min\Bigl\{\frac{1}{s}\max_{t\geq0}\left(tR'-\kappa_{1}\right)-R'-\frac{1}{n}\log2,\nonumber \\
 & \qquad\frac{1}{s}\max_{t\geq0}\left(-tR'-\kappa_{2}\right)\Bigr\}-\delta_{n}\\
 & \geq\min\Bigl\{\frac{1}{s}\max_{t\geq0}\left(tR'-\kappa_{1}\right)-R'-\frac{1}{n}\log2,\nonumber \\
 & \qquad\frac{1}{s}\max_{t\geq0}\left(-tR'-\kappa_{2}\right)-\frac{1}{n}\log2\Bigr\}-\delta_{n}\\
 & =\min\left\{ \frac{1}{s}\max_{t\geq0}\left(tR'-\kappa_{1}\right)-R',\frac{1}{s}\max_{t\geq0}\left(-tR'-\kappa_{2}\right)\right\} \nonumber \\
 & \qquad-\delta_{n}-\delta_{n}'\\
 & =\frac{1}{s}\min\left\{ \max_{t\geq0}\tau\left(R',s,s-t\right),\max_{t\geq0}\tau\left(R',s,t\right)\right\} -\delta_{n}-\delta_{n}'\\
 & =\frac{1}{s}\max_{t\in\left[0,s\right]}\tau\left(R',s,t\right)-\delta_{n}-\delta_{n}'\label{eq:-7-1}\\
 & \geq\frac{1}{s}\max_{t\in\left[0,s\right]}\tau\left(R,s,t\right)-\delta_{n}-\delta_{n}'\label{eq:-125}\\
 & \geq\frac{1}{s}\min_{P_{X}}\max_{t\in\left[0,s\right]}\tau\left(R,s,t\right)-\delta_{n}-\delta_{n}'\\
 & =\Gamma_{1-s}^{\left(n\right)}\left(P_{Y|X},Q_{Y},R\right)-\delta_{n}-\delta_{n}',
\end{align}
where the function $\tau\left(R,s,t\right)$ is defined in \eqref{eq:-124},
\eqref{eq:-7-1} follows from \eqref{eq:-8}, and \eqref{eq:-125}
follows since $\tau\left(R,s,t\right)$ is non-increasing in $R$
for $t\ge0$.

\section{Proof of Theorem \ref{thm:singleletter}}

\label{sec:Proof-of-Theorem-asym-1}

\subsection{Upper Bound for Case $1+s$ with $s\in[0,1]$}

To obtain the upper bound, we set 
\begin{equation}
P\left(x^{n}\right)=\frac{1\big\{ x^{n}\in\mathcal{T}_{\widetilde{T}_{X}}\big\}}{\big|\mathcal{T}_{\widetilde{T}_{X}}\big|}\label{eqn:uniform_type_class}
\end{equation}
and substitute it into the multiletter expression \eqref{eq:Gamma}
in Proposition \ref{thm:multiletter}, where $\widetilde{T}_{X}$
is some type of $n$-length sequences. Define $g(x):=\sum_{y}P^{1+s}(y|x)Q^{-s}(y)$.
Then we obtain 
\begin{align}
 & \frac{1}{n}D_{1+s}\left(P_{X^{n}Y^{n}}\|P_{X^{n}}\times Q_{Y}^{n}\right)\nonumber \\
 & =\frac{1}{ns}\log\sum_{x^{n}}P\left(x^{n}\right)\sum_{y^{n}}P^{1+s}\left(y^{n}|x^{n}\right)Q^{-s}\left(y^{n}\right)\\
 & =\frac{1}{ns}\log\sum_{x^{n}}P\left(x^{n}\right)\prod_{i=1}^{n}\sum_{y_{i}}P^{1+s}\left(y_{i}|x_{i}\right)Q^{-s}\left(y_{i}\right)\\
 & =\frac{1}{ns}\log\sum_{x^{n}}P(x^{n})\prod_{i=1}^{n}g(x_{i})\\
 & =\frac{1}{ns}\log\sum_{T_{X}}P_{X^{n}}(\mathcal{T}_{T_{X}})\e^{n\sum_{x\in\mathcal{X}}T_{X}\left(x\right)\log g(x)}\label{eq:-26}\\
 & ={\displaystyle \frac{1}{s}\sum_{x}\tilde{T}_{X}(x)\log g(x)},\label{eq:-96}
\end{align}
where the last line follows from the definition of $P(x^{n})$. Furthermore,
we also have\footnote{As stated in the notation section (Section \ref{sec:notation}), for
brevity, sometimes we use $T\left(x,y\right)$ to denote the joint
distributions $T\left(x\right)V\left(y|x\right)$ or $T\left(y\right)V\left(x|y\right)$. Furthermore,  for brevity, we use  $\sum_{T_{Y}}$  to denote   $\sum_{T_{Y} \in \mathcal{P}^{\left(n\right)}\left(\mathcal{Y}\right)}$, and    $\sum_{V_{X|Y}}$  to denote   $\sum_{V_{X|Y} \in \mathcal{P}^{\left(n\right)}\left(\mathcal{X}|T_{Y}\right)}$.} \eqref{eq:-101}-\eqref{eq:-110}, where \eqref{eq:-139} follows
from the fact that both the numbers of $n$-types and $n$-conditional
types are polynomial in $n$. 
\begin{figure*}[!t]
\setcounter{mytempeqncnt}{\value{equation}} \setcounter{equation}{198}
\begin{align}
 & \frac{1}{n}D_{1+s}(P_{Y^{n}}\|Q_{Y}^{n})\nonumber \\
 & =\frac{1}{ns}\log\sum_{y^{n}}P^{1+s}\left(y^{n}\right)Q^{-s}\left(y^{n}\right)\label{eq:-101}\\
 & =\frac{1}{ns}\log\left(\sum_{T_{Y}}\sum_{y^{n}\in\mathcal{T}_{T_{Y}}}\Bigl(\sum_{V_{X|Y}}\sum_{x^{n}\in\mathcal{T}_{V_{X|Y}}\left(y^{n}\right)}P\left(x^{n}\right)\e^{n\sum_{x,y}T\left(x,y\right)\log P\left(y|x\right)}\Bigr)^{1+s}\e^{-ns\sum_{x,y}T\left(y\right)\log Q\left(y\right)}\right)\label{eq:-112}\\
 & =\frac{1}{ns}\log\Biggl(\sum_{T_{Y}}\sum_{y^{n}\in\mathcal{T}_{T_{Y}}}\Bigl(\sum_{V_{X|Y}}\sum_{x^{n}\in\mathcal{T}_{V_{X|Y}}\left(y^{n}\right)}\frac{1\big\{ x^{n}\in\mathcal{T}_{\widetilde{T}_{X}}\big\}}{\big|\mathcal{T}_{\widetilde{T}_{X}}\big|}\e^{n\sum_{x,y}T\left(x,y\right)\log P\left(y|x\right)}\Bigr)^{1+s}\e^{-ns\sum_{y}T\left(y\right)\log Q\left(y\right)}\Biggr)\\
 & =\frac{1}{ns}\log\Biggl(\sum_{T_{Y}}\e^{nH\left(T_{Y}\right)}\Bigl(\sum_{V_{X|Y}:V_{X|Y}\circ T_{Y}=\widetilde{T}_{X}}\e^{n\left(H\left(V_{X|Y}|T_{Y}\right)-H\left(\widetilde{T}_{X}\right)+\sum_{x,y}T\left(x,y\right)\log P\left(y|x\right)\right)}\Bigr)^{1+s}\e^{-ns\sum_{y}T\left(y\right)\log Q\left(y\right)}\Biggr)+\delta_{n}\\
 & =\max_{T_{Y},V_{X|Y}:V_{X|Y}\circ T_{Y}=\widetilde{T}_{X}}\frac{1}{s}\Biggl(H\left(T_{Y}\right)+\left(1+s\right)\Bigl(H\left(V_{X|Y}|T_{Y}\right)-H\left(\widetilde{T}_{X}\right)+\sum_{x,y}T\left(x,y\right)\log P\left(y|x\right)\Bigr)\nonumber \\
 & \qquad\qquad\qquad-s\sum_{y}T\left(y\right)\log Q\left(y\right)\Biggr)+\delta_{n}+\delta_{n}'\label{eq:-139}\\
 & =\max_{T_{Y},V_{X|Y}:V_{X|Y}\circ T_{Y}=\widetilde{T}_{X}}\frac{1}{s}\left(\left(1+s\right)\sum_{x,y}T\left(x,y\right)\log\frac{P\left(y|x\right)}{T\left(y|x\right)}+s\sum_{y}T\left(y\right)\log\frac{T\left(y\right)}{Q\left(y\right)}\right)+\delta_{n}+\delta_{n}'\\
 & =\max_{\widetilde{V}_{Y|X}}-\frac{1}{s}\left(\left(1+s\right)\sum_{x,y}\widetilde{T}\left(x,y\right)\log\frac{\widetilde{V}\left(y|x\right)}{P\left(y|x\right)}-s\sum_{y}\widetilde{T}\left(y\right)\log\frac{\widetilde{T}\left(y\right)}{Q\left(y\right)}\right)+\delta_{n}+\delta_{n}',\label{eq:-110}
\end{align}
\setcounter{mytempeqncnt}{\value{equation}} \setcounter{equation}{\value{mytempeqncnt}}
\hrulefill{} 
\end{figure*}

Since $\widetilde{T}_{X}$ is arbitrary, from \eqref{eq:-96} and
\eqref{eq:-110} we have 
\begin{align}
 & \Gamma_{1+s}^{\left(n\right)}\left(P_{Y|X},Q_{Y},R\right)\nonumber \\
 & \leq\min_{\widetilde{T}_{X}}\max\biggl\{\frac{1}{s}\sum_{x}\widetilde{T}\left(x\right)\log(\sum_{y}P^{1+s}\left(y|x\right)Q^{-s}\left(y\right))-R,\nonumber \\
 & \qquad\qquad\max_{\widetilde{V}_{Y|X}}\eta_{1+s}\left(P_{Y|X},Q_{Y},\widetilde{T}_{X},\widetilde{V}_{Y|X}\right)\biggr\}+\delta_{n}+\delta_{n}'.\label{eq:-98}
\end{align}
Note that in \eqref{eq:-98} the minimization and maximization are
taken over the set of types, not the set of general probability mass
functions. To achieve the desired result, we continue upper bounding~\eqref{eq:-98}
to obtain 
\begin{align}
 & \Gamma_{1+s}^{\left(n\right)}\left(P_{Y|X},Q_{Y},R\right)\nonumber \\
 & \leq\min_{\widetilde{T}_{X}}\max\biggl\{\frac{1}{s}\sum_{x}\widetilde{T}\left(x\right)\log(\sum_{y}P^{1+s}\left(y|x\right)Q^{-s}\left(y\right))-R,\nonumber \\
 & \qquad\max_{\widetilde{P}_{Y|X}\in\mathcal{P}\left(\mathcal{Y}|\mathcal{X}\right)}\eta_{1+s}\left(P_{Y|X},Q_{Y},\widetilde{T}_{X},\widetilde{P}_{Y|X}\right)\biggr\}+\delta_{n}+\delta_{n}',\label{eq:-100}
\end{align}
since $\mathcal{P}^{\left(n\right)}\big(\mathcal{Y}|\widetilde{T}_{X}\big)\subseteq\mathcal{P}\left(\mathcal{Y}|\mathcal{X}\right)$.

If the objective function of minimization is continuous, then by Lemma
\ref{lem:minequality} we have 
\begin{align}
 & \Gamma_{1+s}^{\left(n\right)}\left(P_{Y|X},Q_{Y},R\right)\nonumber \\
 & \leq\min_{\widetilde{P}_{X}\in\mathcal{P}\left(\mathcal{X}\right)}\max\biggl\{\frac{1}{s}\sum_{x}\widetilde{P}\left(x\right)\log(\sum_{y}P^{1+s}\left(y|x\right)Q^{-s}\left(y\right))\nonumber \\
 & \quad-R,\:\max_{\widetilde{P}_{Y|X}\in\mathcal{P}\left(\mathcal{Y}|\mathcal{X}\right)}\eta_{1+s}\left(P_{Y|X},Q_{Y},\widetilde{P}_{X},\widetilde{P}_{Y|X}\right)\biggr\}+o\left(1\right).
\end{align}
This completes the proof.

So the rest is to show the continuity of the objective function. To
prove this, we only need to show 
\begin{equation}
\max_{\widetilde{P}_{Y|X}}\eta_{1+s}\left(P_{Y|X},Q_{Y},\widetilde{P}_{X},\widetilde{P}_{Y|X}\right)\label{eq:-105}
\end{equation}
is continuous in $\widetilde{P}_{X}$. Observe that $\mathcal{P}\left(\mathcal{Y}|\mathcal{X}\right)$
is compact, and $\eta_{1+s}\left(P_{Y|X},Q_{Y},\widetilde{P}_{X},\widetilde{P}_{Y|X}\right)$
is (jointly) continuous in $(\widetilde{P}_{X},\widetilde{P}_{Y|X})$.
Hence by the following lemma, we have \eqref{eq:-105} is continuous
in $\widetilde{P}_{X}$. 
\begin{lem}[Lemma~14 in~\cite{tan2011large}]
\label{lemma:continuity} Let $\mathcal{X}$ and $\mathcal{Y}$ be
two metric spaces and let $\calK\subset\mathcal{X}$ be a compact
set. Let $f:\mathcal{X}\times\mathcal{Y}\rightarrow\mathbb{R}$ be
a (jointly) continuous real-valued function. Then the function $g:\mathcal{Y}\rightarrow\mathbb{R}$,
defined as 
\begin{equation}
g(y):=\min_{x\in\calK}\,f(x,y),\quad\forall\,y\in\mathcal{Y},\label{eqn:gy}
\end{equation}
is continuous on $\mathcal{Y}$. 
\end{lem}

\subsection{Lower Bound for Case $1+s$ with $s\in(0,\infty]$}

Observe that \eqref{eq:-26} still holds. That is, 
\begin{align}
 & \frac{1}{n}D_{1+s}\left(P_{X^{n}Y^{n}}\|P_{X^{n}}\times Q_{Y}^{n}\right)\nonumber \\
 & =\frac{1}{ns}\log\sum_{T_{X}}P_{X^{n}}(\mathcal{T}_{T_{X}})\e^{n\sum_{x\in\mathcal{X}}T_{X}\left(x\right)\log\sum_{y}P^{1+s}\left(y|x\right)Q^{-s}\left(y\right)}.\label{eq:-9}
\end{align}
On the other hand, we also have \eqref{eq:-143}-\eqref{eq:-10},
where \eqref{eq:-143} follows from \eqref{eq:-112}, \eqref{eq:-11}
follows from Lemma \ref{lem:norm}, \eqref{eq:-116} follows since
$x\mapsto x^{1+s}$ is a convex function for $s\ge0$, and \eqref{eq:-119}
follows from Lemma \ref{lem:typeequality}.
\begin{figure*}[!t]
\setcounter{mytempeqncnt}{\value{equation}} \setcounter{equation}{215}
\begin{align}
 & \frac{1}{n}D_{1+s}(P_{Y^{n}}\|Q_{Y}^{n})\nonumber \\
 & =\frac{1}{ns}\log\left(\sum_{T_{Y}}\sum_{y^{n}\in\mathcal{T}_{T_{Y}}}\biggl(\sum_{V_{X|Y}}\sum_{x^{n}\in\mathcal{T}_{V_{X|Y}}\left(y^{n}\right)}P\left(x^{n}\right)\e^{n\sum_{x,y}T\left(x,y\right)\log P\left(y|x\right)}\biggr)^{1+s}\e^{-ns\sum_{x,y}T\left(y\right)\log Q\left(y\right)}\right)\label{eq:-143}\\
 & \geq\frac{1}{ns}\log\left(\sum_{T_{Y}}\sum_{y^{n}\in\mathcal{T}_{T_{Y}}}\sum_{V_{X|Y}}P_{X^{n}}^{1+s}\Bigl(\mathcal{T}_{V_{X|Y}}(y^{n})\Bigr)\e^{n\left(1+s\right)\sum_{x,y}T\left(x,y\right)\log P\left(y|x\right)}\e^{-ns\sum_{x,y}T\left(y\right)\log Q\left(y\right)}\right)\label{eq:-11}\\
 & \geq\frac{1}{ns}\log\Biggl(\sum_{T_{Y},V_{X|Y}}\left|\mathcal{T}_{T_{Y}}\right|\biggl(\sum_{y^{n}\in\mathcal{T}_{T_{Y}}}\frac{1}{\left|\mathcal{T}_{T_{Y}}\right|}P_{X^{n}}\Bigl(\mathcal{T}_{V_{X|Y}}(y^{n})\Bigr)\biggr)^{1+s}\e^{n\left(1+s\right)\sum_{x,y}T\left(x,y\right)\log P\left(y|x\right)-ns\sum_{x,y}T\left(y\right)\log Q\left(y\right)}\Biggr)\label{eq:-116}\\
 & =\frac{1}{ns}\log\Biggl(\sum_{T_{Y},V_{X|Y}}\left|\mathcal{T}_{T_{Y}}\right|\biggl(\frac{1}{\left|\mathcal{T}_{T_{Y}}\right|}\e^{nH\left(V_{Y|X}|T_{X}\right)+n\delta_{n}}P_{X^{n}}(\mathcal{T}_{T_{X}})\biggr)^{1+s}\e^{n\left(1+s\right)\sum_{x,y}T\left(x,y\right)\log P\left(y|x\right)-ns\sum_{x,y}T\left(y\right)\log Q\left(y\right)}\Biggr)\label{eq:-119}\\
 & =\frac{1}{ns}\log\Biggl(\sum_{T_{Y},V_{X|Y}}P_{X^{n}}^{1+s}(\mathcal{T}_{T_{X}})\e^{nH\left(T_{Y}\right)-n\left(1+s\right)I\left(T_{X},V_{Y|X}\right)+n\left(1+s\right)\sum_{x,y}T\left(x,y\right)\log P\left(y|x\right)-ns\sum_{x,y}T\left(y\right)\log Q\left(y\right)+n\delta_{n}+ns\delta_{n}'}\Biggr)\label{eq:-10}
\end{align}
\setcounter{mytempeqncnt}{\value{equation}} \setcounter{equation}{\value{mytempeqncnt}}
\hrulefill{} 
\end{figure*}

Since $\sum_{T_{X}}P_{X^{n}}(\mathcal{T}_{T_{X}})=1$ and $\left|\mathcal{P}^{\left(n\right)}\left(\mathcal{X}\right)\right|\leq\left(n+1\right)^{\left|\mathcal{X}\right|}$,
by the pigeonhole principle, we have that there must exist at least
one $\widetilde{T}_{X}$ such that $P_{X^{n}}(\mathcal{T}_{\widetilde{T}_{X}})\geq\left(n+1\right)^{-\left|\mathcal{X}\right|}$.
Therefore, from \eqref{eq:-9} and \eqref{eq:-10}, we have 
\begin{align}
 & \frac{1}{n}D_{1+s}\left(P_{X^{n}Y^{n}}\|P_{X^{n}}\times Q_{Y}^{n}\right)\nonumber \\
 & \geq\frac{1}{ns}\log P_{X^{n}}(\mathcal{T}_{\widetilde{T}_{X}})\e^{n\sum_{x}\widetilde{T}_{X}\left(x\right)\log\sum_{y}P^{1+s}\left(y|x\right)Q^{-s}\left(y\right)}\\
 & =\frac{1}{s}\sum_{x}\widetilde{T}_{X}\left(x\right)\log\left(\sum_{y}P^{1+s}\left(y|x\right)Q^{-s}\left(y\right)\right)+\delta_{n}'',
\end{align}
and 
\begin{align}
 & \frac{1}{n}D_{1+s}(P_{Y^{n}}\|Q_{Y}^{n})\nonumber \\
 & \geq\frac{1}{ns}\log\biggl(\sum_{\widetilde{V}_{Y|X}}P_{X^{n}}^{1+s}(\mathcal{T}_{\widetilde{T}_{X}})\e^{nH\left(\widetilde{T}_{Y}\right)-\left(1+s\right)nI\left(\widetilde{T}_{X},\widetilde{V}_{Y|X}\right)}\nonumber \\
 & \quad\times\e^{n\left(1+s\right)\sum_{x,y}\widetilde{T}\left(x,y\right)\log P\left(y|x\right)-ns\sum_{x,y}\widetilde{T}\left(y\right)\log Q\left(y\right)}\biggr)\nonumber \\
 & \quad+\frac{1}{s}\delta_{n}+\delta_{n}'\\
 & =\max_{\widetilde{V}_{Y|X}}\frac{1}{s}\biggl(H\left(\widetilde{T}_{Y}\right)+\left(1+s\right)I\left(\widetilde{T}_{X},\widetilde{V}_{Y|X}\right)\nonumber \\
 & \quad+\left(1+s\right)\sum_{x,y}\widetilde{T}\left(x,y\right)\log P\left(y|x\right)-s\sum_{y}\widetilde{T}\left(y\right)\log Q\left(y\right)\biggr)\nonumber \\
 & \quad+\frac{1}{s}\delta_{n}+\delta_{n}'+\frac{1+s}{s}\delta_{n}''\label{eq:-140}\\
 & =\max_{\widetilde{V}_{Y|X}}\biggl(-\frac{1+s}{s}\sum_{x,y}\widetilde{T}\left(x,y\right)\log\frac{\widetilde{V}\left(y|x\right)}{P\left(y|x\right)}\nonumber \\
 & \quad+\sum_{y}\widetilde{T}\left(y\right)\log\frac{\widetilde{T}\left(y\right)}{Q\left(y\right)}\biggr)+\frac{1}{s}\delta_{n}+\delta_{n}'+\frac{1+s}{s}\delta_{n}'',
\end{align}
where \eqref{eq:-140} follows from the fact that the number of $n$-conditional
types is polynomial in $n$. Therefore, 
\begin{align}
 & \Gamma_{1+s}^{\left(n\right)}\left(P_{Y|X},Q_{Y},R\right)\nonumber \\
 & \geq\min_{\widetilde{T}_{X}}\max\Biggl\{\frac{1}{s}\sum_{x}\widetilde{T}\left(x\right)\log(\sum_{y}P^{1+s}\left(y|x\right)Q^{-s}\left(y\right))-R,\nonumber \\
 & \qquad\max_{\widetilde{V}_{Y|X}}\eta_{1+s}\left(P_{Y|X},Q_{Y},\widetilde{T}_{X},\widetilde{V}_{Y|X}\right)\Biggr\}+o\left(1\right)\\
 & =\min_{\widetilde{P}_{X}}\max\Biggl\{\frac{1}{s}\sum_{x}\widetilde{P}\left(x\right)\log(\sum_{y}P^{1+s}\left(y|x\right)Q^{-s}\left(y\right))-R,\nonumber \\
 & \qquad\max_{\widetilde{P}_{Y|X}\in\mathcal{P}\left(\mathcal{Y}|\mathcal{X}\right)}\eta_{1+s}\left(P_{Y|X},Q_{Y},\widetilde{P}_{X},\widetilde{P}_{Y|X}\right)\Biggr\}+o\left(1\right),\label{eq:-106}
\end{align}
where \eqref{eq:-106} follows from Lemma \ref{lem:minequality}.

\subsection{Upper Bound for Case $1-s$ with $s\in(0,1)$}

Same as the $1+s$ case, we set $P\left(x^{n}\right)$ as in \eqref{eqn:uniform_type_class}
and substitute it into the multiletter expression \eqref{eq:Gamma2}
in Proposition \ref{thm:multiletter}, where $\widetilde{T}_{X}$
is some type of $n$-length sequences. Then we obtain 
\begin{align}
 & -\frac{1}{ns}\log\sum_{x^{n},y^{n}}P\left(x^{n}\right)P^{1-t}\left(y^{n}|x^{n}\right)P^{t-s}\left(y^{n}\right)Q^{s}\left(y^{n}\right)\nonumber \\
 & =-\frac{1}{ns}\log\sum_{T_{Y}V_{X|Y}}\sum_{y^{n}\in\mathcal{T}_{T_{Y}}}\sum_{x^{n}\in\mathcal{T}_{V_{X|Y}}\left(y^{n}\right)}P(x^{n})P^{1-t}(y^{n}|x^{n})\nonumber \\
 & \qquad\Bigl(\sum_{V_{X|Y}}\sum_{x^{n}\in\mathcal{T}_{V_{X|Y}}\left(y^{n}\right)}P\left(x^{n}\right)P\left(y^{n}|x^{n}\right)\Bigr)^{t-s}Q^{s}\left(y^{n}\right)\\
 & =-\frac{1}{ns}\log\sum_{T_{Y}V_{X|Y}}\sum_{y^{n}\in\mathcal{T}_{T_{Y}}}\sum_{x^{n}\in\mathcal{T}_{V_{X|Y}}\left(y^{n}\right)}P\left(x^{n}\right)\nonumber \\
 & \qquad\Bigl(\sum_{V_{X|Y}}\sum_{x^{n}\in\mathcal{T}_{V_{X|Y}}\left(y^{n}\right)}P\left(x^{n}\right)\e^{n\sum_{x,y}T\left(x,y\right)\log P\left(y|x\right)}\Bigr)^{t-s}\nonumber \\
 & \qquad\e^{n\left(1-t\right)\sum_{x,y}T\left(x,y\right)\log P\left(y|x\right)+ns\sum_{y}T\left(y\right)\log Q\left(y\right)}\\
 & =-\frac{1}{ns}\log\sum_{T_{Y}V_{X|Y}}\sum_{y^{n}\in\mathcal{T}_{T_{Y}}}P_{X^{n}}\Bigl(\mathcal{T}_{V_{X|Y}}\left(y^{n}\right)\Bigr)\nonumber \\
 & \qquad\Bigl(\sum_{V_{X|Y}}P_{X^{n}}\Bigl(\mathcal{T}_{V_{X|Y}}\left(y^{n}\right)\Bigr)\e^{n\sum_{x,y}T\left(x,y\right)\log P\left(y|x\right)}\Bigr)^{t-s}\nonumber \\
 & \qquad\e^{n\left(1-t\right)\sum_{x,y}T\left(x,y\right)\log P\left(y|x\right)+ns\sum_{y}T\left(y\right)\log Q\left(y\right)}.\label{eq:-145}
\end{align}
Observe that for any $y^{n}$ with type $T_{Y}$, we have 
\begin{align}
 & P_{X^{n}}\Bigl(\mathcal{T}_{V_{X|Y}}\left(y^{n}\right)\Bigr)\nonumber \\
 & =\e^{-nI\left(V_{X|Y},T_{Y}\right)+n\delta_{n}}1\left\{ V_{X|Y}\circ T_{Y}=\widetilde{T}_{X}\right\} .
\end{align}
Therefore, we have \eqref{eq:-108}-\eqref{eq:-12}, where \eqref{eq:-139}
follows from the fact that the number of $n$-types $T_{Y}V_{X|Y}$
is polynomial in $n$. Since $\widetilde{T}_{X}$ is arbitrary, by
Proposition \ref{thm:multiletter} and \eqref{eq:-12}, we have \eqref{eq:-109}-\eqref{eqn:Gamma_UB},
where \eqref{eq:-126} follows since for any function $f\left(x,y\right)$,
$\max_{x}\min_{y}f\left(x,y\right)\leq\min_{y}\max_{x}f\left(x,y\right)$,
and \eqref{eq:-107} follows from $\mathcal{P}^{\left(n\right)}\left(\mathcal{Y}|\mathcal{X}\right)\subseteq\mathcal{P}\left(\mathcal{Y}|\mathcal{X}\right)$,
Lemma \ref{lem:minequality}, and the continuity of the objective
function of $\min_{\widetilde{P}_{XY}\in\mathcal{P}\left(\mathcal{X}\times\mathcal{Y}\right)}$
(the continuity can be shown by Lemma \ref{lemma:continuity}). 
\begin{figure*}[!t]
\setcounter{mytempeqncnt}{\value{equation}} \setcounter{equation}{231}
\begin{align}
 & -\frac{1}{ns}\log\sum_{x^{n},y^{n}}P\left(x^{n}\right)P^{1-t}\left(y^{n}|x^{n}\right)P^{t-s}\left(y^{n}\right)Q^{s}\left(y^{n}\right)\nonumber \\
 & \leq-\frac{1}{ns}\log\sum_{T_{Y}}\sum_{y^{n}\in\mathcal{T}_{T_{Y}}}\sum_{V_{X|Y}}\e^{-nI\left(V_{X|Y},T_{Y}\right)+n\delta_{n}}1\left\{ V_{X|Y}\circ T_{Y}=\widetilde{T}_{X}\right\} \nonumber \\
 & \qquad\times\e^{n\left(1-t\right)\sum_{x,y}T\left(x,y\right)\log P\left(y|x\right)+ns\sum_{y}T\left(y\right)\log Q\left(y\right)}\nonumber \\
 & \qquad\times\left(\sum_{V_{X|Y}}\e^{-nI\left(V_{X|Y},T_{Y}\right)+n\delta_{n}}1\left\{ V_{X|Y}\circ T_{Y}=\widetilde{T}_{X}\right\} \e^{n\sum_{x,y}T\left(x,y\right)\log P\left(y|x\right)}\right)^{t-s}\label{eq:-108}\\
 & =-\frac{1}{ns}\log\max_{T_{Y},V_{X|Y}:V_{X|Y}\circ T_{Y}=\widetilde{T}_{X}}\Biggl\{\e^{nH\left(T_{Y}\right)-nI\left(V_{X|Y},T_{Y}\right)+n\left(1-t\right)\sum_{x,y}T\left(x,y\right)\log P\left(y|x\right)+ns\sum_{y}T\left(y\right)\log Q\left(y\right)}\nonumber \\
 & \qquad\times\left(\max_{V_{X|Y}:V_{X|Y}\circ T_{Y}=\widetilde{T}_{X}}\e^{-nI\left(V_{X|Y}T_{Y}\right)}\e^{n\sum_{x,y}T\left(y\right)V\left(x|y\right)\log P\left(y|x\right)}\right)^{t-s}\Biggr\}+\left(1+t-s\right)\delta_{n}+\delta_{n}'\label{eq:-141}\\
 & =-\frac{1}{s}\max_{\widetilde{V}_{Y|X}}\Biggl\{ H\left(\widetilde{T}_{Y}\right)-I\left(\widetilde{V}_{Y|X},\widetilde{T}_{X}\right)+\left(1-t\right)\sum_{x,y}\widetilde{T}\left(x,y\right)\log P\left(y|x\right)+s\sum_{y}\widetilde{T}\left(y\right)\log Q\left(y\right)\nonumber \\
 & \qquad+\left(t-s\right)\left(\max_{\widehat{V}{}_{Y|X}:\widehat{V}{}_{Y|X}\circ\widetilde{T}_{X}=\widetilde{V}_{Y|X}\circ\widetilde{T}_{X}}-H\left(\widetilde{V}_{Y|X}\circ\widetilde{T}_{X}\right)-\sum_{x,y}\widetilde{T}\left(x\right)\widehat{V}\left(y|x\right)\log\frac{\widehat{V}\left(y|x\right)}{P\left(y|x\right)}\right)\Biggr\}+\left(1+t-s\right)\delta_{n}+\delta_{n}'\\
 & =-\frac{1}{s}\max_{\widetilde{V}_{Y|X}}\Biggl\{-\sum_{x,y}\widetilde{T}\left(x,y\right)\log\frac{\widetilde{V}\left(y|x\right)}{P\left(y|x\right)}+t\sum_{x,y}\widetilde{T}\left(x,y\right)\log\frac{\widetilde{T}\left(y\right)}{P\left(y|x\right)}-s\sum_{y}\widetilde{T}\left(y\right)\log\frac{\widetilde{T}\left(y\right)}{Q\left(y\right)}\nonumber \\
 & \qquad+\left(s-t\right)\min_{\widehat{V}{}_{Y|X}:\widehat{V}{}_{Y|X}\circ\widetilde{T}_{X}=\widetilde{V}_{Y|X}\circ\widetilde{T}_{X}}\sum_{x,y}\widetilde{T}\left(x\right)\widehat{V}\left(y|x\right)\log\frac{\widehat{V}\left(y|x\right)}{P\left(y|x\right)}\Biggr\}+\left(1+t-s\right)\delta_{n}+\delta_{n}',\label{eq:-12}
\end{align}
\setcounter{mytempeqncnt}{\value{equation}} \setcounter{equation}{\value{mytempeqncnt}}
\hrulefill{} 
\end{figure*}
 
\begin{figure*}[!t]
\setcounter{mytempeqncnt}{\value{equation}} \setcounter{equation}{235}
\begin{align}
 & \Gamma_{1-s}^{\left(n\right)}\left(P_{Y|X},Q_{Y},R\right)\nonumber \\
 & \leq\min_{\widetilde{T}_{X}}\max_{t\in\left[0,s\right]}\min_{\widetilde{V}_{Y|X}}\Biggl\{-\frac{t}{s}R+\frac{1}{s}\sum_{x,y}\widetilde{T}\left(x,y\right)\log\frac{\widetilde{V}\left(y|x\right)}{P\left(y|x\right)}-\frac{t}{s}\sum_{x,y}\widetilde{T}\left(x,y\right)\log\frac{\widetilde{T}\left(y\right)}{P\left(y|x\right)}+\sum_{y}\widetilde{T}\left(y\right)\log\frac{\widetilde{T}\left(y\right)}{Q\left(y\right)}\nonumber \\
 & \qquad-\left(1-\frac{t}{s}\right)\min_{\widehat{V}{}_{Y|X}:\widehat{V}{}_{Y|X}\circ\widetilde{T}_{X}=\widetilde{V}_{Y|X}\circ\widetilde{T}_{X}}\sum_{x,y}\widetilde{T}\left(x\right)\widehat{V}\left(y|x\right)\log\frac{\widehat{V}\left(y|x\right)}{P\left(y|x\right)}\Biggr\}+o(1)\label{eq:-109}\\
 & \leq\min_{\widetilde{T}_{X}}\min_{\widetilde{V}_{Y|X}}\max_{t\in\left[0,s\right]}\Biggl\{-\frac{t}{s}R+\frac{1}{s}\sum_{x,y}\widetilde{T}\left(x,y\right)\log\frac{\widetilde{V}\left(y|x\right)}{P\left(y|x\right)}-\frac{t}{s}\sum_{x,y}\widetilde{T}\left(x,y\right)\log\frac{\widetilde{T}\left(y\right)}{P\left(y|x\right)}+\sum_{y}\widetilde{T}\left(y\right)\log\frac{\widetilde{T}\left(y\right)}{Q\left(y\right)}\nonumber \\
 & \qquad-\left(1-\frac{t}{s}\right)\min_{\widehat{V}{}_{Y|X}:\widehat{V}{}_{Y|X}\circ\widetilde{T}_{X}=\widetilde{V}_{Y|X}\circ\widetilde{T}_{X}}\sum_{x,y}\widetilde{T}\left(x\right)\widehat{V}\left(y|x\right)\log\frac{\widehat{V}\left(y|x\right)}{P\left(y|x\right)}\Biggr\}+o(1)\label{eq:-126}\\
 & =\min_{\widetilde{T}_{X}}\min_{\widetilde{V}_{Y|X}}\max\biggl\{-R+\sum_{x,y}\widetilde{T}\left(x,y\right)\log\frac{P\left(y|x\right)}{Q\left(y\right)}+\frac{1}{s}\sum_{x,y}\widetilde{T}\left(x,y\right)\log\frac{\widetilde{V}\left(y|x\right)}{P\left(y|x\right)},\nonumber \\
 & \qquad\frac{1}{s}\sum_{x,y}\widetilde{T}\left(x,y\right)\log\frac{\widetilde{V}\left(y|x\right)}{P\left(y|x\right)}+\sum_{y}\widetilde{T}\left(y\right)\log\frac{\widetilde{T}\left(y\right)}{Q\left(y\right)}-\min_{\widetilde{V}'_{Y|X}:\widetilde{V}'_{Y|X}\circ\widetilde{T}_{X}=\widetilde{V}_{Y|X}\circ\widetilde{T}_{X}}\sum_{x,y}\widetilde{T}\left(x\right)\widetilde{V}'\left(y|x\right)\log\frac{\widetilde{V}'\left(y|x\right)}{P\left(y|x\right)}\biggr\}+o(1)\\
 & \leq\min_{\widetilde{P}_{XY}\in\mathcal{P}\left(\mathcal{X}\times\mathcal{Y}\right)}\max\biggl\{\left(\frac{1}{s}-1\right)\sum_{x,y}\widetilde{P}\left(x,y\right)\log\frac{\widetilde{P}\left(y|x\right)}{P\left(y|x\right)}+\sum_{x,y}\widetilde{P}\left(x,y\right)\log\frac{\widetilde{P}\left(y|x\right)}{Q\left(y\right)}-R,\nonumber \\
 & \qquad\frac{1}{s}\sum_{x,y}\widetilde{P}\left(x,y\right)\log\frac{\widetilde{P}\left(y|x\right)}{P\left(y|x\right)}+\sum_{x,y}\widetilde{P}\left(y\right)\log\frac{\widetilde{P}\left(y\right)}{Q\left(y\right)}\nonumber \\
 & \qquad\qquad-\min_{\widehat{P}_{Y|X}\in\mathcal{P}\left(\mathcal{Y}|\mathcal{X}\right):\widehat{P}_{Y|X}\circ\widetilde{P}_{X}=\widetilde{P}_{Y|X}\circ\widetilde{P}_{X}}\sum_{x,y}\widetilde{P}\left(x\right)\widehat{P}\left(y|x\right)\log\frac{\widehat{P}\left(y|x\right)}{P\left(y|x\right)}\biggr\}+o(1)\label{eq:-107}\\
 & =\Gamma_{1-s}^{\mathsf{UB}}\left(P_{Y|X},Q_{Y},R\right)+o(1).\label{eqn:Gamma_UB}
\end{align}
\setcounter{mytempeqncnt}{\value{equation}} \setcounter{equation}{\value{mytempeqncnt}}
\hrulefill{} 
\end{figure*}

\subsection{Lower Bound for Case $1-s$ with $s\in[0,1)$}

Observe that \eqref{eq:-144}-\eqref{eq:-102} hold, where \eqref{eq:-144}
follows from \eqref{eq:-145}, \eqref{eq:-128} follows from that
$x\mapsto x^{1+t-s}$ with $0\leq t\leq s<1$ is a concave function,
\eqref{eq:-45} follows from Lemma \ref{lem:typeequality} and the
fact $P_{X^{n}}(\mathcal{T}_{T_{X}})\leq1$, and \eqref{eq:-142}
follows from the fact that the number of $n$-types $T_{Y}V_{X|Y}$
is polynomial in $n$. 
\begin{figure*}[!t]
\setcounter{mytempeqncnt}{\value{equation}} \setcounter{equation}{240}
\begin{align}
 & -\frac{1}{ns}\log\sum_{x^{n},y^{n}}P\left(x^{n},y^{n}\right)P^{-t}\left(y^{n}|x^{n}\right)P^{t-s}\left(y^{n}\right)Q^{s}\left(y^{n}\right)\nonumber \\
 & =-\frac{1}{ns}\log\sum_{T_{Y}}\sum_{y^{n}\in\mathcal{T}_{T_{Y}}}\sum_{V_{X|Y}}P_{X^{n}}\Bigl(\mathcal{T}_{V_{X|Y}}\left(y^{n}\right)\Bigr)\e^{n\left(1-t\right)\sum_{x,y}T\left(x,y\right)\log P\left(y|x\right)+ns\sum_{y}T\left(y\right)\log Q\left(y\right)}\nonumber \\
 & \qquad\times\left(\sum_{V_{X|Y}}P_{X^{n}}\Bigl(\mathcal{T}_{V_{X|Y}}\left(y^{n}\right)\Bigr)\e^{n\sum_{x,y}T\left(y\right)V\left(x|y\right)\log P\left(y|x\right)}\right)^{t-s}\label{eq:-144}\\
 & \geq-\frac{1}{ns}\log\sum_{T_{Y}}\sum_{y^{n}\in\mathcal{T}_{T_{Y}}}\sum_{V_{X|Y}}P_{X^{n}}^{1+t-s}\Bigl(\mathcal{T}_{V_{X|Y}}\left(y^{n}\right)\Bigr)\e^{n\left(1-s\right)\sum_{x,y}T\left(x,y\right)\log P\left(y|x\right)+ns\sum_{y}T\left(y\right)\log Q\left(y\right)}\\
 & \geq-\frac{1}{ns}\log\sum_{T_{Y},V_{X|Y}}\left|\mathcal{T}_{T_{Y}}\right|\Biggl(\sum_{y^{n}\in\mathcal{T}_{T_{Y}}}\frac{1}{\left|\mathcal{T}_{T_{Y}}\right|}P_{X^{n}}\Bigl(\mathcal{T}_{V_{X|Y}}\left(y^{n}\right)\Bigr)\Biggr)^{1+t-s}\e^{n\left(1-s\right)\sum_{x,y}T\left(x,y\right)\log P\left(y|x\right)+ns\sum_{y}T\left(y\right)\log Q\left(y\right)}\label{eq:-128}\\
 & \geq-\frac{1}{ns}\log\sum_{T_{Y},V_{X|Y}}\left|\mathcal{T}_{T_{Y}}\right|\left(\frac{\e^{nH\left(V_{Y|X}|T_{X}\right)+n\delta_{n}}}{\left|\mathcal{T}_{T_{Y}}\right|}\right)^{1+t-s}\e^{n\left(1-s\right)\sum_{x,y}T\left(x,y\right)\log P\left(y|x\right)+ns\sum_{y}T\left(y\right)\log Q\left(y\right)}\label{eq:-45}\\
 & \geq-\frac{1}{ns}\log\sum_{T_{Y},V_{X|Y}}\e^{n\left(s-t\right)H\left(T_{Y}\right)}\e^{n\left(1+t-s\right)H\left(V_{Y|X}|T_{X}\right)+n\left(1-s\right)\sum_{x,y}T\left(x,y\right)\log P\left(y|x\right)+ns\sum_{y}T\left(y\right)\log Q\left(y\right)}\nonumber \\
 & \qquad-\frac{1+t-s}{s}\delta_{n}+\delta_{n}'\\
 & =-\frac{1}{ns}\log\max_{T_{Y},V_{X|Y}}\e^{n\left(s-t\right)H\left(T_{Y}\right)+n\left(1+t-s\right)H\left(V_{Y|X}|T_{X}\right)+n\left(1-s\right)\sum_{x,y}T\left(x,y\right)\log P\left(y|x\right)+ns\sum_{y}T\left(y\right)\log Q\left(y\right)}\nonumber \\
 & \qquad-\frac{1+t-s}{s}\delta_{n}+\delta_{n}'+\delta_{n}''\label{eq:-142}\\
 & =\min_{T_{Y},V_{X|Y}}-\frac{1}{s}\left(H\left(T_{Y}\right)-\left(1+t-s\right)I\left(V_{Y|X},T_{X}\right)+\left(1-s\right)\sum_{x,y}T\left(x,y\right)\log P\left(y|x\right)+s\sum_{y}T\left(y\right)\log Q\left(y\right)\right)\nonumber \\
 & \qquad-\frac{1+t-s}{s}\delta_{n}+\delta_{n}'+\delta_{n}''\\
 & =\min_{T_{X},V_{Y|X}}\frac{t}{s}I\left(V_{Y|X},T_{X}\right)+\left(\frac{1}{s}-1\right)\sum_{x,y}T\left(x,y\right)\log\frac{V\left(y|x\right)}{P\left(y|x\right)}+\sum_{y}T\left(y\right)\log\frac{T\left(y\right)}{Q\left(y\right)}\nonumber \\
 & \qquad-\frac{1+t-s}{s}\delta_{n}+\delta_{n}'+\delta_{n}'',\label{eq:-102}
\end{align}
\setcounter{mytempeqncnt}{\value{equation}} \setcounter{equation}{\value{mytempeqncnt}}
\hrulefill{} 
\end{figure*}

Therefore, from Proposition \ref{thm:multiletter} we have 
\begin{align}
 & \Gamma_{1-s}^{\left(n\right)}\left(P_{Y|X},Q_{Y},R\right)\nonumber \\
 & \geq\min_{P_{X^{n}}}\max_{t\in\left[0,s\right]}\min_{T_{X},V_{Y|X}}-\frac{t}{s}R+\frac{t}{s}I\left(V_{Y|X},T_{X}\right)\nonumber \\
 & \qquad+\left(\frac{1}{s}-1\right)\sum_{x,y}T\left(x,y\right)\log\frac{V\left(y|x\right)}{P\left(y|x\right)}\nonumber \\
 & \qquad+\sum_{y}T\left(y\right)\log\frac{T\left(y\right)}{Q\left(y\right)}+o\left(1\right)\\
 & =\max_{t\in\left[0,s\right]}\min_{T_{X},V_{Y|X}}-\frac{t}{s}R+\frac{t}{s}I\left(V_{Y|X},T_{X}\right)\nonumber \\
 & \qquad+\left(\frac{1}{s}-1\right)\sum_{x,y}T\left(x,y\right)\log\frac{V\left(y|x\right)}{P\left(y|x\right)}\nonumber \\
 & \qquad+\sum_{y}T\left(y\right)\log\frac{T\left(y\right)}{Q\left(y\right)}+o\left(1\right)\\
 & \geq\max_{t\in\left[0,s\right]}\min_{\widetilde{P}_{XY}\in\mathcal{P}\left(\mathcal{X\times Y}\right)}-\frac{t}{s}R+\frac{t}{s}I\left(\widetilde{P}_{Y|X},\widetilde{P}_{X}\right)\nonumber \\
 & \qquad+\left(\frac{1}{s}-1\right)\sum_{x,y}\widetilde{P}\left(x,y\right)\log\frac{\widetilde{P}\left(y|x\right)}{P\left(y|x\right)}\nonumber \\
 & \qquad+\sum_{y}\widetilde{P}\left(y\right)\log\frac{\widetilde{P}\left(y\right)}{Q\left(y\right)}+o\left(1\right)\\
 & =\min_{\widetilde{P}_{XY}\in\mathcal{P}\left(\mathcal{X\times Y}\right)}\max_{t\in\left[0,s\right]}-\frac{t}{s}R+\frac{t}{s}I\left(\widetilde{P}_{Y|X},\widetilde{P}_{X}\right)\nonumber \\
 & \qquad+\left(\frac{1}{s}-1\right)\sum_{x,y}\widetilde{P}\left(x,y\right)\log\frac{\widetilde{P}\left(y|x\right)}{P\left(y|x\right)}\nonumber \\
 & \qquad+\sum_{y}\widetilde{P}\left(y\right)\log\frac{\widetilde{P}\left(y\right)}{Q\left(y\right)}+o\left(1\right)\label{eq:-47}\\
 & =\min_{\widetilde{P}_{XY}\in\mathcal{P}\left(\mathcal{X\times Y}\right)}\max\biggl\{\left(\frac{1}{s}-1\right)\sum_{x,y}\widetilde{P}\left(x,y\right)\log\frac{\widetilde{P}\left(y|x\right)}{P\left(y|x\right)}\nonumber \\
 & \qquad+\sum_{x,y}\widetilde{P}\left(x,y\right)\log\frac{\widetilde{P}\left(y|x\right)}{Q\left(y\right)}-R,\nonumber \\
 & \quad\left(\frac{1}{s}-1\right)\sum_{x,y}\widetilde{P}\left(x,y\right)\log\frac{\widetilde{P}\left(y|x\right)}{P\left(y|x\right)}\nonumber \\
 & \qquad+\sum_{x,y}\widetilde{P}\left(y\right)\log\frac{\widetilde{P}\left(y\right)}{Q\left(y\right)}\biggr\}+o\left(1\right)\label{eq:-99}\\
 & =\Gamma_{1-s}^{\mathsf{LB}}\left(P_{Y|X},Q_{Y},R\right)+o\left(1\right),
\end{align}
where the swapping of min and max in \eqref{eq:-47} follows from
the fact that the objective function, equal to 
\begin{align}
 & -\frac{t}{s}R+\frac{t}{s}\sum_{x,y}\widetilde{P}\left(x,y\right)\log\frac{\widetilde{P}\left(y|x\right)}{\widetilde{P}\left(y\right)}\nonumber \\
 & \qquad+\left(\frac{1}{s}-1\right)\sum_{x,y}\widetilde{P}\left(x,y\right)\log\frac{\widetilde{P}\left(y|x\right)}{P\left(y|x\right)}\nonumber \\
 & \qquad+\sum_{y}\widetilde{P}\left(y\right)\log\frac{\widetilde{P}\left(y\right)}{Q\left(y\right)}\nonumber \\
 & =-\frac{t}{s}R+\left(\frac{1+t}{s}-1\right)\sum_{x,y}\widetilde{P}\left(x,y\right)\log\widetilde{P}\left(y|x\right)\nonumber \\
 & \qquad+\left(1-\frac{t}{s}\right)\sum_{y}\widetilde{P}\left(y\right)\log\widetilde{P}\left(y\right)\nonumber \\
 & \qquad-\left(\frac{1}{s}-1\right)\sum_{x,y}\widetilde{P}\left(x,y\right)\log P\left(y|x\right)\nonumber \\
 & \qquad-\sum_{y}\widetilde{P}\left(y\right)\log Q\left(y\right),
\end{align}
is convex and concave in $\widetilde{P}_{XY}$ and $t$ respectively,
$\widetilde{P}_{XY}$ resides in a compact, convex set (the probability
simplex) and $t$ resides in a convex set $\left[0,s\right]$ (Sion's
minimax theorem \cite{Sion}). %, and \eqref{eq:-99} follows
%from the fact $\mathcal{P}^{\left(n\right)}\left(\mathcal{X}\times\mathcal{Y}\right)\subseteq\mathcal{P}\left(\mathcal{X}\times\mathcal{Y}\right)$.

\section{\label{sec:Proof-of-Theorem-normalized}Proof of Theorem \ref{thm:Resolvability}}

Since the unnormalized R\'enyi resolvability is not smaller than normalized
one, we only need prove the converse part for normalized case and
the achievability part for unnormalized case.

\subsection{Converse for Normalized Case with $1+s,\;s\in(0,\infty]$}

We first consider the case $s\in(0,1]$. By Theorem \ref{thm:singleletter},
$\lim_{n\to\infty}\frac{1}{n}\inf_{f_{\mathcal{C}_{n}}}D_{1+s}(P_{Y^{n}\mathcal{C}_{n}}\|Q_{Y}^{n}P_{\mathcal{C}_{n}})=0$
if and only if there exists a $\widetilde{P}_{X}$ such that 
\begin{align}
\max_{\widetilde{P}_{Y|X}}\eta_{1+s}\left(P_{Y|X},Q_{Y},\widetilde{P}_{X},\widetilde{P}_{Y|X}\right) & \leq0,\label{eq:-20}\\
\frac{1}{s}\sum_{x}\widetilde{P}\left(x\right)\log\left(\sum_{y}P^{1+s}\left(y|x\right)Q^{-s}\left(y\right)\right)-R & \leq0.\label{eq:-14}
\end{align}

On one hand, 
\begin{align}
 & \max_{\widetilde{P}_{Y|X}}\eta_{1+s}\left(P_{Y|X},Q_{Y},\widetilde{P}_{X},\widetilde{P}_{Y|X}\right)\nonumber \\
 & \geq\sum_{x,y}\widetilde{P}\left(x\right)P\left(y|x\right)\log\frac{\sum_{x}\widetilde{P}\left(x\right)P\left(y|x\right)}{Q\left(y\right)}.
\end{align}
Therefore, \eqref{eq:-20} implies 
\begin{equation}
\sum_{x}\widetilde{P}\left(x\right)P\left(y|x\right)=Q\left(y\right),
\end{equation}
i.e., 
\begin{equation}
\widetilde{P}_{X}\in\mathcal{P}\left(P_{Y|X},Q_{Y}\right).\label{eq:-1-1}
\end{equation}

On the other hand, if $\widetilde{P}_{X}\in\mathcal{P}\left(P_{Y|X},Q_{Y}\right)$,
then 
\begin{align}
 & \max_{\widetilde{P}_{Y|X}}\eta_{1+s}\left(P_{Y|X},Q_{Y},\widetilde{P}_{X},\widetilde{P}_{Y|X}\right)\nonumber \\
 & =\max_{\widetilde{P}_{Y|X}}\biggl\{\left(-\frac{1}{s}-1\right)\sum_{x,y}\widetilde{P}\left(x,y\right)\log\frac{\widetilde{P}\left(y|x\right)}{P\left(y|x\right)}\nonumber \\
 & \qquad+\sum_{x,y}\widetilde{P}\left(y\right)\log\frac{\widetilde{P}\left(y\right)}{Q\left(y\right)}\biggr\}\\
 & \leq\max_{\widetilde{P}_{Y|X}}\biggl\{\left(-\frac{1}{s}-1\right)\sum_{x,y}\widetilde{P}\left(x,y\right)\log\frac{\sum_{x}\widetilde{P}\left(y|x\right)\widetilde{P}\left(x\right)}{\sum_{x}P\left(y|x\right)\widetilde{P}\left(x\right)}\nonumber \\
 & \qquad+\sum_{x,y}\widetilde{P}\left(y\right)\log\frac{\widetilde{P}\left(y\right)}{Q\left(y\right)}\biggr\}\label{eq:-14-1}\\
 & =\max_{\widetilde{P}_{Y|X}}\left\{ -\frac{1}{s}\sum_{x,y}\widetilde{P}\left(y\right)\log\frac{\widetilde{P}\left(y\right)}{Q\left(y\right)}\right\} \\
 & \leq0,
\end{align}
where follows from the log-sum inequality \cite{Cover}. Therefore,
\eqref{eq:-20} is equivalent to \eqref{eq:-1-1}.

Combining \eqref{eq:-14-1} and \eqref{eq:-1-1} we have 
\begin{align}
 & \inf\left\{ R:\lim_{n\to\infty}\Gamma_{1+s}^{\left(n\right)}\left(P_{Y|X},Q_{Y},R\right)=0\right\} \nonumber \\
 & =\inf_{P_{X}\in\mathcal{P}\left(P_{Y|X},Q_{Y}\right)}\frac{1}{s}\sum_{x}P\left(x\right)\log\sum_{y}P^{1+s}\left(y|x\right)Q^{-s}\left(y\right).
\end{align}

Now we consider the case $s\in(-1,0]$. That is, we need to prove
for $s\in(0,1]$, 
\begin{align}
 & \inf\left\{ R:\frac{1}{n}\inf_{f_{\mathcal{C}_{n}}}D_{1-s}(P_{Y^{n}\mathcal{C}_{n}}\|Q_{Y}^{n}P_{\mathcal{C}_{n}})\rightarrow0\right\} \nonumber \\
 & \ge\min_{P_{X}\in\mathcal{P}\left(P_{Y|X},Q_{Y}\right)}I\left(X;Y\right).
\end{align}

By Theorem \ref{thm:singleletter}, we have 
\begin{align}
 & \inf\left\{ R:\Gamma_{1-s}^{\mathsf{LB}}\left(P_{Y|X},Q_{Y},R\right)=0\right\} \nonumber \\
 & \leq\inf\left\{ R:\frac{1}{n}\inf_{f_{\mathcal{C}_{n}}}D_{1-s}(P_{Y^{n}\mathcal{C}_{n}}\|Q_{Y^{n}}P_{\mathcal{C}_{n}})\rightarrow0\right\} .
\end{align}

Furthermore, $\Gamma_{1-s}^{\mathsf{LB}}\left(P_{Y|X},Q_{Y},R\right)=0$
is equivalent to that there exist $\widetilde{P}_{X},\widetilde{P}_{Y|X}$
such that 
\begin{align}
\left(\frac{1}{s}-1\right)\sum_{x,y}\widetilde{P}\left(x,y\right)\log\frac{\widetilde{P}\left(y|x\right)}{P\left(y|x\right)}+\sum_{x,y}\widetilde{P}\left(y\right)\log\frac{\widetilde{P}\left(y\right)}{Q\left(y\right)} & \leq0,\label{eq:-48}\\
\left(\frac{1}{s}-1\right)\sum_{x,y}\widetilde{P}\left(x,y\right)\log\frac{\widetilde{P}\left(y|x\right)}{P\left(y|x\right)}\qquad\qquad\qquad\nonumber \\
+\sum_{x,y}\widetilde{P}\left(x,y\right)\log\frac{\widetilde{P}\left(y|x\right)}{Q\left(y\right)}-R & \leq0.\label{eq:-2-1}
\end{align}
Note that \eqref{eq:-48} is equivalent to 
\begin{align}
\widetilde{P}\left(y|x\right)=P\left(y|x\right),\quad\mbox{and}\quad\widetilde{P}\left(y\right)=Q\left(y\right).
\end{align}
Hence \eqref{eq:-1-1} also holds. Combining \eqref{eq:-2-1} and
\eqref{eq:-1-1} we have 
\begin{align}
 & \inf\left\{ R:\Gamma_{1-s}^{\mathsf{LB}}\left(P_{Y|X},Q_{Y},R\right)=0\right\} \nonumber \\
 & =\inf_{\widetilde{P}_{X}\in\mathcal{P}\left(P_{Y|X},Q_{Y}\right)}\sum_{x,y}\widetilde{P}\left(x\right)P\left(y|x\right)\log\frac{P\left(y|x\right)}{Q\left(y\right)}\\
 & =\inf_{P_{X}\in\mathcal{P}\left(P_{Y|X},Q_{Y}\right)}I\left(X;Y\right).
\end{align}

\subsection{Achievability for Unnormalized Case with $1+s,\;s\in(-1,1]\cup\{\infty\}$ }

Next we focus on the achievability part. Since the result for $s\in(-1,0]$
can be obtained from existing works (see Remark \ref{rmk:unnormRenyi}),
we only need to prove the case $s\in(0,1]\cup\{\infty\}$.

\subsubsection{Case $s\in(0,1]$}

We first consider the case $s\in(0,1]$. For this case, by Lemmas
\ref{lem:oneshotach} and \ref{lem:oneshotcon}, we deduce that 
\begin{align}
 & \inf\left\{ R:\inf_{f_{\mathcal{C}_{n}}}D_{1+s}(P_{Y^{n}\mathcal{C}_{n}}\|Q_{Y}^{n}P_{\mathcal{C}_{n}})\rightarrow0\right\} \nonumber \\
 & =\inf_{\left\{ P_{X^{n}}\right\} :D_{1+s}(P_{Y^{n}}\|Q_{Y}^{n})\rightarrow0}\limsup_{n\to\infty}\frac{1}{n}D_{1+s}\left(P_{X^{n}Y^{n}}\|P_{X^{n}}Q_{Y}^{n}\right).\label{eq:-81}
\end{align}
Set $P_{X^{n}}\left(x^{n}\right)\propto Q_{X}^{n}\left(x^{n}\right)1\left\{ x^{n}\in\mathcal{T}_{\epsilon}^{n}\left(Q_{X}\right)\right\} $
for some $Q_{X}\in\mathcal{P}\left(P_{Y|X},Q_{Y}\right)$. On one
hand, 
\begin{align}
 & D_{1+s}(P_{X^{n}}\|Q_{X}^{n})\nonumber \\
 & =\frac{1}{s}\log\sum_{x^{n}}\left(\frac{Q_{X}^{n}\left(x^{n}\right)1\left\{ x^{n}\in\mathcal{T}_{\epsilon}^{n}\right\} }{Q_{X}^{n}\left(\mathcal{T}_{\epsilon}^{n}\right)}\right)^{1+s}\left(Q_{X}^{n}\left(x^{n}\right)\right)^{-s}\label{eq:-46}\\
 & =\frac{1}{s}\log\sum_{x^{n}\in\mathcal{T}_{\epsilon}^{n}}\left(\frac{1}{Q_{X}^{n}\left(\mathcal{T}_{\epsilon}^{n}\right)}\right)^{1+s}Q_{X}^{n}\left(x^{n}\right)\\
 & =\log\frac{1}{Q_{X}^{n}\left(\mathcal{T}_{\epsilon}^{n}\right)}\label{eq:-52}\\
 & \rightarrow0,\label{eq:-15}
\end{align}
where \eqref{eq:-15} follows from the fact that $Q_{X}^{n}\left(\mathcal{T}_{\epsilon}^{n}\right)\rightarrow1$.
By the data processing inequality \cite{Erven}, we have 
\begin{equation}
D_{1+s}(P_{Y^{n}}\|Q_{Y}^{n})\leq D_{1+s}(P_{X^{n}}\|Q_{X}^{n}).
\end{equation}
Hence $D_{1+s}(P_{Y^{n}}\|Q_{Y}^{n})\rightarrow0$ as well.

On the other hand, 
\begin{align}
 & \frac{1}{n}D_{1+s}\left(P_{X^{n}Y^{n}}\|P_{X^{n}}Q_{Y}^{n}\right)\nonumber \\
 & =\frac{1}{ns}\log\sum_{x^{n}}\frac{Q_{X}^{n}\left(x^{n}\right)1\left\{ x^{n}\in\mathcal{T}_{\epsilon}^{n}\right\} }{Q_{X}^{n}\left(\mathcal{T}_{\epsilon}^{n}\right)}\nonumber \\
 & \qquad\times\e^{n\sum_{x}T_{x^{n}}\left(x\right)\log\sum_{y}P^{1+s}\left(y|x\right)Q^{-s}\left(y\right)}\\
 & \leq\frac{1}{ns}\log\sum_{x^{n}}\frac{Q_{X}^{n}\left(x^{n}\right)1\left\{ x^{n}\in\mathcal{T}_{\epsilon}^{n}\right\} }{Q_{X}^{n}\left(\mathcal{T}_{\epsilon}^{n}\right)}\max_{\substack{T_{X}:\\
\forall x:\left|T_{X}\left(x\right)-Q_{X}\left(x\right)\right|\leq\epsilon Q_{X}\left(x\right)
}
}\nonumber \\
 & \qquad\e^{n\sum_{x}T_{X}\left(x\right)\log\sum_{y}P^{1+s}\left(y|x\right)Q^{-s}\left(y\right)}\\
 & =\max_{\substack{T_{X}:\\
\forall x:\left|T_{X}\left(x\right)-Q_{X}\left(x\right)\right|\leq\epsilon Q_{X}\left(x\right)
}
}\frac{1}{s}\sum_{x}T_{X}\left(x\right)\nonumber \\
 & \qquad\times\log\sum_{y}P^{1+s}\left(y|x\right)Q^{-s}\left(y\right)\\
 & \leq\left(1+\epsilon\right)\frac{1}{s}\sum_{x}Q\left(x\right)\log\sum_{y}P^{1+s}\left(y|x\right)Q^{-s}\left(y\right).\label{eq:-49}
\end{align}

By letting $n\to\infty$ and $\epsilon\rightarrow0$, we have 
\begin{align}
 & \limsup_{n\rightarrow\infty}\frac{1}{n}D_{1+s}\left(P_{X^{n}Y^{n}}\|P_{X^{n}}Q_{Y}^{n}\right)\nonumber \\
 & \leq\frac{1}{s}\sum_{x}Q\left(x\right)\log\sum_{y}P^{1+s}\left(y|x\right)Q^{-s}\left(y\right).
\end{align}
Furthermore, since $Q_{X}\in\mathcal{P}\left(P_{Y|X},Q_{Y}\right)$
is arbitrary, 
\begin{align}
 & \limsup_{n\rightarrow\infty}\frac{1}{n}D_{1+s}\left(P_{X^{n}Y^{n}}\|P_{X^{n}}Q_{Y}^{n}\right)\nonumber \\
 & \leq\inf_{Q_{X}\in\mathcal{P}\left(P_{Y|X},Q_{Y}\right)}\frac{1}{s}\sum_{x}Q\left(x\right)\log\sum_{y}P^{1+s}\left(y|x\right)Q^{-s}\left(y\right).
\end{align}
Combining this with \eqref{eq:-81} we have the achievability part
for the case of $s\in(0,1]$.

\subsubsection{\label{subsec:Case}Case $s=\infty$}

Let $\epsilon>0$ be such that 
\begin{equation}
R>\left(1+\epsilon\right)\sum_{x}Q_{X}(x)D_{\infty}\left(P_{Y|X}\left(\cdot|x\right)\|Q_{Y}\right)+\epsilon.
\end{equation}
Here 
\begin{equation}
Q_{X}:=\argmin_{P_{X}\in\mathcal{P}\left(P_{Y|X},Q_{Y}\right)}\sum_{x}P_{X}\left(x\right)D_{\infty}\left(P_{Y|X}\left(\cdot|x\right)\|Q_{Y}\right).
\end{equation}

We set the random code to be $\mathcal{C}_{n}=\left\{ X^{n}\left(m\right)\right\} _{m\in\calM_{n}}$
with $X^{n}\left(m\right),m\in\calM_{n}$ drawn independently for
different $m$'s and according to the same distribution $P_{X^{n}}$
such that $P_{X^{n}}\left(x^{n}\right)\propto Q_{X}^{n}\left(x^{n}\right)1\left\{ x^{n}\in\mathcal{T}_{\epsilon}^{n}\left(Q_{X}\right)\right\} $.
%Consider the random code $f_{\mathcal{C}_{n}}\left(m\right)=X^{n}\left(m\right)$.
Next we prove that such a sequence of random codes satisfies $\mathbb{E}_{\mathcal{C}_{n}}\left[D_{\infty}(P_{Y^{n}|\mathcal{C}_{n}}\|Q_{Y}^{n})\right]\to0$
as $n\to\infty$.

For brevity, in the following we denote $M=\e^{nR}$. According to
the definition of the R\'enyi divergence, we first have 
\begin{align}
 & \e^{\mathbb{E}_{\mathcal{C}_{n}}\left[D_{\infty}(P_{Y^{n}|\mathcal{C}_{n}}\|Q_{Y}^{n})\right]}\nonumber \\
 & \leq\mathbb{E}_{\mathcal{C}_{n}}\left[\e^{D_{\infty}(P_{Y^{n}|\mathcal{C}_{n}}\|Q_{Y}^{n})}\right]\\
 & =\mathbb{E}_{\mathcal{C}_{n}}\left[\max_{y^{n}}\frac{P_{Y^{n}|\mathcal{C}_{n}}\left(y^{n}|\mathcal{C}_{n}\right)}{Q_{Y}^{n}\left(y^{n}\right)}\right]\\
 & =\mathbb{E}_{\mathcal{C}_{n}}\left[\max_{y^{n}}\widetilde{g}(\mathcal{C}_{n},y^{n})\right],\label{eq:-151}
\end{align}
where $\widetilde{g}(\mathcal{C}_{n},y^{n}):=\sum_{m\in\calM_{n}}g(X^{n}(m),y^{n})/M$
with $g(x^{n},y^{n}):={P_{Y|X}^{n}\left(y^{n}|x^{n}\right)}/{Q_{Y}^{n}\left(y^{n}\right)}$.
Obviously, for any $x^{n}\in\mathcal{T}_{\epsilon}^{n}\left(Q_{X}\right)$,
its type $T_{x^{n}}$ satisfies that $\left|T_{x^{n}}\left(x\right)-Q_{X}\left(x\right)\right|\leq\epsilon Q_{X}\left(x\right),\forall x$.
Therefore, for any $x^{n}\in\mathcal{T}_{\epsilon}^{n}\left(Q_{X}\right)$
and any $y^{n}\in\mathcal{Y}^{n}$, we have 
\begin{align}
 & g(x^{n},y^{n})\nonumber \\
 & =\e^{n\sum_{x,y}T_{x^{n}y^{n}}(x,y)\log\frac{P_{Y|X}\left(y|x\right)}{Q_{Y}\left(y\right)}}\\
 & \leq\max_{T_{XY}:\forall x:\left|T_{X}\left(x\right)-Q_{X}\left(x\right)\right|\leq\epsilon Q_{X}\left(x\right)}\e^{n\sum_{x,y}T_{XY}(x,y)\log\frac{P_{Y|X}\left(y|x\right)}{Q_{Y}\left(y\right)}}\\
 & \leq\e^{n\max_{T_{X}:\forall x:\left|T_{X}\left(x\right)-Q_{X}\left(x\right)\right|\leq\epsilon Q_{X}\left(x\right)}\sum_{x}T_{X}(x)\max_{y}\log\frac{P_{Y|X}\left(y|x\right)}{Q_{Y}\left(y\right)}}\\
 & \leq\e^{n\left(1+\epsilon\right)\sum_{x}Q_{X}(x)D_{\infty}\left(P_{Y|X}\left(\cdot|x\right)\|Q_{Y}\right)}\\
 & =:A_{n}.
\end{align}
Continuing \eqref{eq:-151}, we get for any $\epsilon'>0$, 
\begin{align}
 & \e^{\mathbb{E}_{\mathcal{C}_{n}}\left[D_{\infty}(P_{Y^{n}|\mathcal{C}_{n}}\|Q_{Y}^{n})\right]}\nonumber \\
 & =\mathbb{E}_{\mathcal{C}_{n}}\left[\max_{y^{n}}\widetilde{g}(\mathcal{C}_{n},y^{n})1\left\{ \max_{y^{n}}\widetilde{g}(\mathcal{C}_{n},y^{n})\geq1+\epsilon'\right\} \right]\nonumber \\
 & \qquad+\mathbb{E}_{\mathcal{C}_{n}}\left[\max_{y^{n}}\widetilde{g}(\mathcal{C}_{n},y^{n})1\left\{ \max_{y^{n}}\widetilde{g}(\mathcal{C}_{n},y^{n})<1+\epsilon'\right\} \right]\\
 & \leq\mathbb{E}_{\mathcal{C}_{n}}\left[A_{n}\cdot1\left\{ \max_{y^{n}}\widetilde{g}(\mathcal{C}_{n},y^{n})\geq1+\epsilon'\right\} \right]+1+\epsilon'\\
 & =A_{n}\mathbb{P}_{\mathcal{C}_{n}}\left(\max_{y^{n}}\widetilde{g}(\mathcal{C}_{n},y^{n})\geq1+\epsilon'\right)+1+\epsilon'\\
 & \leq A_{n}\left|\mathcal{Y}\right|^{n}\max_{y^{n}}\mathbb{P}_{\mathcal{C}_{n}}\left(\widetilde{g}(\mathcal{C}_{n},y^{n})\geq1+\epsilon'\right)+1+\epsilon',\label{eq:-152}
\end{align}
where \eqref{eq:-152} follows from the union bound. Obviously, both
$A_{n}$ and $\left|\mathcal{Y}\right|^{n}$ are only exponentially
growing. Therefore, if the probability $\mathbb{P}_{\mathcal{C}_{n}}\left(\widetilde{g}(\mathcal{C}_{n},y^{n})\geq1+\epsilon'\right)$
vanishes doubly exponentially fast, then $\mathbb{E}_{\mathcal{C}_{n}}\left[D_{\infty}(P_{Y^{n}|\mathcal{C}_{n}}\|Q_{Y}^{n})\right]\to\log\left(1+\epsilon'\right)$
as $n\to\infty$. To this end, we use Bernstein's inequality \cite{boucheron2013concentration}
to bound the probability uniformly over all $y^{n}$. Observe that
$g(X^{n}(m),y^{n}),m\in\calM_{n}$ are i.i.d.\ random variables with
mean 
\begin{align}
\mu_{\epsilon,n} & :=\mathbb{E}_{X^{n}}\left[g(X^{n},y^{n})\right]\\
 & =\sum_{x^{n}}\frac{Q_{X}^{n}\left(x^{n}\right)1\left\{ x^{n}\in\mathcal{T}_{\epsilon}^{n}\left(Q_{X}\right)\right\} }{Q_{X}^{n}\left(\mathcal{T}_{\epsilon}^{n}\left(Q_{X}\right)\right)}\frac{P_{Y|X}^{n}\left(y^{n}|x^{n}\right)}{Q_{Y}^{n}\left(y^{n}\right)}\\
 & \leq\frac{1}{Q_{X}^{n}\left(\mathcal{T}_{\epsilon}^{n}\left(Q_{X}\right)\right)}\\
 & \to1,\textrm{ as }n\to\infty,
\end{align}
and variance 
\begin{align}
\mathrm{Var}_{X^{n}}\left[g(X^{n},y^{n})\right] & \leq\mathbb{E}_{X^{n}}\left[g(X^{n},y^{n})^{2}\right]\\
 & \leq A_{n}\mu_{\epsilon,n}.
\end{align}
Then we get 
\begin{align}
 & \mathbb{P}_{\mathcal{C}_{n}}\left(\widetilde{g}(\mathcal{C}_{n},y^{n})\geq1+\epsilon'\right)\nonumber \\
 & =\mathbb{P}_{\mathcal{C}_{n}}\biggl(\sum_{m\in\calM_{n}}g(X^{n}(m),y^{n})-\mu_{\epsilon,n}M\nonumber \\
 & \qquad\qquad\qquad\geq(1+\epsilon'-\mu_{\epsilon,n})M\biggr)\\
 & \leq\exp\left(-\frac{\frac{1}{2}\left(1+\epsilon'-\mu_{\epsilon,n}\right)^{2}M^{2}}{MA_{n}\mu_{\epsilon,n}+\frac{1}{3}\left(1+\epsilon'-\mu_{\epsilon,n}\right)MA_{n}}\right)\\
 & \leq\exp\left(-\frac{3\left(1+\epsilon'-\mu_{\epsilon,n}\right)^{2}M}{2\left(1+\epsilon'+2\mu_{\epsilon,n}\right)A_{n}}\right).\label{eq:-153-1}
\end{align}
Since $\mu_{\epsilon,n}\to1$ as $n\to\infty$, we have that for any
$\epsilon'>0$, there exists a sufficiently large $n_{0}$ such that
$\mu_{\epsilon,n}\leq1+\frac{\epsilon'}{2}$ for $n\ge n_{0}$. Hence
for $n\ge n_{0}$, \eqref{eq:-153-1} is further upper bounded by
 $\exp\left(-\frac{3\epsilon'^{2}}{8\left(3+2\epsilon'\right)}\mathrm{e}^{n\epsilon}\right)$,
which converges to zero doubly exponentially fast. Therefore, $\mathbb{E}_{\mathcal{C}_{n}}\left[D_{\infty}(P_{Y^{n}|\mathcal{C}_{n}}\|Q_{Y}^{n})\right]\to\log\left(1+\epsilon'\right)$
as $n\to\infty$. Since $\epsilon'>0$ is arbitrary, $\mathbb{E}_{\mathcal{C}_{n}}\left[D_{\infty}(P_{Y^{n}|\mathcal{C}_{n}}\|Q_{Y}^{n})\right]\to0$
as $n\to\infty$. 

Note that here we have proven that if the code rate 
\begin{align}
R>\min\limits _{P_{X}\in\mathcal{P}\left(P_{Y|X},Q_{Y}\right)}\sum_{x}P_{X}\left(x\right)D_{\infty}\left(P_{Y|X}\left(\cdot|x\right)\|Q_{Y}\right),
\end{align}
then 
\begin{align}
\mathbb{E}_{\mathcal{C}_{n}}\left[D_{\infty}(P_{Y^{n}|\mathcal{C}_{n}}\|Q_{Y}^{n})\right]\rightarrow0.\label{eqn:ECDinfty}
\end{align}
The convergence in \eqref{eqn:ECDinfty} implies that there exists
a sequence of deterministic codebooks $\left\{ c_{n}\right\} $ with
rate $R$ such that $D_{\infty}(P_{Y^{n}|\mathcal{C}_{n}=c_{n}}\|Q_{Y}^{n})\rightarrow0$.
If we set the random mapping $\mathcal{C}_{n}'$ to be the deterministic
codebook/mapping $c_{n}$, i.e., $\mathcal{C}_{n}'=c_{n}$, then $D_{\infty}(P_{Y^{n}\mathcal{C}_{n}'}\|Q_{Y}^{n}P_{\mathcal{C}_{n}'})=D_{\infty}(P_{Y^{n}|\mathcal{C}_{n}=c_{n}}\|Q_{Y}^{n})$.
Therefore, we have $D_{\infty}(P_{Y^{n}\mathcal{C}_{n}'}\|Q_{Y}^{n}P_{\mathcal{C}_{n}'})\rightarrow0$
as desired.

\section{\label{sec:Proof-of-Theorem-expiid}Proof of Theorem \ref{thm:exponentforiid}}

\emph{Achievability:} We first consider $s\in(0,1]$ case. Since $P_{X}\in\mathcal{P}\left(P_{Y|X},Q_{Y}\right)$,
$D_{1+s}(P_{Y}\|Q_{Y})=0$. By Lemma \ref{lem:oneshotach}, we obtain
\begin{align}
 & \e^{sD_{1+s}(P_{Y^{n}\mathcal{C}_{n}}\|Q_{Y}^{n}\times P_{\mathcal{C}_{n}})}\nonumber \\
 & \leq\e^{n\log\sum_{x,y}P\left(x\right)P^{1+s}\left(y|x\right)Q^{-s}\left(y\right)-nsR}+\e^{nsD_{1+s}(P_{Y}\|Q_{Y})}\\
 & =\e^{n\log\sum_{x,y}P\left(x\right)P^{1+s}\left(y|x\right)Q^{-s}\left(y\right)-nsR}+1.
\end{align}
Take $\log$'s, 
\begin{align}
 & sD_{1+s}(P_{Y^{n}\mathcal{C}_{n}}\|Q_{Y}^{n}\times P_{\mathcal{C}_{n}})\nonumber \\
 & \leq\log\left(\e^{n\log\sum_{x,y}P\left(x\right)P^{1+s}\left(y|x\right)Q^{-s}\left(y\right)-nsR}+1\right)\\
 & \leq\e^{-ns\left(R-\frac{1}{s}\log\sum_{x,y}P\left(x\right)P^{1+s}\left(y|x\right)Q^{-s}\left(y\right)\right)}\\
 & =\e^{-ns\left(R-D_{1+s}\left(P_{XY}\|P_{X}\times Q_{Y}\right)\right)}.
\end{align}
Hence 
\begin{align}
 & -\frac{1}{n}\log D_{1+s}(P_{Y^{n}\mathcal{C}_{n}}\|Q_{Y}^{n}\times P_{\mathcal{C}_{n}})\nonumber \\
 & \geq s\left(R-D_{1+s}\left(P_{XY}\|P_{X}\times Q_{Y}\right)\right)+\delta_{n}.\label{eq:-4-2}
\end{align}
This implies $D_{1+s}(P_{Y^{n}\mathcal{C}_{n}}\|Q_{Y}^{n}\times P_{\mathcal{C}_{n}})$
vanishes at least exponentially fast for $s\in(0,1]$. Now we refine
the exponential rate of decay. Denote $t_{1}^{*}\in[s,1]$ as the
maximizer of $\max_{t\in[s,1]}t\left(R-D_{1+t}(P_{XY}\|P_{X}\times Q_{Y})\right)$.
Since \eqref{eq:-4-2} holds for any $s\in(0,1]$, we have for $s\in(0,1]$,
\begin{align}
 & \liminf_{n\to\infty}-\frac{1}{n}\log D_{1+s}(P_{Y^{n}\mathcal{C}_{n}}\|Q_{Y}^{n}\times P_{\mathcal{C}_{n}})\nonumber \\
 & \geq\liminf_{n\to\infty}-\frac{1}{n}\log D_{1+t_{1}^{*}}(P_{Y^{n}\mathcal{C}_{n}}\|Q_{Y}^{n}\times P_{\mathcal{C}_{n}})\label{eq:-60}\\
 & \geq t_{1}^{*}\left(R-D_{1+t_{1}^{*}}(P_{XY}\|P_{X}\times Q_{Y})\right)\label{eq:-61}\\
 & =\max_{t\in[s,1]}t\left(R-D_{1+t}(P_{XY}\|P_{X}\times Q_{Y})\right).
\end{align}

As for $s\in(-1,0]$ case, denote $t_{2}^{*}\in[0,1]$ as the maximizer
of $\max_{t\in[0,1]}t\left(R-D_{1+t}(P_{XY}\|P_{X}\times Q_{Y})\right)$.
Then similarly we can have 
\begin{align}
 & \liminf_{n\to\infty}-\frac{1}{n}\log D_{1+s}(P_{Y^{n}\mathcal{C}_{n}}\|Q_{Y}^{n}\times P_{\mathcal{C}_{n}})\nonumber \\
 & \geq\liminf_{n\to\infty}-\frac{1}{n}\log D_{1+t_{2}^{*}}(P_{Y^{n}\mathcal{C}_{n}}\|Q_{Y}^{n}\times P_{\mathcal{C}_{n}})\label{eq:-62}\\
 & \geq t_{2}^{*}\left(R-D_{1+t_{2}^{*}}(P_{XY}\|P_{X}\times Q_{Y})\right)\label{eq:-63}\\
 & =\max_{t\in[0,1]}t\left(R-D_{1+t}(P_{XY}\|P_{X}\times Q_{Y})\right).
\end{align}

\emph{Converse for $s\in(0,1]$ case:} For the converse part, we follow
steps similar to the proof in \cite{Parizi}. Let 
\begin{equation}
L\left(y^{n}\right):=\begin{cases}
\frac{P_{Y^{n}}\left(y^{n}\right)}{Q_{Y}^{n}\left(y^{n}\right)} & \text{if \ensuremath{Q_{Y}^{n}\left(y^{n}\right)>0}},\\
1 & \text{otherwise},
\end{cases}\label{eq:ldef}
\end{equation}
denote the (random) likelihood ratio of each sequence $y^{n}\in\cY^{n}$.
Note that $P\left(y^{n}\right)$ is a random probability distribution,
since the codebook is random. Since $P_{X}\in\mathcal{P}\left(P_{Y|X},Q_{Y}\right)$,
by the construction of the codebook, we have 
\begin{equation}
\bbE_{\mathcal{C}_{n}}[L\left(y^{n}\right)]=1,\qquad\forall y^{n}\in\cY^{n}.
\end{equation}

Denote 
\begin{equation}
\ell(T):=\frac{P_{Y|X}^{n}(\tilde{y}^{n}|\tilde{x}^{n})}{Q_{Y}^{n}(\tilde{y}^{n})}\qquad\text{for some \ensuremath{(\tilde{x}^{n},\tilde{y}^{n})\in\cT_{T}}}.
\end{equation}
Denote 
\begin{equation}
N_{T}(y^{n}):=\left|\bigl\{ x^{n}\in\mathcal{C}_{n}:(x^{n},y^{n})\in\cT_{T}\bigl\}\right|
\end{equation}
as the number of codewords in $\mathcal{C}_{n}$ that have the joint
type $T$ with $y^{n}$. Then $\{N_{T}(y^{n})\colon T\in\cP^{(n)}(\cX\times\cY)\}$
is a collection of $M$ random variables with multinomial distributions
and success probabilities 
\begin{equation}
p_{T}(y^{n})=\bbE_{\mathcal{C}_{n}}\left[\frac{N_{T}(y^{n})}{M}\right].\label{eq:p}
\end{equation}
For brevity, here and in the following we denote $M=\e^{nR}$.

Partition $\cP^{(n)}(\cX\times\cY)=\cP_{1}\cup\cP_{2}$ and split
$L(y^{n})=L_{1}(y^{n})+L_{2}(y^{n})$, where 
\begin{align}
\cP_{1} & :=\{T\in\cP^{(n)}(\cX\times\cY):\ell(T)\le\e^{2}M\},\\
\cP_{2} & :=\{T\in\cP^{(n)}(\cX\times\cY):\ell(T)>\e^{2}M\},
\end{align}
and 
\begin{align}
L_{1}(y^{n}) & :=\frac{1}{M}\sum_{T\in\cP_{1}}N_{T}(y^{n})\ell(T),\\
L_{2}(y^{n}) & :=\frac{1}{M}\sum_{T\in\cP_{2}}N_{T}(y^{n})\ell(T).
\end{align}
Hence 
\begin{equation}
\mathbb{E}L_{1}(y^{n})+\mathbb{E}L_{2}(y^{n})=1,\qquad\forall y^{n}\in\cY^{n}.
\end{equation}
Also define 
\begin{align}
\nu(y^{n}) & :=\var\bigl(L_{1}(y^{n})\bigr)+\frac{1}{M}\bbE^{2}[L_{1}(y^{n})],\text{ and}\label{eq:nudef}\\
\mu(y^{n}) & :=\bbE[L_{2}(y^{n})].
\end{align}
As in \cite{Parizi}, by elementary properties of multinomial distribution
one can show that 
\begin{align}
\nu(y^{n}) & =\frac{1}{M}\sum_{T\in\cP_{1}}\ell(T)^{2}p_{T}(y^{n}),\label{eq:nuval}\\
\mu(y^{n}) & =\sum_{T\in\cP_{2}}\ell(T)p_{T}(y^{n}).\label{eq:muval}
\end{align}

Based on the above considerations, we have 
\begin{align}
 & D_{1+s}(P_{Y^{n}\mathcal{C}_{n}}\|Q_{Y}^{n}\times P_{\mathcal{C}_{n}})\nonumber \\
 & =\frac{1}{s}\log\left(\bbE_{\mathcal{C}_{n}}\sum_{y^{n}}Q\left(y^{n}\right)L^{1+s}\left(y^{n}\right)\right)\label{eq:-134}\\
 & \geq\frac{1}{s}\log\left(\bbE_{\mathcal{C}_{n}}\sum_{y^{n}}Q\left(y^{n}\right)\left(L_{1}^{1+s}\left(y^{n}\right)+L_{2}^{1+s}\left(y^{n}\right)\right)\right)\label{eq:-133}\\
 & =\frac{1}{s}\log\biggl(1+\bbE_{\mathcal{C}_{n}}\sum_{y^{n}}Q\left(y^{n}\right)\bigl(L_{1}^{1+s}\left(y^{n}\right)-L_{1}\left(y^{n}\right)\nonumber \\
 & \qquad+L_{2}^{1+s}\left(y^{n}\right)-L_{2}\left(y^{n}\right)\bigr)\biggr)\\
 & \doteq\frac{1}{s}\bbE_{\mathcal{C}_{n}}\sum_{y^{n}}Q\left(y^{n}\right)\bigl(L_{1}^{1+s}\left(y^{n}\right)-L_{1}\left(y^{n}\right)\nonumber \\
 & \qquad+L_{2}^{1+s}\left(y^{n}\right)-L_{2}\left(y^{n}\right)\bigr)\label{eq:-111}\\
 & \geq\bbE_{\mathcal{C}_{n}}\sum_{y^{n}}Q\left(y^{n}\right)L_{1}\left(y^{n}\right)\log L_{1}\left(y^{n}\right)\nonumber \\
 & \qquad+\frac{1}{s}\bbE_{\mathcal{C}_{n}}\sum_{y^{n}}Q\left(y^{n}\right)\left(L_{2}^{1+s}\left(y^{n}\right)-L_{2}\left(y^{n}\right)\right)\label{eq:-64}\\
 & =\sum_{y^{n}}Q\left(y^{n}\right)\Bigl(\bbE_{\mathcal{C}_{n}}L_{1}\left(y^{n}\right)\log L_{1}\left(y^{n}\right)\nonumber \\
 & \qquad+\bbE_{\mathcal{C}_{n}}\frac{1}{s}\left(L_{2}^{1+s}\left(y^{n}\right)-L_{2}\left(y^{n}\right)\right)\Bigr)\label{eq:-21}\\
 & \geq\sum_{y^{n}}Q\left(y^{n}\right)\Bigl(\frac{1}{M}\sum_{T\in\mathcal{P}_{1}}\ell^{2}\left(T\right)p_{T}\left(y^{n}\right)-\bbE_{\mathcal{C}_{n}}L_{2}\left(y^{n}\right)-\frac{1}{M}\nonumber \\
 & \qquad+\bbE_{\mathcal{C}_{n}}\frac{1}{s}\left(L_{2}^{1+s}\left(y^{n}\right)-L_{2}\left(y^{n}\right)\right)\Bigr),\label{eq:-22}
\end{align}
where \eqref{eq:-133} follows from Lemma \ref{lem:norm}, \eqref{eq:-111}
follows from $\lim_{x\to0}\frac{\log\left(1+x\right)}{x}=1$ and 
\begin{align}
 & \bbE_{\mathcal{C}_{n}}\sum_{y^{n}}Q\left(y^{n}\right)\Bigl(L_{1}^{1+s}\left(y^{n}\right)-L_{1}\left(y^{n}\right)\nonumber \\
 & \qquad\qquad+L_{2}^{1+s}\left(y^{n}\right)-L_{2}\left(y^{n}\right)\Bigr)\rightarrow0
\end{align}
(this is obtained from the achievability part, where we have $D_{1+s}(P_{Y^{n}\mathcal{C}_{n}}\|Q_{Y}^{n}\times P_{\mathcal{C}_{n}})\rightarrow0$),
\eqref{eq:-64} follows from $\frac{1}{s}\left(x^{1+s}-x\right)\geq x\log x$
(i.e., $\frac{1}{s}\left(x^{s}-1\right)\geq\log x$) for $s>0$ and
$x\geq0$ ($0\log0:=0$), and \eqref{eq:-22} follows from $\bbE_{\mathcal{C}_{n}}L_{1}\left(y^{n}\right)\log L_{1}\left(y^{n}\right)\geq\frac{1}{M}\sum_{T\in\mathcal{P}_{1}}\ell^{2}\left(T\right)p_{T}\left(y^{n}\right)-\bbE_{\mathcal{C}_{n}}L_{2}\left(y^{n}\right)-\frac{1}{M}$
(which was proven in \cite[Section~V-C]{Parizi}).

Considering the last term in the bracket of \eqref{eq:-22}, we have
\begin{align}
 & \frac{1}{s}\left(L_{2}^{1+s}\left(y^{n}\right)-L_{2}\left(y^{n}\right)\right)\nonumber \\
 & =\frac{\alpha+1-\alpha}{s}\left(L_{2}^{1+s}\left(y^{n}\right)-L_{2}\left(y^{n}\right)\right)\\
 & \geq\alpha L_{2}\left(y^{n}\right)\log L_{2}\left(y^{n}\right)+\frac{1-\alpha}{s}\left(L_{2}^{1+s}\left(y^{n}\right)-L_{2}\left(y^{n}\right)\right)\label{eq:-65}\\
 & \geq2\alpha L_{2}\left(y^{n}\right)+\frac{1-\alpha}{s}\left(L_{2}^{1+s}\left(y^{n}\right)-L_{2}\left(y^{n}\right)\right)\label{eq:-66}\\
 & =\left(2\alpha-1-\frac{1-\alpha}{s}\right)L_{2}\left(y^{n}\right)+\frac{1-\alpha}{s}L_{2}^{1+s}\left(y^{n}\right),\label{eq:-122}
\end{align}
where $\alpha\in[0,1]$ is an arbitrary number, \eqref{eq:-65} follows
from $\frac{1}{s}\left(x^{1+s}-x\right)\geq x\log x$, and \eqref{eq:-66}
follows from $L_{2}\left(y^{n}\right)\geq\e^{2}$.

Substitute \eqref{eq:-122} into \eqref{eq:-22}, then we get 
\begin{align}
 & D_{1+s}(P_{Y^{n}\mathcal{C}_{n}}\|Q_{Y}^{n}\times P_{\mathcal{C}_{n}})\nonumber \\
 & \geq\sum_{y^{n}}Q\left(y^{n}\right)\biggl(\frac{1}{M}\sum_{T\in\mathcal{P}_{1}}\ell^{2}\left(T\right)p_{T}\left(y^{n}\right)\nonumber \\
 & \qquad+\left(2\alpha-1-\frac{1-\alpha}{s}\right)\bbE_{\mathcal{C}_{n}}L_{2}\left(y^{n}\right)\nonumber \\
 & \qquad+\bbE_{\mathcal{C}_{n}}\frac{1-\alpha}{s}L_{2}^{1+s}\left(y^{n}\right)-\frac{1}{M}\biggr).
\end{align}

Choose $\alpha=\frac{1+s}{1+2s}$, then the second term above vanishes.
Hence we have 
\begin{align}
 & D_{1+s}(P_{Y^{n}\mathcal{C}_{n}}\|Q_{Y}^{n}\times P_{\mathcal{C}_{n}})+\frac{1}{M}\nonumber \\
= & \sum_{y^{n}}Q\left(y^{n}\right)\biggl(\frac{1}{M}\sum_{T\in\mathcal{P}_{1}}\ell^{2}\left(T\right)p_{T}\left(y^{n}\right)\nonumber \\
 & \qquad+\bbE_{\mathcal{C}_{n}}\frac{1-\alpha}{s}L_{2}^{1+s}\left(y^{n}\right)\biggr)\\
\doteq & \sum_{y^{n}}Q\left(y^{n}\right)\biggl(\frac{1}{M}\sum_{T\in\mathcal{P}_{1}}\ell^{2}\left(T\right)p_{T}\left(y^{n}\right)\nonumber \\
 & \qquad+\bbE_{\mathcal{C}_{n}}L_{2}^{1+s}\left(y^{n}\right)\biggr)\\
\geq & \sum_{y^{n}}Q\left(y^{n}\right)\biggl(\frac{1}{M}\sum_{T\in\mathcal{P}_{1}}\ell^{2}\left(T\right)p_{T}\left(y^{n}\right)\nonumber \\
 & \qquad+\bbE_{\mathcal{C}_{n}}\sum_{T\in\mathcal{P}_{2}}\left(\frac{\ell\left(T\right)}{M}\right)^{1+s}N_{T}\left(y^{n}\right)\biggr)\label{eq:-56}\\
= & \sum_{y^{n}}Q\left(y^{n}\right)\biggl(\frac{1}{M}\sum_{T\in\mathcal{P}_{1}}\ell^{2}\left(T\right)p_{T}\left(y^{n}\right)\nonumber \\
 & \qquad+\sum_{T\in\mathcal{P}_{2}}\left(\frac{\ell\left(T\right)}{M}\right)^{s}\ell\left(T\right)p_{T}\left(y^{n}\right)\biggr)\\
\geq & \sum_{y^{n}}Q\left(y^{n}\right)\biggl(\sum_{T\in\cP^{(n)}(\cX\times\cY)}\ell\left(T\right)p_{T}\left(y^{n}\right)\nonumber \\
 & \qquad\times\min\left\{ \frac{\ell\left(T\right)}{M},\left(\frac{\ell\left(T\right)}{M}\right)^{s}\right\} \biggr).
\end{align}
where \eqref{eq:-56} follows from that 
\begin{align}
L_{2}^{1+s}\left(y^{n}\right) & =\left(\sum_{T\in\mathcal{P}_{2}}\frac{\ell\left(T\right)N_{T}\left(y^{n}\right)}{M}\right)^{1+s}\\
 & =\sum_{T\in\mathcal{P}_{2}}\frac{\ell\left(T\right)N_{T}\left(y^{n}\right)}{M}\left(\sum_{T'\in\mathcal{P}_{2}}\frac{\ell\left(T'\right)N_{T'}\left(y^{n}\right)}{M}\right)^{s}\\
 & \geq\sum_{T\in\mathcal{P}_{2}}\left(\frac{\ell\left(T\right)N_{T}\left(y^{n}\right)}{M}\right)^{1+s}\\
 & \geq\sum_{T\in\mathcal{P}_{2}}\left(\frac{\ell\left(T\right)}{M}\right)^{1+s}N_{T}\left(y^{n}\right).
\end{align}

Following steps similar to (111)-(121) of \cite{Parizi}, we can get
\begin{align}
 & D_{1+s}(P_{Y^{n}\mathcal{C}_{n}}\|Q_{Y}^{n}\times P_{\mathcal{C}_{n}})\nonumber \\
 & \dotge\sum_{T\in\cP^{(n)}(\cX\times\cY)}\e^{-nD\left(T\|T_{X}\times P_{Y|X}\right)}P_{X^{n}}\left(\mathcal{T}_{T_{X}}\right)\nonumber \\
 & \qquad\times\min\left\{ \frac{\ell\left(T\right)}{M},\left(\frac{\ell\left(T\right)}{M}\right)^{s}\right\} .\label{eq:-67}
\end{align}

Note that \eqref{eq:-67} holds for all random codes such that 
\begin{equation}
\bbE[P\left(y^{n}\right)]=Q\left(y^{n}\right),\qquad\forall y^{n}\in\cY^{n}.
\end{equation}
Moreover, for the ensemble of i.i.d. random codes, we have 
\begin{equation}
P_{X^{n}}\left(\mathcal{T}_{T_{X}}\right)\doteq\e^{-nD\left(T_{X}\|P_{X}\right)},
\end{equation}
and 
\begin{align}
 & \min\left\{ \frac{\ell\left(T\right)}{M},\left(\frac{\ell\left(T\right)}{M}\right)^{s}\right\} \nonumber \\
 & \doteq\e^{-n\max\left\{ R-f\left(T\|P_{XY}\right),s\left(R-f\left(T\|P_{XY}\right)\right)\right\} },
\end{align}
where 
\begin{equation}
f(P\|P'):=\sum_{(x,y)\in\cX\times\cY}P(x,y)\log\frac{P'(x,y)}{P'_{X}(y)P'_{Y}(y)},\label{eq:fdef}
\end{equation}
for any two distributions $P,P'\in\cP(\cX\times\cY)$. Therefore,
\begin{align}
 & D_{1+s}(P_{Y^{n}\mathcal{C}_{n}}\|Q_{Y}^{n}\times P_{\mathcal{C}_{n}})\nonumber \\
 & \dotge\exp\Bigl\{-n\min_{T}\bigl\{ D\left(T\|T_{X}\times P_{Y|X}\right)+D\left(T_{X}\|P_{X}\right)\nonumber \\
 & \qquad+\max\left\{ R-f\left(T\|P_{XY}\right),s\left(R-f\left(T\|P_{XY}\right)\right)\right\} \bigr\}\Bigr\}\\
 & =\e^{-n\min_{T}\left\{ D\left(T\|P_{XY}\right)+\max\left\{ R-f\left(T\|P_{XY}\right),s\left(R-f\left(T\|P_{XY}\right)\right)\right\} \right\} }.
\end{align}
Furthermore, we can get 
\begin{align}
 & \min_{T\in\cP^{(n)}(\cX\times\cY)}\Bigl\{ D\left(T\|P_{XY}\right)\nonumber \\
 & \qquad+\max\left\{ R-f\left(T\|P_{XY}\right),s\left(R-f\left(T\|P_{XY}\right)\right)\right\} \Bigr\}\nonumber \\
 & =\min_{\widetilde{P}\in\cP(\cX\times\cY)}\Bigl\{ D\left(\widetilde{P}\|P_{XY}\right)\nonumber \\
 & \qquad+\max\{R-f(\widetilde{P}\|P_{XY}),s(R-f(\widetilde{P}\|P_{XY}))\}\Bigr\}+\delta_{n}\label{eq:-53}\\
 & =\max_{t\in[s,1]}t\left(R-D_{1+t}\left(Q_{XY}\|Q_{X}Q_{Y}\right)\right)+\delta_{n},\label{eq:-129}
\end{align}
where \eqref{eq:-53} follows from Lemma \ref{lem:minequality}, and
\eqref{eq:-129} is obtained by following steps similar to the proof
in Appendix B-D of \cite{Parizi}. Hence we have for i.i.d. codes,
\begin{align}
 & \limsup_{n\to\infty}-\frac{1}{n}\log D_{1+s}(P_{Y^{n}\mathcal{C}_{n}}\|Q_{Y}^{n}P_{\mathcal{C}_{n}})\nonumber \\
 & \leq\max_{t\in[s,1]}t\left(R-D_{1+t}\left(Q_{XY}\|Q_{X}Q_{Y}\right)\right).
\end{align}

\emph{Converse for $s\in(-1,0]$ case:} For this case, we need to
prove for $s\in[0,1)$, 
\begin{align}
 & \limsup_{n\to\infty}-\frac{1}{n}\log D_{1-s}(P_{Y^{n}\mathcal{C}_{n}}\|Q_{Y}^{n}\times P_{\mathcal{C}_{n}})\nonumber \\
 & \leq\max_{t\in[0,1]}t\left(R-D_{1+t}\left(Q_{XY}\|Q_{X}\times Q_{Y}\right)\right).
\end{align}
We also follow steps similar to the proof in \cite{Parizi}, and still
use the notations \eqref{eq:ldef}\textendash \eqref{eq:muval}, but
we need to instead choose 
\begin{align}
\cP_{1} & :=\{T\in\cP^{(n)}(\cX\times\cY):\ell(T)\le\beta M\},\\
\cP_{2} & :=\{T\in\cP^{(n)}(\cX\times\cY):\ell(T)>\beta M\},
\end{align}
for some $\beta>0$. Then we have 
\begin{align}
 & D_{1-s}(P_{Y^{n}\mathcal{C}_{n}}\|Q_{Y}^{n}\times P_{\mathcal{C}_{n}})\nonumber \\
 & \dotge-\frac{1}{s}\bbE_{\mathcal{C}_{n}}\sum_{y^{n}}Q\left(y^{n}\right)\bigl(L_{1}^{1-s}\left(y^{n}\right)-L_{1}\left(y^{n}\right)\nonumber \\
 & \qquad+L_{2}^{1-s}\left(y^{n}\right)-L_{2}\left(y^{n}\right)\bigr)\label{eq:-69}\\
 & \geq\sum_{y^{n}}Q\left(y^{n}\right)\biggl(-\bbE_{\mathcal{C}_{n}}\left(\frac{L_{1}^{1-s}\left(y^{n}\right)-L_{1}\left(y^{n}\right)}{s}\right)\nonumber \\
 & \qquad-\frac{1}{s}\bbE_{\mathcal{C}_{n}}\left(\beta^{-s}L_{2}\left(y^{n}\right)-L_{2}\left(y^{n}\right)\right)\biggr)\label{eq:-70}\\
 & =\sum_{y^{n}}Q\left(y^{n}\right)\left(\bbE_{\mathcal{C}_{n}}\frac{L_{1}-L_{1}^{1-s}}{s}+\frac{1-\beta^{-s}}{s}\bbE_{\mathcal{C}_{n}}L_{2}\right).\label{eq:-25}
\end{align}
where \eqref{eq:-69} is obtained by following steps similar to \eqref{eq:-134}-\eqref{eq:-111},
and \eqref{eq:-70} follows from $L_{2}\left(y^{n}\right)\geq\beta$.

To continue the proof, we need the following lemma. The proof is similar
as that of~\cite[Lemma~7]{Parizi}, and hence omitted here. 
\begin{lem}
\label{lem:alna} Let $U$ be an arbitrary non-negative random variable
with $\bbE[U]=1$. Then, for any $\theta>0$, 
\begin{equation}
c(\theta)\Bigl[\var(U)-\tau_{\theta}(U)\Bigr]\le\bbE\bigg[\frac{U-U^{1-s}}{s}\bigg]\le\var(U)\label{eq:alna}
\end{equation}
where 
\begin{align}
\tau_{\theta}(U) & :=\theta^{2}\mathbb{P}\{U>(\theta+1)\}+2\int_{\theta}^{+\infty}v\mathbb{P}\{U>v+1\}\rd v,\label{eq:taudef}
\end{align}
and 
\begin{equation}
c(\theta):=\frac{1}{\theta^{2}}\left({\frac{\theta+1-\left(\theta+1\right)^{1-s}}{s}-\theta}\right).\label{eq:cthetadef}
\end{equation}
\end{lem}
Using this lemma, we have for all $\theta>0$, 
\begin{align}
 & \bbE\bigg[\frac{L_{1}-L_{1}^{1-s}}{s}\bigg]\nonumber \\
 & =\bbE\bigg[\frac{L_{1}-\bbE\left[L_{1}\right]^{s}L_{1}^{1-s}+\bbE\left[L_{1}\right]^{s}L_{1}^{1-s}-L_{1}^{1-s}}{s}\bigg]\\
 & =\bbE\left[L_{1}\right]\bbE\bigg[\frac{1}{s}\left(\frac{L_{1}}{\mathbb{E}L_{1}}-\left(\frac{L_{1}}{\mathbb{E}L_{1}}\right)^{1-s}\right)\bigg]\nonumber \\
 & \qquad+\frac{\left(\bbE\left[L_{1}\right]^{s}-1\right)\bbE\left[L_{1}^{1-s}\right]}{s}\label{eq:-27}\\
 & \geq\bbE\left[L_{1}\right]c(\theta)\Bigl[\var(U_{1})-\tau_{\theta}(U_{1})\Bigr]-\bbE[L_{2}],\label{eq:-24}
\end{align}
where $U_{1}:=\frac{L_{1}}{\mathbb{E}L_{1}}$ and \eqref{eq:-24}
follows from the lemma above and the following inequalities. 
\begin{align}
 & \frac{\left(\bbE\left[L_{1}\right]^{s}-1\right)\bbE\left[L_{1}^{1-s}\right]}{s}\nonumber \\
 & \geq\frac{\left(\bbE\left[L_{1}\right]^{s}-1\right)\bbE\left[L_{1}\right]^{1-s}}{s}\label{eq:-71}\\
 & =\frac{\left(\bbE\left[L_{1}\right]-\bbE\left[L_{1}\right]^{1-s}\right)}{s}\\
 & \geq\bbE\left[L_{1}\right]-1\label{eq:-23}\\
 & =-\bbE[L_{2}],\nonumber 
\end{align}
where \eqref{eq:-71} follows from the fact that $x\mapsto x^{1-s}$
is a concave function, and $\mathbb{E}L_{1}\leq1$, and \eqref{eq:-23}
follows since $\frac{1}{s}\left(x-x^{1-s}\right)\geq x-1$ for $s\in[0,1)$
and $x\in[0,1]$.

Using \eqref{eq:-25} and \eqref{eq:-24} we obtain that $\forall\theta>0\colon$
\begin{align}
 & D_{1-s}(P_{Y^{n}\mathcal{C}_{n}}\|Q_{Y}^{n}\times P_{\mathcal{C}_{n}})\nonumber \\
 & \ge\bbE\left[L_{1}\right]c(\theta)\Bigl[\var(U_{1})-\tau_{\theta}(U_{1})\Bigr]+\frac{1-\beta^{-s}-s}{s}\bbE[L_{2}].
\end{align}
Furthermore, choose $\beta>\left(\frac{1}{1-s}\right)^{s}$, then
$\frac{1-\beta^{-s}-s}{s}>0$. Hence 
\begin{align}
 & D_{1-s}(P_{Y^{n}\mathcal{C}_{n}}\|Q_{Y}^{n}\times P_{\mathcal{C}_{n}})\nonumber \\
 & \dotge\bbE\left[L_{1}\right]c(\theta)\Bigl[\var(U_{1})-\tau_{\theta}(U_{1})\Bigr]+\bbE[L_{2}].
\end{align}
%Also as shown in \cite{Parizi}, $\tau_{\theta}(U_{1})$ is upper-bounded
%as
%\begin{align}
%\tau_{\theta}(U_{1}) & \dotle\frac{\nu\abs{\cP{}_{1}}^{4}}{\theta^{2}},\label{eq:-26}
%\end{align}
%which implies $\tau_{\theta}(U_{1})\le d(n)\abs{\cP{}_{1}}^{4}\nu/\theta^{2}$
%for some sub-exponentially increasing sequence $d(n)$. Hence by
%choosing
%\begin{equation}
%\theta_{n}:=2\sqrt{d(n)}\abs{\cP{}_{1}}^{2},\label{eq:theta}
%\end{equation}
%we get
%\begin{equation}
%\tau_{\theta_{n}}(U_{1})\le\frac{1}{4}\nu.\label{eq:taus}
%\end{equation}
%Substitute \eqref{eq:nudef} and \eqref{eq:taus} into \eqref{eq:lower:split}
%then we get
%\begin{align}
%D_{1-s}(P_{Y^{n}\mathcal{C}_{n}}\|Q_{Y}^{n}\times P_{\mathcal{C}_{n}}) & \ge\bbE\left[L_{1}\right]c(\theta)\Bigl[\var(U_{1})-\tau_{\theta}(U_{1})\Bigr]+\bbE[L_{2}]\\
% & \ge\bbE\left[L_{1}\right]c(\theta_{n})\Bigl[\frac{\nu}{\E^{2}[L_{1}]}-\frac{1}{M}-\frac{1}{4}\nu\Bigr]+\bbE[L_{2}]\\
% & \ge c(\theta_{n})\Bigl[\frac{3}{4}\nu-\frac{1}{M}\Bigr]+\bbE[L_{2}],\label{eq:lower:split2}
%\end{align}
%where \eqref{eq:lower:split2} follows since $\bbE[L_{1}]\le1$. Moreover,
%for $\theta>0$, $c(\theta)\le c(0)=\frac{1}{2}<1$, hence we can
%further lower bound \eqref{eq:lower:split2} as
%\begin{align}
%D_{1-s}(P_{Y^{n}\mathcal{C}_{n}}\|Q_{Y}^{n}\times P_{\mathcal{C}_{n}}) & \ge\frac{3}{4}c(\theta_{n})\nu-\frac{1}{M}+\bbE[L_{2}].\label{eq:lower:split3}
%\end{align}
%Moreover, $c(\theta_{n})$ is also a sub-exponential sequence since
%$d(n)$ is sub-exponential. Therefore, we get

Then we can get %\begin{equation}
%\dotge\nu+\mu.\label{eq:lower:sum}
%\end{equation}
%i.e.,
%\begin{align}
% & D_{1-s}(P_{Y^{n}\mathcal{C}_{n}}\|Q_{Y}^{n}\times P_{\mathcal{C}_{n}})+\frac{1}{M}\nonumber \\
% & \dotge\sum_{y^{n}}Q\left(y^{n}\right)\left(\frac{1}{M}\sum_{T\in\mathcal{P}_{1}}\ell^{2}\left(T\right)p_{T}\left(y^{n}\right)+\sum_{T\in\mathcal{P}_{2}}\ell\left(T\right)p_{T}\left(y^{n}\right)\right)\\
% & =\sum_{y^{n}}Q\left(y^{n}\right)\left(\sum_{T\in\cP^{(n)}(\cX\times\cY)}\ell\left(T\right)p_{T}\left(y^{n}\right)\min\left\{ \frac{\ell\left(T\right)}{M},1\right\} \right)\\
% & =\sum_{T\in\cP^{(n)}(\cX\times\cY)}\e^{-nD\left(T\|T_{X}\times P_{Y|X}\right)}P_{X^{n}}\left(\mathcal{T}_{T_{X}}^{n}\right)\min\left\{ \frac{\ell\left(T\right)}{M},1\right\} .\label{eq:-112}
%\end{align}
%On the other hand, for the ensemble of i.i.d.\ random codes, \cite{Parizi} showed
\begin{align}
 & D_{1-s}(P_{Y^{n}\mathcal{C}_{n}}\|Q_{Y}^{n}\times P_{\mathcal{C}_{n}})+\frac{1}{M}\nonumber \\
 & \dotge\nu+\mu\label{eqn:-72}\\
 & \doteq\e^{-n\underset{T\in\cP^{(n)}(\cX\times\cY)}{\min}\left(D\left(T\|T_{X}P_{Y|X}\right)+D\left(T_{X}\|P_{X}\right)+\left[R-f\left(T\|P_{XY}\right)\right]^{+}\right)}\label{eq:-72}\\
 & =\e^{-n\underset{T\in\cP^{(n)}(\cX\times\cY)}{\min}\left(D\left(T\|P_{XY}\right)+\left[R-f\left(T\|P_{XY}\right)\right]^{+}\right)}\\
 & \doteq\e^{-n\underset{\widetilde{P}\in\cP(\cX\times\cY)}{\min}\left(D\left(\widetilde{P}\|P_{XY}\right)+\left[R-f\left(\widetilde{P}\|P_{XY}\right)\right]^{+}\right)}\label{eq:-74}\\
 & =\e^{-n\underset{t\in[0,1]}{\max}t\left(R-D_{1+t}\left(Q_{XY}\|Q_{X}\times Q_{Y}\right)\right)},\label{eq:-73}
\end{align}
where \eqref{eqn:-72} is obtained by following steps similar to (101)-(125)
of \cite{Parizi}, $f\left(\cdot\right)$ is defined in \eqref{eq:fdef},
\eqref{eq:-72} follows from \cite[Eqns.~(122)--(125)]{Parizi}, \eqref{eq:-74}
follows from Lemma \ref{lem:minequality} (or \cite[Appendix~B-A]{Parizi}),
and \eqref{eq:-73} follows from \cite[Appendix~B-D]{Parizi}.

Since the exponent of $\frac{1}{M}$ is $R$, which is larger than
the exponent in \eqref{eq:-73}, the exponent in \eqref{eq:-73} is
the dominant exponent. Hence \eqref{eq:-73} implies the converse
part.

\section{\label{sec:Proof-of-Theorem-exp}Proof of Theorem \ref{thm:exponent}}

The achievability of $\mathsf{E_{iid}}\left(P_{X},P_{Y|X},Q_{Y}\right)$
has been proven in Theorem \ref{thm:exponentforiid}, hence we only
need to prove the achievability of $\mathsf{E_{ts}}\left(P_{X},P_{Y|X},Q_{Y}\right)$.

For the case of $s\in(-1,0]$, the exponent $\sup_{\epsilon\in(0,1]}\min\big\{\frac{\epsilon^{2}P_{\mathsf{min}}}{3},\theta\left(0,\epsilon,P_{X}\right)\big\}$
is obtained from the exponent for $s\in(0,1]$ by letting $s\to0$.
Hence we only need to focus on the case $s\in(0,1]$. %Next we consider the case $s\in(0,1]$.
We use the random code given in the proof of Theorem~\ref{thm:Resolvability}.
For this code, $P_{X^{n}}\left(x^{n}\right)\propto Q_{X}^{n}\left(x^{n}\right)1\left\{ x^{n}\in\mathcal{T}_{\epsilon}^{n}\left(Q_{X}\right)\right\} $
for some $Q_{X}\in\mathcal{P}\left(P_{Y|X},Q_{Y}\right)$.

By Lemma \ref{lem:oneshotach}, we obtain 
\begin{align}
 & \e^{sD_{1+s}(P_{Y^{n}\mathcal{C}_{n}}\|Q_{Y}^{n}\times P_{\mathcal{C}_{n}})}\nonumber \\
 & \leq\e^{sD_{1+s}(P_{X^{n}Y^{n}}\|P_{X^{n}}Q_{Y}^{n})-nsR}+\e^{sD_{1+s}(P_{Y^{n}}\|Q_{Y}^{n})}\\
 & =\e^{sD_{1+s}(P_{Y^{n}}\|Q_{Y}^{n})}\nonumber \\
 & \qquad\times\left(\e^{sD_{1+s}(P_{X^{n}Y^{n}}\|P_{X^{n}}Q_{Y}^{n})-nsR-sD_{1+s}(P_{Y^{n}}\|Q_{Y}^{n})}+1\right).\label{eq:-50}
\end{align}
Take $\log$'s, 
\begin{align}
 & D_{1+s}(P_{Y^{n}\mathcal{C}_{n}}\|Q_{Y}^{n}\times P_{\mathcal{C}_{n}})\nonumber \\
 & =D_{1+s}(P_{Y^{n}}\|Q_{Y}^{n})\nonumber \\
 & \:+\frac{1}{s}\log\left(\e^{sD_{1+s}(P_{X^{n}Y^{n}}\|P_{X^{n}}Q_{Y}^{n})-nsR-sD_{1+s}(P_{Y^{n}}\|Q_{Y}^{n})}+1\right)\\
 & \leq D_{1+s}(P_{Y^{n}}\|Q_{Y}^{n})\nonumber \\
 & \:+\frac{1}{s}\e^{sD_{1+s}(P_{X^{n}Y^{n}}\|P_{X^{n}}Q_{Y}^{n})-nsR-sD_{1+s}(P_{Y^{n}}\|Q_{Y}^{n})}.\label{eq:-54}
\end{align}

On the other hand, 
\begin{align}
D_{1+s}(P_{Y^{n}}\|Q_{Y}^{n}) & \leq D_{1+s}(P_{X^{n}}\|Q_{X}^{n})\\
 & =\log\frac{1}{Q_{X}^{n}\left(\mathcal{T}_{\epsilon}^{n}\right)}\label{eq:-51}\\
 & \leq\frac{1}{Q_{X}^{n}\left(\mathcal{T}_{\epsilon}^{n}\right)}-1\\
 & \doteq Q_{X}^{n}\left((\mathcal{T}_{\epsilon}^{n})^{c}\right),\label{eq:-58}
\end{align}
where $(\mathcal{T}_{\epsilon}^{n})^{c}:=\mathcal{X}^{n}\backslash\mathcal{T}_{\epsilon}^{n}$,
and \eqref{eq:-51} follows from \eqref{eq:-46}-\eqref{eq:-52}.
Now we bound $Q_{X}^{n}\left((\mathcal{T}_{\epsilon}^{n})^{c}\right)$
using the Chernoff bound~\cite{Mitzenmacher} as 
\begin{align}
Q_{X}^{n}\left((\mathcal{T}_{\epsilon}^{n})^{c}\right) & \leq2\left|\mathcal{X}\right|\e^{-\frac{\epsilon^{2}nQ_{\mathsf{min}}}{3}},\label{eq:-59}
\end{align}
where $Q_{\mathsf{min}}:=\min_{x}Q_{X}\left(x\right).$ Substituting
\eqref{eq:-59} into \eqref{eq:-58}, we obtain 
\begin{equation}
D_{1+s}(P_{Y^{n}}\|Q_{Y}^{n})\dotleq2\left|\mathcal{X}\right|\e^{-\frac{\epsilon^{2}nQ_{\mathsf{min}}}{3}}.\label{eq:-15-3}
\end{equation}

By \eqref{eq:-15-3} we can bound the exponent of the second term
of \eqref{eq:-54} as 
\begin{align}
 & sR-\frac{1}{n}sD_{1+s}(P_{X^{n}Y^{n}}\|P_{X^{n}}Q_{Y}^{n})+\frac{1}{n}sD_{1+s}(P_{Y^{n}}\|Q_{Y}^{n})\nonumber \\
 & =sR+\delta_{n}-\frac{1}{n}\log\sum_{x^{n}\in\mathcal{T}_{\epsilon}^{n}}P\left(x^{n}\right)\nonumber \\
 & \qquad\times\e^{n\sum_{x\in\mathcal{X}}T_{x^{n}}\left(x\right)\log\sum_{y}P^{1+s}\left(y|x\right)Q^{-s}\left(y\right)}\\
 & =sR+\delta_{n}-\frac{1}{n}\log\sum_{T_{X}:\forall x:\left|T_{X}\left(x\right)-Q\left(x\right)\right|\leq\epsilon Q\left(x\right)}P_{X^{n}}(\mathcal{T}_{T_{X}})\nonumber \\
 & \qquad\times\e^{n\sum_{x\in\mathcal{X}}T_{X}\left(x\right)\log\sum_{y}P^{1+s}\left(y|x\right)Q^{-s}\left(y\right)}\\
 & \geq sR+\delta_{n}-\max_{T_{X}:\forall x:\left|T_{X}\left(x\right)-Q\left(x\right)\right|\leq\epsilon Q\left(x\right)}\sum_{x}T_{X}\left(x\right)\nonumber \\
 & \qquad\times\log\sum_{y}P^{1+s}\left(y|x\right)Q^{-s}\left(y\right)\label{eq:-16}\\
 & \geq sR+\delta_{n}-\left(1+\epsilon\right)\sum_{x}Q\left(x\right)\log\sum_{y}P^{1+s}\left(y|x\right)Q^{-s}\left(y\right),\label{eq:-4-1}
\end{align}
where $\delta_{n}$ is a term vanishing as $n\to\infty$, and \eqref{eq:-16}
follows since $\sum_{T_{X}:\forall x:\left|T_{X}\left(x\right)-Q\left(x\right)\right|\leq\epsilon Q\left(x\right)}P_{X^{n}}(\mathcal{T}_{T_{X}})\leq1$
and for any $T_{X}$ such that for all $x$, $\left|T_{X}\left(x\right)-Q\left(x\right)\right|\leq\epsilon Q\left(x\right)$,
it holds that 
\begin{align}
 & \sum_{x\in\mathcal{X}}T_{X}\left(x\right)\log\left(\sum_{y}P^{1+s}\left(y|x\right)Q^{-s}\left(y\right)\right)\nonumber \\
 & \leq\max_{T_{X}:\forall x:\left|T_{X}\left(x\right)-Q\left(x\right)\right|\leq\epsilon Q\left(x\right)}\sum_{x\in\mathcal{X}}T_{X}\left(x\right)\nonumber \\
 & \qquad\times\log\sum_{y}P^{1+s}\left(y|x\right)Q^{-s}\left(y\right).
\end{align}

Substituting \eqref{eq:-15-3} and \eqref{eq:-4-1} into \eqref{eq:-54},
we have 
\begin{align}
 & \liminf_{n\to\infty}-\frac{1}{n}\log D_{1+s}(P_{Y^{n}\mathcal{C}_{n}}\|Q_{Y}^{n}\times P_{\mathcal{C}_{n}})\nonumber \\
 & \geq\min\Bigl\{\frac{\epsilon^{2}Q_{\mathsf{min}}}{3},\nonumber \\
 & \qquad sR-\left(1+\epsilon\right)\sum_{x}Q\left(x\right)\log\sum_{y}P^{1+s}\left(y|x\right)Q^{-s}\left(y\right)\Bigr\}.\label{eq:-6-1}
\end{align}
Note that the second term of minimization is {\em not} $\theta\left(s,\epsilon,P_{X}\right)$.
To obtain the desired result, by using the fact that the R\'enyi divergence
is non-decreasing in its parameter, we get 
\begin{align}
 & \liminf_{n\to\infty}-\frac{1}{n}\log D_{1+s}(P_{Y^{n}\mathcal{C}_{n}}\|Q_{Y}^{n}\times P_{\mathcal{C}_{n}})\nonumber \\
 & \geq\sup_{t\in[s,1]}\liminf_{n\to\infty}-\frac{1}{n}\log D_{1+t}(P_{Y^{n}\mathcal{C}_{n}}\|Q_{Y}^{n}\times P_{\mathcal{C}_{n}})\\
 & \geq\sup_{t\in[s,1]}\min\Bigl\{\frac{\epsilon^{2}Q_{\mathsf{min}}}{3},\nonumber \\
 & \qquad tR-\left(1+\epsilon\right)\sum_{x}Q\left(x\right)\log\sum_{y}P^{1+t}\left(y|x\right)Q^{-t}\left(y\right)\Bigr\}\\
 & =\min\left\{ \frac{\epsilon^{2}Q_{\mathsf{min}}}{3},\theta\left(s,\epsilon,P_{X}\right)\right\} .\label{eq:-55}
\end{align}
Since $\epsilon\in(0,1]$ is arbitrary, we can optimize \eqref{eq:-55}
over all possible $\epsilon$. This concludes the proof.

\section{\label{sec:Proof-of-Theorem-DetEnc}Proof of Theorem \ref{thm:DetEnc}}

\emph{Achievability:} We use random coding to prove the achievability
part. Generate $\mathcal{C}_{n}=\left\{ X^{n}\left(m_{0},m_{1}\right)\right\} _{\left(m_{0},m_{1}\right)\in\mathcal{M}_{0}\times\mathcal{M}_{1}}$
with $X^{n}\left(m_{0},m_{1}\right)\sim P_{X^{n}}$ and set the encoder
as $f_{\mathcal{C}_{n}}\left(m_{0},m_{1}\right)=X^{n}\left(m_{0},m_{1}\right)$.
This constitutes our random code. Moreover, we set $P_{X^{n}}\left(x^{n}\right)\propto P_{X}^{n}\left(x^{n}\right)1\left\{ x^{n}\in\mathcal{T}_{\epsilon}^{n}\left(P_{X}\right)\right\} $
for some $P_{X}\in\mathcal{P}\left(P_{Z|X},Q_{Z}\right)$. At the
legitimate user side, the standard joint-typicality decoder is adopted.

For this random code, by the standard proof \cite[Section 3.1.2]{Gamal}\footnote{Although here $P_{X^{n}}$ is not an i.i.d. distribution, it satisfies
$P_{X^{n}}\left(x^{n}\right)=\e^{-n\left(H\left(P_{X}\right)+\delta_{n}\right)}$.
Hence the joint typicality lemma \cite{Gamal} still holds, which
further guarantees that the standard proof for channel coding works
for our case.}, it is easy to verify that 
\begin{align}
 & \mathbb{P}\left(\left(M_{0},M_{1}\right)\neq(\widehat{M}_{0},\widehat{M}_{1})\right)\nonumber \\
 & =\bbE_{\mathcal{C}_{n}}\left[\mathbb{P}\left(\left(M_{0},M_{1}\right)\neq(\widehat{M}_{0},\widehat{M}_{1})|\mathcal{C}_{n}\right)\right]\\
 & \rightarrow0\label{eq:-30-1-1}
\end{align}
if $R_{0}+R_{1}\leq I\left(X;Y\right)$. Therefore, the error constraint
is satisfied.

By the codebook generation procedure, $\mathcal{C}_{n}$ is independent
of $M_{1}$, and the subcodebooks $\mathcal{C}_{n}\left(m_{1}\right):=\left\{ X^{n}\left(m_{0},m_{1}\right)\right\} _{m_{0}\in\mathcal{M}_{0}}$
for different $m_{1}$ have the same distribution (which implies $P_{\mathcal{C}_{n}(m_{1})}(c_{n})=P_{\mathcal{C}_{n}(M_{1})}(c_{n})$
for any $m_{1}$). Hence $\mathcal{C}_{n}\left(M_{1}\right)$ is independent
of $M_{1}$.\footnote{Indeed, we have $P_{M_{1},\mathcal{C}_{n}(M_{1})}(m_{1},c_{n})=P_{M_{1}}(m_{1})P_{\mathcal{C}_{n}(m_{1})|M_{1}}(c_{n}|m_{1})=P_{M_{1}}(m_{1})P_{\mathcal{C}_{n}(m_{1})}(c_{n})=P_{M_{1}}(m_{1})P_{\mathcal{C}_{n}(M_{1})}(c_{n})$.}
Furthermore, from our result for the channel resolvability problem
(Theorem~\ref{thm:Resolvability}), given $M_{1}=m_{1}$ and $P_{X}\in\mathcal{P}\left(P_{Z|X},Q_{Z}\right)$,
for $s\in[-1,1]$, the random code constructed above satisfies 
\begin{equation}
D_{1+s}(P_{Z^{n}\mathcal{C}_{n}\left(m_{1}\right)|M_{1}=m_{1}}\|Q_{Z}^{n}\times P_{\mathcal{C}_{n}\left(m_{1}\right)})\rightarrow0
\end{equation}
if $R_{0}>\widetilde{R}_{1+s}\left(P_{X},P_{Z|X},Q_{Z}\right)$. Therefore,
\begin{align}
 & \e^{sD_{1+s}(P_{Z^{n}M_{1}\mathcal{C}_{n}}\|Q_{Z}^{n}\times P_{M_{1}\mathcal{C}_{n}})}\nonumber \\
 & =\bbE_{M_{1}\mathcal{C}_{n}}\left[\e^{sD_{1+s}(P_{Z^{n}|M_{1}\mathcal{C}_{n}}\|Q_{Z}^{n})}\right]\\
 & =\bbE_{M_{1},\mathcal{C}_{n}\left(M_{1}\right)}\left[\e^{sD_{1+s}(P_{Z^{n}|M_{1},\mathcal{C}_{n}\left(M_{1}\right)}\|Q_{Z}^{n})}\right]\label{eq:-130}\\
 & =\bbE_{M_{1}}\left[\e^{sD_{1+s}(P_{Z^{n}\mathcal{C}_{n}\left(M_{1}\right)|M_{1}}\|Q_{Z}^{n}\times P_{\mathcal{C}_{n}\left(M_{1}\right)})}\right]\label{eq:-131}\\
 & \rightarrow1,
\end{align}
where \eqref{eq:-130} follows since $\mathcal{C}_{n}\rightarrow\left(M_{1},\mathcal{C}_{n}\left(M_{1}\right)\right)\to Z^{n}$
forms a Markov chain (this results from the encoding process\textemdash the
transmitted codeword is chosen from $\mathcal{C}_{n}\left(M_{1}\right)$),
and \eqref{eq:-131} follows since $M_{1}$ and $\mathcal{C}_{n}\left(M_{1}\right)$
are independent. On the other hand, 
\begin{align}
 & \e^{sD_{1+s}(P_{Z^{n},M_{1},\mathcal{C}_{n}}\|Q_{Z}^{n}\times P_{M_{1},\mathcal{C}_{n}})}\nonumber \\
 & =\bbE_{\mathcal{C}_{n}}\left[\e^{sD_{1+s}(P_{Z^{n}M_{1}}\|Q_{Z}^{n}\times P_{M_{1}})}\right].
\end{align}
Hence 
\begin{equation}
\bbE_{\mathcal{C}_{n}}\left[\e^{sD_{1+s}(P_{Z^{n}M_{1}}\|Q_{Z}^{n}\times P_{M_{1}})}\right]\rightarrow1.\label{eq:-29}
\end{equation}

Applying the selection lemma \cite[Lem. 2.2]{Bloch2011} to \eqref{eq:-30-1-1}
and \eqref{eq:-29} we deduce that there exists one sequence of realizations
$\left\{ c_{n}\right\} _{n}$ such that given $\mathcal{C}_{n}=c_{n}$,
\begin{equation}
\lim_{n\rightarrow\infty}\mathbb{P}\left(\left(M_{0},M_{1}\right)\neq(\widehat{M}_{0},\widehat{M}_{1})\right)=0,\label{eq:-30-1-2}
\end{equation}
and 
\begin{equation}
{\displaystyle \lim_{n\rightarrow\infty}D_{1+s}(P_{M_{1}Z^{n}}\|P_{M_{1}}Q_{Z}^{n})=0}.\label{eq:-31-4}
\end{equation}
Hence $f_{\mathcal{C}_{n}=c_{n}}$ is the desired encoder. The proof
of the achievability part for $s\in[-1,1]$ is complete. For $s=\infty$,
the achievability part can be proven similarly. 

\emph{Converse:} By the data processing inequality \cite{Erven},
we have 
\begin{align}
R_{0}+R_{1} & \leq\frac{1}{n}I\left(X^{n};Y^{n}\right)\\
 & \leq I\left(X_{J};Y_{J}\right),\label{eq:-33}
\end{align}
where $J\sim\textrm{Unif}\left[1:n\right]$ denotes a time index variable,
independent of $X^{n},Y^{n}$. It is easy to verify that the distribution
of $\left(X_{J},Y_{J}\right)$ induced by an $n$-length code satisfies
\begin{align}
P_{X_{J}Y_{J}}^{(n)}\left(x,y\right) & =\bbE_{X^{n}Y^{n}}\left[T_{X^{n}Y^{n}}\left(x,y\right)\right]\\
 & =\bbE_{X^{n}}\left[T_{X^{n}}\left(x\right)\right]P\left(y|x\right).
\end{align}
Now, Pinsker's inequality for R\'enyi parameter $1+s\in(0,1]$~\cite{Erven}
implies that, 
\begin{equation}
\left|P-Q\right|\leq\sqrt{\frac{2}{1+s}D_{1+s}\left(P\|Q\right)},\label{eq:-75}
\end{equation}
and for R\'enyi parameter $1+s\in(1,\infty]$, we also have 
\begin{equation}
\left|P-Q\right|\leq\sqrt{2D\left(P\|Q\right)}\leq\sqrt{2D_{1+s}\left(P\|Q\right)}.\label{eq:-76}
\end{equation}
Applying \eqref{eq:-75} and \eqref{eq:-76} to $P_{M_{1}Z^{n}}$
and $P_{M_{1}}Q_{Z}^{n}$, we obtain $\left|P_{M_{1}Z^{n}}-P_{M_{1}}Q_{Z}^{n}\right|\rightarrow0$
and hence $\left|P_{Z^{n}}-Q_{Z}^{n}\right|\rightarrow0$ (by the
data processing inequality of TV distance $\left|P_{X}-Q_{X}\right|\leq\left|P_{XY}-Q_{XY}\right|$),
regardless of $1+s\in(0,1]$ or $(1,2]$.

Observe that $\left|T_{Z^{n}}-Q_{Z}\right|$ is a function of $Z^{n}$
and upper-bounded by 1, and $\Ebb_{Q_{Z}^{n}}\left|T_{Z^{n}}-Q_{Z}\right|\rightarrow0$
as $n\to\infty$. Hence by the property 
\begin{equation}
\sup_{f:\mathcal{X}\rightarrow[0,1]}\big|\Ebb_{P}f(X)-\Ebb_{Q}f(X)\big|=\left|P-Q\right|,\label{eq:tvcontinuous-1}
\end{equation}
we have 
\begin{equation}
{\displaystyle \lim_{n\rightarrow\infty}\bbE_{P_{Z^{n}}}\left|T_{Z^{n}}-Q_{Z}\right|=\lim_{n\rightarrow\infty}\Ebb_{Q_{Z}^{n}}\left|T_{Z^{n}}-Q_{Z}\right|=0},\label{eq:-31-4-1-2-1-1}
\end{equation}
which further implies 
\begin{equation}
{\displaystyle \lim_{n\rightarrow\infty}\left|\bbE_{P_{Z^{n}}}\left[T_{Z^{n}}\right]-Q_{Z}\right|=0},\label{eq:-31-4-1-2-1-1-2}
\end{equation}
i.e., 
\begin{equation}
{\displaystyle \lim_{n\rightarrow\infty}\left|P_{Z_{J}}^{(n)}-Q_{Z}\right|=0},\label{eq:-31-4-1-2-1-1-2-1}
\end{equation}
or equivalently 
\begin{equation}
{\displaystyle \lim_{n\rightarrow\infty}\left|P_{X_{J}}^{(n)}\circ P_{Z|X}-Q_{Z}\right|=0}.\label{eq:-32}
\end{equation}

Since $\mathcal{P}(\mathcal{X})$ is compact, there must exist some
increasing sequence $\left\{ n_{k}\right\} _{k=1}^{\infty}$ such
that $P_{X_{J}}^{(n_{k})}$ converges to some distribution $\widetilde{P}_{X}$.
From \eqref{eq:-32}, $\widetilde{P}_{X}\in\mathcal{P}\left(P_{Z|X},Q_{Z}\right)$
holds.

We first consider the case of $s\in(0,\infty]$. By the one-shot bound
in Lemma \ref{lem:oneshotcon}, 
\begin{equation}
{\displaystyle \lim_{n\rightarrow\infty}D_{1+s}(P_{M_{1}Z^{n}}\|P_{M_{1}}Q_{Z}^{n})=0}\label{eq:-31-4-1}
\end{equation}
implies 
\begin{equation}
R_{0}\geq\limsup_{n\rightarrow\infty}\frac{1}{n}D_{1+s}\left(P_{X^{n}Z^{n}}\|P_{X^{n}}\times Q_{Z}^{n}\right).
\end{equation}
On the other hand, 
\begin{align}
 & \frac{1}{n}D_{1+s}\left(P_{X^{n}Z^{n}}\|P_{X^{n}}\times Q_{Z}^{n}\right)\nonumber \\
 & =\frac{1}{ns}\log\sum_{T_{X}}P_{X^{n}}\left(\mathcal{T}_{T_{X}}\right)\e^{n\sum_{x\in\mathcal{X}}T_{X}\left(x\right)\log\sum_{z}P^{1+s}\left(z|x\right)Q^{-s}\left(z\right)}\\
 & \geq\frac{1}{s}\sum_{T_{X}}P_{X^{n}}\left(\mathcal{T}_{T_{X}}\right)\sum_{x\in\mathcal{X}}T_{X}\left(x\right)\log\sum_{z}P^{1+s}\left(z|x\right)Q^{-s}\left(z\right)\\
 & =\frac{1}{s}\sum_{x\in\mathcal{X}}\bbE_{X^{n}}\left[T_{X^{n}}\left(x\right)\right]\log\sum_{z}P^{1+s}\left(z|x\right)Q^{-s}\left(z\right)\\
 & =\frac{1}{s}\sum_{x\in\mathcal{X}}P_{X_{J}}^{(n)}\left(x\right)\log\sum_{z}P^{1+s}\left(z|x\right)Q^{-s}\left(z\right).\label{eq:-34}
\end{align}
Hence 
\begin{equation}
R_{0}\geq\limsup_{n\rightarrow\infty}\frac{1}{s}\sum_{x\in\mathcal{X}}P_{X_{J}}^{(n)}\left(x\right)\log\sum_{z}P^{1+s}\left(z|x\right)Q^{-s}\left(z\right).\label{eq:-132}
\end{equation}

Consider the blocklengths $\left\{ n_{k}\right\} _{k=1}^{\infty}$.
Since $P_{X_{J}}^{(n_{k})}$ converges to $\widetilde{P}_{X}$, \eqref{eq:-33}
and \eqref{eq:-132} respectively imply 
\begin{align}
R_{0}+R_{1} & \leq I_{\widetilde{P}}\left(X;Y\right),
\end{align}
and 
\begin{align}
R_{0} & \geq\frac{1}{s}\sum_{x\in\mathcal{X}}\widetilde{P}_{X}\left(x\right)\log\sum_{z}P^{1+s}\left(z|x\right)Q^{-s}\left(z\right).\label{eq:-9-4-1}
\end{align}
Therefore, $\widetilde{P}_{X}$ is the desired distribution $P_{X}$
in \eqref{eq:-40}. The proof for $s\in(0,\infty]$ case is complete.

Next we consider the case of $s\in(-1,0]$. This case can be proved
by following steps similar to the proof of the traditional channel
resolvability problem (or distributed channel synthesis problem) \cite{Cuff}.
Observe 
\begin{align}
R_{0} & \geq\frac{1}{n}I\left(M_{0};Z^{n}|M_{1}\right)\\
 & =\frac{1}{n}I\left(X^{n};Z^{n}|M_{1}\right)\\
 & =\frac{1}{n}I\left(X^{n};Z^{n}\right)-\frac{1}{n}I\left(M_{1};Z^{n}\right)\\
 & =\frac{1}{n}I\left(X^{n};Z^{n}\right)-\delta_{n}\label{eq:-77}\\
 & =\frac{1}{n}H\left(Z^{n}\right)-\frac{1}{n}H\left(Z^{n}|X^{n}\right)-\delta_{n}\\
 & =H_{Q}\left(Z\right)-H\left(Z_{J}|X_{J}\right)-\delta_{n}+\delta_{n}',\label{eq:-78}
\end{align}
where \eqref{eq:-77} and \eqref{eq:-78} follow from the facts$\left|P_{M_{1}Z^{n}}-P_{M_{1}}Q_{Z}^{n}\right|\rightarrow0$
and $\left|P_{Z^{n}}-Q_{Z}^{n}\right|\rightarrow0$, respectively.

Furthermore, since there exist some sequence $\left\{ n_{k}\right\} _{k=1}^{\infty}$
such that $P_{X_{J}}^{(n_{k})}$ converges to some distribution $\widetilde{P}_{X}$
such that $\widetilde{P}_{X}\in\mathcal{P}\left(P_{Z|X},Q_{Z}\right)$,
we have 
\begin{align}
R_{0} & \geq H_{Q}\left(Z\right)-H\left(Z_{J}|X_{J}\right)\\
 & =I_{\widetilde{P}}\left(X;Z\right).\label{eq:-33-2-1}
\end{align}
On the other hand, 
\begin{align}
R_{0}+R_{1} & \leq I_{\widetilde{P}}\left(X;Y\right).\label{eq:-80}
\end{align}
Combining \eqref{eq:-33-2-1} and \eqref{eq:-80} gives the desired
result.

\section{\label{sec:Proof-of-Theorem-StocEnc}Proof of Theorem \ref{thm:StochasticEncoder}}

We first prove \eqref{eq:-39}.

\emph{Achievability:} We only consider the case $s\in(0,1]$. The
achievability result for $s\in(-1,0]$ can be obtained from the result
for the case $s\in(0,1]$ by letting $s\downarrow0$.

We use a similar random code as the one given in Lemma \ref{lem:oneshotach}.
That is, we set $\mathcal{C}_{n}=\left\{ W^{n}\left(m\right)\right\} _{m\in\calM}$
with $W^{n}\left(m\right)\sim P_{W^{n}},m\in\calM$, and set the encoder
as $f_{\mathcal{C}_{n}}\left(m\right)=W^{n}\left(m\right)$. We insert
a random mapping (virtual channel) between the encoder $f_{\mathcal{C}_{n}}\left(m\right)$
and the channel, which is denoted as $P_{X^{n}|W^{n}}$. For this
cascaded code, we set the distributions $P_{W^{n}}\left(w^{n}\right)\propto\widetilde{P}_{W}^{n}\left(w^{n}\right)1\left\{ w^{n}\in\mathcal{T}_{\epsilon'}^{n}\right\} $
and $P_{X^{n}|W^{n}}\left(x^{n}|w^{n}\right)\propto\widetilde{P}_{X|W}^{n}\left(x^{n}|w^{n}\right)1\left\{ \left(w^{n},x^{n}\right)\in\mathcal{T}_{\epsilon}^{n}\right\} $
for some $\widetilde{P}_{WX}$ such that $\widetilde{P}_{X}\in\mathcal{P}\left(P_{Z|X},Q_{Z}\right)$,
where $\epsilon'<\epsilon$, and $\mathcal{T}_{\epsilon'}^{n},\mathcal{T}_{\epsilon}^{n}$,
and $\mathcal{T}_{\epsilon}^{n}\left(w^{n}\right)$ respectively denote
the typical set respect to $\widetilde{P}_{W}$, as well as the jointly
typical set and conditional typical set respect to $\widetilde{P}_{WX}$.

Then by the method of types, we obtain \eqref{eq:-103}-\eqref{eq:-3},
where $[(V_{X|WZ}\circ V_{Z|W})T_{W}](x,w):=\sum_{z}V_{X|WZ}(x|w,z)V_{Z|W}(z|x)T_{W}(w)$,
\eqref{eq:-127} follows since, by the law of large numbers, ${\widetilde{P}_{W}^{n}\left(\mathcal{T}_{\epsilon'}^{n}\right)}\rightarrow1$
and ${\widetilde{P}_{X|W}^{n}\left(\mathcal{T}_{\epsilon}^{n}\left(w^{n}\right)|w^{n}\right)}\rightarrow1$
uniformly for all $w^{n}\in\mathcal{T}_{\epsilon'}^{n}\left(\widetilde{P}_{W}\right)$
(this can be shown by following steps similar to the proof of conditional
typicality lemma in \cite{Gamal}, and hence the proof is omitted
here), \eqref{eqn:delta_n} follows from the fact that the number
of $n$-types $T_{Y}V_{X|Y}$ is polynomial in $n$, and in \eqref{eq:-3}
the arguments of maximization are replaced by $T_{WX},V_{Z|WX}$ (this
is feasible since both $(T_{W},V_{Z|W},V_{X|WZ})$ in \eqref{eq:-44}
and $(T_{WX},V_{Z|WX})$ in \eqref{eq:-3} run through all the types
of sequences in $\mathcal{X}^{n}\times\mathcal{Y}^{n}\times\mathcal{Z}^{n}$).
\begin{figure*}[!t]
\setcounter{mytempeqncnt}{\value{equation}} \setcounter{equation}{448}
\begin{align}
 & \frac{1}{n}D_{1+s}\left(P_{W^{n}Z^{n}}\|P_{W^{n}}\times Q_{Z}^{n}\right)\nonumber \\
 & =\frac{1}{ns}\log\sum_{w^{n},z^{n}}P\left(w^{n}\right)P^{1+s}\left(z^{n}|w^{n}\right)Q^{-s}\left(z^{n}\right)\label{eq:-103}\\
 & =\frac{1}{ns}\log\sum_{w^{n},z^{n}}P\left(w^{n}\right)\Bigl(\sum_{x^{n}}P\left(x^{n}|w^{n}\right)P\left(z^{n}|x^{n}\right)\Bigr)^{1+s}Q^{-s}\left(z^{n}\right)\\
 & =\frac{1}{ns}\log\sum_{T_{W}}\sum_{w^{n}\in\mathcal{T}_{T_{W}}}\sum_{V_{Z|W}}\sum_{z^{n}\in\mathcal{T}_{V_{Z|W}}\left(w^{n}\right)}\frac{\widetilde{P}\left(w^{n}\right)1\left\{ w^{n}\in\mathcal{T}_{\epsilon'}^{n}\right\} }{\widetilde{P}_{W}^{n}\left(\mathcal{T}_{\epsilon'}^{n}\right)}\nonumber \\
 & \qquad\times\Bigl(\sum_{V_{X|WZ}}\sum_{x^{n}\in\mathcal{T}_{V_{X|WZ}}\left(w^{n},z^{n}\right)}\frac{\widetilde{P}\left(x^{n}|w^{n}\right)1\left\{ \left(w^{n},x^{n}\right)\in\mathcal{T}_{\epsilon}^{n}\right\} }{\widetilde{P}_{X|W}^{n}\left(\mathcal{T}_{\epsilon}^{n}\left(w^{n}\right)|w^{n}\right)}\e^{n\sum_{x,z}T\left(x,z\right)\log P\left(z|x\right)}\Bigr)^{1+s}\e^{-ns\sum_{z}T\left(z\right)\log Q\left(z\right)}\\
 & =\delta_{n}+\frac{1}{ns}\log\sum_{T_{W}}\sum_{w^{n}\in\mathcal{T}_{T_{W}}}\sum_{V_{Z|W}}\sum_{z^{n}\in\mathcal{T}_{V_{Z|W}}\left(w^{n}\right)}\widetilde{P}\left(w^{n}\right)1\left\{ w^{n}\in\mathcal{T}_{\epsilon'}^{n}\right\} \nonumber \\
 & \qquad\times\Bigl(\sum_{V_{X|WZ}}\sum_{x^{n}\in\mathcal{T}_{V_{X|WZ}}\left(w^{n},z^{n}\right)}\widetilde{P}\left(x^{n}|w^{n}\right)1\left\{ \left(w^{n},x^{n}\right)\in\mathcal{T}_{\epsilon}^{n}\right\} \e^{n\sum_{x,z}T\left(x,z\right)\log P\left(z|x\right)}\Bigr)^{1+s}\e^{-ns\sum_{z}T\left(z\right)\log Q\left(z\right)}\label{eq:-127}\\
 & \leq\delta_{n}+\delta_{n}'+\frac{1}{ns}\log\max_{T_{W}:\left|T_{W}-\widetilde{P}_{W}\right|\leq\epsilon'}\max_{V_{Z|W}}\sum_{w^{n}\in\mathcal{T}_{T_{W}}}\sum_{z^{n}\in\mathcal{T}_{V_{Z|W}}\left(w^{n}\right)}\widetilde{P}\left(w^{n}\right)\nonumber \\
 & \qquad\times\Bigl(\max_{\substack{V_{X|WZ}:\\
\left|\left(V_{X|WZ}\circ V_{Z|W}\right)T_{W}-\widetilde{P}_{WX}\right|\leq\epsilon
}
}\sum_{x^{n}\in\mathcal{T}_{V_{X|WZ}}\left(w^{n},z^{n}\right)}\widetilde{P}\left(x^{n}|w^{n}\right)\e^{n\sum_{x,z}T\left(x,z\right)\log P\left(z|x\right)}\Bigr)^{1+s}\e^{-ns\sum_{z}T\left(z\right)\log Q\left(z\right)}\label{eqn:delta_n}\\
 & =\max_{\substack{T_{W},V_{Z|W},V_{X|WZ}:\left|T_{W}-\widetilde{P}_{W}\right|\leq\epsilon',\\
\left|\left(V_{X|WZ}\circ V_{Z|W}\right)T_{W}-\widetilde{P}_{WX}\right|\leq\epsilon
}
}\frac{1}{s}\Bigl(H(V_{Z|W}\times T_{W})+\sum_{w}T\left(w\right)\log\widetilde{P}\left(w\right)\Bigr)\nonumber \\
 & \qquad+\frac{1+s}{s}\biggl(H\left(V_{X|WZ}|T_{W}V_{Z|W}\right)+\sum_{w,x}T\left(w,x\right)\log\widetilde{P}\left(x|w\right)+\sum_{x,z}T\left(x,z\right)\log P\left(z|x\right)\biggr)\nonumber \\
 & \qquad-\sum_{z}T\left(z\right)\log Q\left(z\right)+\delta_{n}+\delta_{n}'+\delta_{n}''\label{eq:-44}\\
 & =\max_{\substack{T_{WX},V_{Z|WX}:\left|T_{W}-\widetilde{P}_{W}\right|\leq\epsilon',\\
\left|T_{WX}-\widetilde{P}_{WX}\right|\leq\epsilon
}
}\biggl\{\frac{1+s}{s}\sum_{w,x,z}T\left(w,x,z\right)\log\frac{\widetilde{P}\left(w,x\right)P\left(z|x\right)}{T\left(w,x,z\right)}+\sum_{w,z}T\left(w,z\right)\log\frac{T\left(w,z\right)}{\widetilde{P}\left(w\right)Q\left(z\right)}\biggr\}\nonumber \\
 & \qquad+\delta_{n}+\delta_{n}'+\delta_{n}'',\label{eq:-3}
\end{align}
\setcounter{mytempeqncnt}{\value{equation}} \setcounter{equation}{\value{mytempeqncnt}}
\hrulefill{} 
\end{figure*}

Observe that in \eqref{eq:-3} $T_{WX}$ is restricted to being close
to $\widetilde{P}_{WX}$ but there is no restriction on $V_{Z|WX}$.
Actually Lemma \ref{lem:minequality} implies that as $n\rightarrow\infty$
and $\epsilon,\epsilon'\rightarrow0$, \eqref{eq:-3} asymptotically
equals 
\begin{align}
 & \max_{\widetilde{P}_{Z|WX}}\biggl\{-\frac{1+s}{s}\sum_{w,x,z}\widetilde{P}\left(w,x,z\right)\log\frac{\widetilde{P}\left(z|w,x\right)}{P\left(z|x\right)}\nonumber \\
 & \qquad\qquad\qquad+\sum_{w,z}\widetilde{P}\left(w,z\right)\log\frac{\widetilde{P}\left(z|w\right)}{Q\left(z\right)}\biggr\},\label{eq:-83}
\end{align}
in the sense that the difference between~\eqref{eq:-3} and \eqref{eq:-83}
vanishes as $n\to\infty$. That is, we can replace the (conditional)
types with their corresponding (conditional) distributions. %The detailed proof is omitted here as it is standard.
Hence $\frac{1}{n}D_{1+s}\left(P_{W^{n}Z^{n}}\|P_{W^{n}}\times Q_{Z}^{n}\right)\to$~\eqref{eq:-83}
as $n\to\infty$. Comparing \eqref{eq:-83} to the definition of $\widetilde{R}'_{1+s}\left(\widetilde{P}_{W|X}\widetilde{P}_{X},P_{Z|X},Q_{Z}\right)$
in \eqref{eq:-6}, we can find that they are equal for the case of
$s\in(0,1]$. Hence 
\begin{align}
 & \lim_{n\rightarrow\infty}\frac{1}{n}D_{1+s}\left(P_{W^{n}Z^{n}}\|P_{W^{n}}\times Q_{Z}^{n}\right)\nonumber \\
 & =\widetilde{R}'_{1+s}\left(\widetilde{P}_{W|X}\widetilde{P}_{X},P_{Z|X},Q_{Z}\right).
\end{align}

Furthermore, observe 
\begin{align}
 & P_{X^{n}}\left(x^{n}\right)\nonumber \\
 & =\sum_{w^{n}}\frac{\widetilde{P}\left(w^{n}\right)1\left\{ w^{n}\in\mathcal{T}_{\epsilon'}^{n}\right\} }{\widetilde{P}_{W}^{n}\left(\mathcal{T}_{\epsilon'}^{n}\right)}\frac{\widetilde{P}\left(x^{n}|w^{n}\right)1\left\{ \left(w^{n},x^{n}\right)\in\mathcal{T}_{\epsilon}^{n}\right\} }{\widetilde{P}_{X|W}^{n}\left(\mathcal{T}_{\epsilon}^{n}\left(w^{n}\right)|w^{n}\right)}\\
 & \leq\sum_{w^{n}}\frac{\widetilde{P}\left(w^{n}\right)1\left\{ w^{n}\in\mathcal{T}_{\epsilon'}^{n}\right\} \widetilde{P}\left(x^{n}|w^{n}\right)1\left\{ \left(w^{n},x^{n}\right)\in\mathcal{T}_{\epsilon}^{n}\right\} }{1-\delta_{n}}\label{eq:-43}\\
 & \leq\frac{\widetilde{P}\left(x^{n}\right)1\left\{ x^{n}\in\mathcal{T}_{\epsilon}^{n}\right\} }{1-\delta_{n}},
\end{align}
where \eqref{eq:-43} follows since as $n\rightarrow\infty$, $\widetilde{P}_{W}^{n}\left(\mathcal{T}_{\epsilon'}^{n}\right)$
converges to 1 and $\widetilde{P}_{X|W}^{n}\left(\mathcal{T}_{\epsilon}^{n}\left(w^{n}\right)|w^{n}\right)$
uniformly converges to 1 for all $w^{n}\in\mathcal{T}_{\epsilon'}^{n}$.
Therefore, 
\begin{align}
 & D_{1+s}(P_{X^{n}}\|\widetilde{P}_{X^{n}})\nonumber \\
 & \leq\frac{1}{s}\log\sum_{x^{n}}\left(\frac{\widetilde{P}\left(x^{n}\right)1\left\{ x^{n}\in\mathcal{T}_{\epsilon}^{n}\right\} }{1-\delta_{n}}\right)^{1+s}\widetilde{P}^{-s}\left(x^{n}\right)\\
 & =\frac{1}{s}\log\frac{\widetilde{P}_{X}^{n}\left(\mathcal{T}_{\epsilon}^{n}\right)}{\left(1-\delta_{n}\right)^{1+s}}\\
 & \rightarrow0,\label{eq:-15-4}
\end{align}
where \eqref{eq:-15-4} follows since $\widetilde{P}_{X}^{n}\left(\mathcal{T}_{\epsilon}^{n}\right)$
converges to 1 as $n\rightarrow\infty$. Since $P_{Z^{n}}$ and $Q_{Z}^{n}$
are respectively the distributions of the channel output induced by
the input $P_{X^{n}}$ and $\widetilde{P}_{X}^{n}$, by the data processing
inequality~\cite{Erven}, we have 
\begin{equation}
D_{1+s}(P_{Z^{n}}\|Q_{Z}^{n})\leq D_{1+s}(P_{X^{n}}\|\widetilde{P}_{X}^{n}).
\end{equation}
Hence $D_{1+s}(P_{Z^{n}}\|Q_{Z}^{n})\rightarrow0$ as well.

Finally, by Lemma \ref{lem:oneshotach}, we obtain 
\begin{align}
 & \e^{sD_{1+s}(P_{M_{1}Z^{n}}\|P_{M_{1}}Q_{Z}^{n})}\nonumber \\
 & \leq\e^{sD_{1+s}\left(P_{W^{n}Z^{n}}\|P_{W^{n}}Q_{Z}^{n}\right)-nsR_{0}}+\e^{sD_{1+s}(P_{Z^{n}}\|Q_{Z}^{n})}\label{eq:-3-2}\\
 & \rightarrow1,\label{eq:-82}
\end{align}
where \eqref{eq:-82} holds for $s\in(0,1]$ if 
\begin{equation}
R_{0}>\widetilde{R}'_{1+s}\bigl(\widetilde{P}_{W|X}\widetilde{P}_{X},P_{Z|X},Q_{Z}\bigr)
\end{equation}
by \eqref{eq:-83} with a small enough $\epsilon>0$. Hence the secrecy
constraint is satisfied.

Moreover, using standard joint typicality decoding, we have that error
constraint 
\begin{equation}
\mathbb{P}\left(\left(M_{0},M_{1}\right)\neq(\widehat{M}_{0},\widehat{M}_{1})\right)\rightarrow0\label{eq:-30-1-1-1}
\end{equation}
is satisfied as well if $R_{0}+R_{1}\leq I_{\widetilde{P}}\left(W;Y\right)$.
The proof of the achievability part for $s\in[-1,1]$ is complete.
For $s=\infty$, the achievability part can be proven by similar steps
to those in Appendix \ref{rem:By-checking-our}. 

\emph{Converse:} Set $W=(M_{0},M_{1})$. By the data processing inequality,
\begin{align}
R_{0}+R_{1} & \leq\frac{1}{n}I\left(W;Y^{n}\right)\leq I\left(W;Y_{J}\right),\label{eq:-33-2}
\end{align}
where $J\sim\mathsf{Unif}\left[1:n\right]$ denotes a time index variable,
independent of $(W,Y^{n})$. It is easy to verify that 
\begin{align}
 & P_{WX_{J}Y_{J}}\left(w,x,y\right)\nonumber \\
 & =P_{W}\left(w\right)\frac{1}{n}\sum_{j=1}^{n}\mathbb{P}\left\{ \left(X_{j},Y_{j}\right)=\left(x,y\right)|W=w\right\} \label{eq:-86}\\
 & =P_{W}\left(w\right)\bbE_{X^{n}Y^{n}|W=w}[T_{X^{n}Y^{n}}\left(x,y\right)],\label{eq:-87}
\end{align}
and 
\begin{align}
 & P_{WX_{J}Y_{J}}\left(w,x,y\right)\nonumber \\
 & =P_{WX_{J}}\left(w,x\right)P\left(y|x\right)\\
 & =P_{W}\left(w\right)\bbE_{X^{n}|W=w}\left[T_{X^{n}}\left(x\right)\right]P\left(y|x\right),\label{eq:-85}
\end{align}
where \eqref{eq:-85} is obtained similarly to~\eqref{eq:-86}\textendash \eqref{eq:-87}.

We first consider the case $s\in(0,\infty]$. Observe $M_{1}$ is
independent of $M_{0}$. Hence if we consider $M_{1}$ as $\mathcal{C}$
and $M_{0}$ as $M$, then the wiretap channel problem turns into
the channel resolvability problem. By Lemma \ref{lem:oneshotcon},
we obtain 
\begin{align}
 & D_{1+s}(P_{M_{1}Z^{n}}\|P_{M_{1}}Q_{Z}^{n})\nonumber \\
 & \geq\max\Bigl\{ D_{1+s}\left(P_{M_{0}M_{1}Z^{n}}\|P_{M_{0}M_{1}}\times Q_{Z}^{n}\right)-nR_{0},\nonumber \\
 & \qquad D_{1+s}(P_{Z^{n}}\|Q_{Z}^{n})\Bigr\}\\
 & =\max\Bigl\{ D_{1+s}\left(P_{WZ^{n}}\|P_{W}\times Q_{Z}^{n}\right)-nR_{0},\nonumber \\
 & \qquad D_{1+s}(P_{Z^{n}}\|Q_{Z}^{n})\Bigr\}.\label{eq:-4}
\end{align}

Define $\widetilde{P}_{Z|WX}$ as the maximizing distribution of 
\begin{align}
 & \max_{\widetilde{P}_{Z|WX}\in\mathcal{P}\left(\mathcal{Z}|\mathcal{W\times X}\right)}\Bigl\{-\frac{1+s}{s}\sum_{w,x,z}P\left(w\right)P_{X_{J}|W}\left(x|w\right)\nonumber \\
 & \qquad\times\widetilde{P}\left(z|w,x\right)\log\frac{\widetilde{P}\left(z|w,x\right)}{P\left(z|x\right)}\nonumber \\
 & \quad+\sum_{w,x,z}P\left(w\right)P_{X_{J}|W}\left(x|w\right)\widetilde{P}\left(z|w,x\right)\nonumber \\
 & \qquad\times\log\frac{\sum_{x}P_{X_{J}|W}\left(x|w\right)\widetilde{P}\left(z|w,x\right)}{Q\left(z\right)}\Bigr\},\label{eq:-36}
\end{align}
where $P_{WX_{J}Z_{J}}$ is the distribution of $W,X_{J},Z_{J}$ induced
by the code. Note that $\widetilde{P}_{Z|WX}$ is determined by the
code, the channel $P_{Z|X}$, and the target distribution $Q_{Z}$.

From Lemma \ref{lem:typecovering} we know that for any $w\in\mathcal{W}$
and any $T_{X}\in\mathcal{P}^{\left(n\right)}\left(\mathcal{X}\right)$,
we can find a conditional type $V_{Z|X}^{\left(w\right)}\in\mathcal{P}^{\left(n\right)}\left(\mathcal{X}|T_{X}\right)$
such that 
\begin{equation}
{\displaystyle \left|T_{X}\widetilde{P}_{Z|XW}\left(\cdot|\cdot,w\right)-T_{X}V_{Z|X}^{\left(w\right)}\right|\leq\frac{\left|\mathcal{X}\right|\left|\mathcal{Z}\right|}{2n}}=O\left(\frac{1}{n}\right).\label{eq:-30}
\end{equation}

Consider the first term of the maximization in \eqref{eq:-4}, then
we obtain \eqref{eq:-104}-\eqref{eq:-95}, where~\eqref{eq:-79}
follows from Lemma \ref{lem:norm}, \eqref{eq:-88} and \eqref{eq:-91}
follow since $x\mapsto x^{1+s}$ is a convex function for $s\geq0$,
\eqref{eq:-89} follows from Lemma \ref{lem:typeequality}, in \eqref{eq:-90}
$V_{Z|X}^{\left(w\right)}:\mathcal{W}\rightarrow\mathcal{P}^{\left(n\right)}\left(\mathcal{Z}|T_{X}\right)$
is the conditional type above satisfying \eqref{eq:-30},\footnote{Note that the choice of $V_{Z|X}^{\left(w\right)}$ and not necessarily
an optimal one for the lower bound \eqref{eq:-90}, since the optima
should be independent of $w$. However, it is, in fact, optimal for
the final lower bound \eqref{eq:-94}.} \eqref{eq:-17} follows from \eqref{eq:-30} and \cite[Lem.~8]{Yassaee},
\eqref{eq:-92} follows since the number of types in $\mathcal{P}^{\left(n\right)}\left(\mathcal{X}\right)$
is polynomial in $n$, \eqref{eq:-93} follows since $x\mapsto\log x$
is a concave function, \eqref{eq:-93a} follows since $P_{X^{n}|W}\left(\mathcal{T}_{T_{X}}|w\right)=\sum_{x^{n}\in\mathcal{T}_{T_{X}}}P_{X^{n}|W}\left(x^{n}|w\right)$
and $\mathcal{T}_{T_{X}}\subseteq\calX^{n}$ runs through all the
sequences in $\mathcal{X}^{n}$, \eqref{eq:-94} follows since $x\mapsto x\log x$
is a convex function, and \eqref{eq:-95} follows since $\bbE_{X^{n}|W=w}\left[T_{X^{n}}\left(x\right)\right]=P_{X_{J}|W}\left(x|w\right)$;
see \eqref{eq:-85}. 
\begin{figure*}[!t]
\setcounter{mytempeqncnt}{\value{equation}} \setcounter{equation}{477}
\begin{align}
 & \frac{1}{n}D_{1+s}\left(P_{WZ^{n}}\|P_{W}\times Q_{Z}^{n}\right)\nonumber \\
 & =\frac{1}{ns}\log\sum_{w\in\mathcal{W}}\sum_{T_{Z}}\sum_{z^{n}\in\mathcal{T}_{T_{Z}}}P\left(w\right)\e^{-ns\sum_{z}T\left(z\right)\log Q\left(z\right)}\left(\sum_{V_{X|Z}}\sum_{x^{n}\in\mathcal{T}_{V_{X|Z}}\left(z^{n}\right)}P\left(x^{n}|w\right)\e^{n\sum_{x,z}T\left(x,z\right)\log P\left(z|x\right)}\right)^{1+s}\label{eq:-104}\\
 & =\frac{1}{ns}\log\sum_{w\in\mathcal{W}}\sum_{T_{Z}}\sum_{z^{n}\in\mathcal{T}_{T_{Z}}}P\left(w\right)\e^{-ns\sum_{z}T\left(z\right)\log Q\left(z\right)}\left(\sum_{V_{X|Z}}P_{X^{n}|W}\left(\mathcal{T}_{V_{X|Z}}\left(z^{n}\right)|w\right)\e^{n\sum_{x,z}T\left(x,z\right)\log P\left(z|x\right)}\right)^{1+s}\\
 & \geq\frac{1}{ns}\log\sum_{w\in\mathcal{W}}\sum_{T_{Z}}\sum_{z^{n}\in\mathcal{T}_{T_{Z}}}P\left(w\right)\sum_{V_{X|Z}}P_{X^{n}|W}^{1+s}\left(\mathcal{T}_{V_{X|Z}}\left(z^{n}\right)|w\right)\e^{n\left(1+s\right)\sum_{x,z}T\left(x,z\right)\log P\left(z|x\right)-ns\sum_{z}T\left(z\right)\log Q\left(z\right)}\label{eq:-79}\\
 & \geq\frac{1}{ns}\log\sum_{w,T_{Z},V_{X|Z}}\left|\mathcal{T}_{T_{Z}}\right|P\left(w\right)\left(\sum_{z^{n}\in\mathcal{T}_{T_{Z}}}\frac{1}{\left|\mathcal{T}_{T_{Z}}\right|}P_{X^{n}|W}\left(\mathcal{T}_{V_{X|Z}}\left(z^{n}\right)|w\right)\right)^{1+s}\e^{n\left(1+s\right)\sum_{x,z}T\left(x,z\right)\log P\left(z|x\right)-ns\sum_{z}T\left(z\right)\log Q\left(z\right)}\label{eq:-88}\\
 & =\delta_{n}+\frac{1}{ns}\log\sum_{w,T_{Z},V_{X|Z}}P\left(w\right)P_{X^{n}|W}^{1+s}\left(\mathcal{T}_{T_{X}}|w\right)\e^{-nsH\left(T_{Z}\right)+n\left(1+s\right)H\left(V_{Z|X}|T_{X}\right)}\e^{n\left(1+s\right)\sum_{x,z}T\left(x,z\right)\log P\left(z|x\right)-ns\sum_{z}T\left(z\right)\log Q\left(z\right)}\label{eq:-89}\\
 & \geq\delta_{n}+\frac{1}{ns}\log\sum_{w,T_{X}}P\left(w\right)P_{X^{n}|W}^{1+s}\left(\mathcal{T}_{T_{X}}|w\right)\e^{-nsH\left(V_{Z|X}^{\left(w\right)}\circ T_{X}\right)+n\left(1+s\right)\left(H\left(V_{Z|X}^{\left(w\right)}|T_{X}\right)+\sum_{x,z}T\left(x\right)V_{Z|X}^{\left(w\right)}\left(z|x\right)\log P\left(z|x\right)\right)}\nonumber \\
 & \qquad\times\e^{-ns\sum_{z}\left[V_{Z|X}^{\left(w\right)}\circ T_{X}\right]\left(z\right)\log Q\left(z\right)}\label{eq:-90}\\
 & =\delta_{n}+\frac{1}{ns}\log\sum_{w,T_{X}}P\left(w\right)P_{X^{n}|W}^{1+s}\left(\mathcal{T}_{T_{X}}|w\right)\e^{-nsH\left(\widetilde{P}_{Z|WX}\circ T_{X}\right)+n\left(1+s\right)\left(H\left(\widetilde{P}_{Z|WX}|T_{X}\right)+\sum_{x,z}T\left(x\right)\widetilde{P}_{Z|WX}\left(z|x\right)\log P\left(z|x\right)\right)}\nonumber \\
 & \qquad\times\e^{-ns\sum_{z}\left[\widetilde{P}_{Z|WX}\circ T_{X}\right]\left(z\right)\log Q\left(z\right)+n\cdot\delta_{n}'}\label{eq:-17}\\
 & \geq\delta_{n}+\delta_{n}'+\frac{1}{ns}\log|\mathcal{P}^{\left(n\right)}\left(\mathcal{X}\right)|\biggl(\sum_{w,T_{X}}\frac{1}{\left|\mathcal{P}^{\left(n\right)}\left(\mathcal{X}\right)\right|}P\left(w\right)P_{X^{n}|W}\left(\mathcal{T}_{T_{X}}|w\right)\nonumber \\
 & \qquad\times\e^{-n\sum_{x,z}T\left(x\right)\widetilde{P}\left(z|w,x\right)\log\frac{\widetilde{P}\left(z|w,x\right)}{P\left(z|x\right)}+\frac{ns}{1+s}\sum_{x,z}T\left(x\right)\widetilde{P}\left(z|w,x\right)\log\frac{\sum_{x}T\left(x\right)\widetilde{P}\left(z|w,x\right)}{Q\left(z\right)}}\biggr)^{1+s}\label{eq:-91}\\
 & \geq\delta_{n}+\delta_{n}'+\delta_{n}''+\frac{1+s}{ns}\log\sum_{w,T_{X}}P\left(w\right)P_{X^{n}|W}\left(\mathcal{T}_{T_{X}}|w\right)\nonumber \\
 & \qquad\times\e^{-n\sum_{x,z}T\left(x\right)\widetilde{P}\left(z|w,x\right)\log\frac{\widetilde{P}\left(z|w,x\right)}{P\left(z|x\right)}+\frac{ns}{1+s}\sum_{x,z}T\left(x\right)\widetilde{P}\left(z|w,x\right)\log\frac{\sum_{x}T\left(x\right)\widetilde{P}\left(z|w,x\right)}{Q\left(z\right)}}\label{eq:-92}\\
 & \geq\delta_{n}+\delta_{n}'+\delta_{n}''+\sum_{w,T_{X}}P\left(w\right)P_{X^{n}|W}\left(\mathcal{T}_{T_{X}}|w\right)\biggl(-\frac{1+s}{s}\sum_{x,z}T_{X}\left(x\right)\widetilde{P}\left(z|w,x\right)\log\frac{\widetilde{P}\left(z|w,x\right)}{P\left(z|x\right)}\nonumber \\
 & \qquad+\sum_{x,z}T_{X}\left(x\right)\widetilde{P}\left(z|w,x\right)\log\frac{\sum_{x}T_{X}\left(x\right)\widetilde{P}\left(z|w,x\right)}{Q\left(z\right)}\biggr)\label{eq:-93}\\
 & =\delta_{n}+\delta_{n}'+\delta_{n}''+\sum_{w\in\mathcal{W}}P\left(w\right)\Bigl(-\frac{1+s}{s}\sum_{x,z}\bbE_{X^{n}|W=w}\left[T_{X^{n}}\left(x\right)\right]\widetilde{P}\left(z|w,x\right)\log\frac{\widetilde{P}\left(z|w,x\right)}{P\left(z|x\right)}\nonumber \\
 & \qquad+\sum_{x,z}\bbE_{X^{n}|W=w}\Big[T_{X^{n}}\left(x\right)\widetilde{P}\left(z|w,x\right)\log\frac{\sum_{x}T_{X^{n}}\left(x\right)\widetilde{P}\left(z|w,x\right)}{Q\left(z\right)}\Big]\Bigr)\label{eq:-93a}\\
 & \geq\delta_{n}+\delta_{n}'+\delta_{n}''+\sum_{w\in\mathcal{W}}P\left(w\right)\Bigl(-\frac{1+s}{s}\sum_{x,z}\bbE_{X^{n}|W=w}\left[T_{X^{n}}\left(x\right)\right]\widetilde{P}\left(z|w,x\right)\log\frac{\widetilde{P}\left(z|w,x\right)}{P\left(z|x\right)}\nonumber \\
 & \qquad+\sum_{x,z}\bbE_{X^{n}|W=w}\left[T_{X^{n}}\left(x\right)\right]\widetilde{P}\left(z|w,x\right)\log\frac{\sum_{x}\bbE_{X^{n}|W=w}\left[T_{X^{n}}\left(x\right)\right]\widetilde{P}\left(z|w,x\right)}{Q\left(z\right)}\Bigr)\label{eq:-94}
\end{align}
\setcounter{mytempeqncnt}{\value{equation}} \setcounter{equation}{\value{mytempeqncnt}}
\end{figure*}

\begin{figure*}[!t]
\setcounter{mytempeqncnt}{\value{equation}} \setcounter{equation}{489}  
\begin{align}
 & =\delta_{n}+\delta_{n}'+\delta_{n}''+\sum_{w\in\mathcal{W}}P\left(w\right)\Bigl(-\frac{1+s}{s}\sum_{x,z}P_{X_{J}|W}\left(x|w\right)\widetilde{P}\left(z|w,x\right)\log\frac{\widetilde{P}\left(z|w,x\right)}{P\left(z|x\right)}\nonumber \\
 & \qquad+\sum_{x,z}P_{X_{J}|W}\left(x|w\right)\widetilde{P}\left(z|w,x\right)\log\frac{\sum_{x}P_{X_{J}|W}\left(x|w\right)\widetilde{P}\left(z|w,x\right)}{Q\left(z\right)}\Bigr). \qquad \qquad \qquad \qquad \qquad \qquad \qquad \qquad \qquad\label{eq:-95} 
\end{align}  
\setcounter{mytempeqncnt}{\value{equation}} \setcounter{equation}{\value{mytempeqncnt}}
\hrulefill{} 
\end{figure*}

By the choice of $\widetilde{P}_{Z|WX}$, from \eqref{eq:-95} we
have 
\begin{align}
 & \lim_{n\rightarrow\infty}\frac{1}{n}D_{1+s}\left(P_{WZ^{n}}\|P_{W}\times Q_{Z}^{n}\right)\geq\eqref{eq:-36}.\label{eq:-19}
\end{align}
Furthermore, it is easy to verify 
\begin{equation}
{\displaystyle \lim_{n\rightarrow\infty}\left|P_{X_{J}}^{(n)}\circ P_{Z|X}-Q_{Z}\right|=0},\label{eq:-32-2}
\end{equation}
since $D_{1+s}(P_{Z^{n}}\|Q_{Z^{n}})\to0$ (see \eqref{eq:-4}).

Since $\mathcal{P}\left(\mathcal{X}\right)$ is compact, for each
$w$, there must exist some sequence of increasing integers $\left\{ n_{k}\right\} _{k=1}^{\infty}$
such that $P_{X_{J}|W=w}^{\left(n_{k}\right)}$ converges to some
distribution $\widetilde{P}_{X|W=w}$. By \eqref{eq:-32-2}, $\bbE_{W}[\widetilde{P}_{X|W}(\cdot|W)]\in\mathcal{P}\left(P_{Z|X},Q_{Z}\right)$
holds. Moreover, \eqref{eq:-33-2} and \eqref{eq:-19} respectively
imply 
\begin{align}
R_{0}+R_{1} & \leq I_{\widetilde{P}}\left(W;Y\right),
\end{align}
and 
\begin{align}
R_{0}\geq & \max_{\widetilde{P}_{Z|WX}}\biggl\{-\frac{1+s}{s}\sum_{w,x,z}\widetilde{P}\left(w,x,z\right)\log\frac{\widetilde{P}\left(z|w,x\right)}{P\left(z|x\right)}\nonumber \\
 & \qquad\qquad+\sum_{w,z}\widetilde{P}\left(w,z\right)\log\frac{\widetilde{P}\left(z|w\right)}{Q\left(z\right)}\biggr\}.\label{eq:-7}
\end{align}
Observe that the RHS of \eqref{eq:-7} is just $\widetilde{R}'_{1+s}\left(\widetilde{P}_{W|X}\widetilde{P}_{X},P_{Z|X},Q_{Z}\right)$
with $s\in(0,\infty]$. Hence $R_{0}\geq\widetilde{R}'_{1+s}\left(\widetilde{P}_{W|X}\widetilde{P}_{X},P_{Z|X},Q_{Z}\right)$.

Therefore, $P_{W}\widetilde{P}_{X|W}$ is the desired distribution
$\widetilde{P}_{WX}$ in \eqref{eq:-39}. The proof for the case $s\in(0,\infty]$
is complete.

Next we consider the case $s\in(-1,0]$. This case can be proved by
following steps similar to the proof of traditional channel resolvability
problem \cite{Han} or the distributed channel synthesis problem \cite{Cuff}.
Observe 
\begin{align}
R_{0} & \geq\frac{1}{n}I\left(M_{0};Z^{n}|M_{1}\right)\\
 & =\frac{1}{n}I\left(M_{0}M_{1};Z^{n}|M_{1}\right)\\
 & =\frac{1}{n}I\left(M_{0}M_{1};Z^{n}\right)-\frac{1}{n}I\left(M_{1};Z^{n}\right)\\
 & =\frac{1}{n}H\left(Z^{n}\right)-\frac{1}{n}H\left(Z^{n}|M_{0}M_{1}\right)-\delta_{n}\label{eq:-77-1}\\
 & =H_{Q}\left(Z\right)-H\left(Z_{J}|W\right)-\delta_{n}+\delta_{n}',\label{eq:-78-1}
\end{align}
where \eqref{eq:-77-1} and \eqref{eq:-78-1} follow from the facts
$\left|P_{M_{1}Z^{n}}-P_{M_{1}}Q_{Z}^{n}\right|\rightarrow0$ and
$\left|P_{Z^{n}}-Q_{Z}^{n}\right|\rightarrow0$ respectively.

Furthermore, for each $w$, there exists some increasing sequence
of integers $\left\{ n_{k}\right\} _{k=1}^{\infty}$ such that $P_{X_{J}|W=w}^{(n_{k})}$
converges to some distribution $\widetilde{P}_{X|W=w}$ that satisfies
$\bbE_{W}[\widetilde{P}_{X|W}(\cdot|W)]\in\mathcal{P}\left(P_{Z|X},Q_{Z}\right)$.
Hence letting $n=n_{k}$ and $k\rightarrow\infty$ in \eqref{eq:-78-1},
we get 
\begin{align}
R_{0}\geq H_{\widetilde{P}}\left(Z\right)-H_{\widetilde{P}}\left(Z_{J}|W\right)=I_{\widetilde{P}}\left(W;Z\right).\label{eq:-33-2-1-2}
\end{align}
On the other hand, 
\begin{align}
R_{0}+R_{1} & \leq I_{\widetilde{P}}\left(W;Y\right).\label{eq:-80-1}
\end{align}
Combining \eqref{eq:-33-2-1-2} and \eqref{eq:-80-1} gives the converse
part. Therefore, the proof of \eqref{eq:-39} is complete.

Next we prove \eqref{eq:-120}. By adding an artificial non-secret
message $M_{0}'$ (with rate $R_{0}'$) in the achievability scheme
above, we have the following achievable region. 
\begin{equation}
\bigcup_{\substack{\widetilde{P}_{WX}:\\
\widetilde{P}_{X}\in\mathcal{P}\left(P_{Z|X},Q_{Z}\right)
}
}\left\{ \begin{array}{l}
(R_{0},R_{1}):R_{0}'\geq0,\\
R_{0}'+R_{0}+R_{1}\leq I_{\widetilde{P}}\left(W;Y\right),\\
R_{0}'+R_{0}\geq\\
\widetilde{R}'_{1+s}(\widetilde{P}_{W|X}\widetilde{P}_{X},P_{Z|X},Q_{Z})
\end{array}\right\} .\label{eq:-121}
\end{equation}
Using Fourier\textendash Motzkin Elimination (see \cite[Appendix D]{Gamal}),
we can show that the regions in \eqref{eq:-121} and \eqref{eq:-120}
are the same. Hence \eqref{eq:-120} $\subseteq\mathcal{R}_{1+s}(Q_{Z})$.
On the other hand, comparing the RHSes of \eqref{eq:-39} and~\eqref{eq:-120}
yields that the RHS of~\eqref{eq:-39} $\subseteq$ \eqref{eq:-120}.
In addition, $\mathcal{R}_{1+s}(Q_{Z})=$ the RHS of \eqref{eq:-39}.
Hence $\mathcal{R}_{1+s}(Q_{Z})\subseteq$ \eqref{eq:-120}. Therefore,
$\mathcal{R}_{1+s}(Q_{Z})=$ \eqref{eq:-120}.

Lastly, by standard cardinality bounding techniques \cite[Appendix C]{Gamal},
the alphabet size of $W$ can be limited to $\left|\mathcal{W}\right|\leq\left|\mathcal{X}\right|+1$.

\subsection*{Acknowledgements}

The authors would like to thank Prof.\ Masahito Hayashi for pointing
out the relevance of~\cite[Thm.~14]{Hayashi} to the present work.

%The authors are supported by a Singapore National Research Foundation
%(NRF) National Cybersecurity R\&D Grant (R-263-000-C74-281 and NRF2015NCR-NCR003-006).
%The first author is also supported by a National Natural Science Foundation
%of China (NSFC) under Grant (61631017).

The authors are extremely grateful to the Associate Editor Prof.\ Matthieu
Bloch and the two reviewers for their extensive, constructive and
helpful feedback to improve the manuscript.

\bibliographystyle{unsrt}
\bibliography{ref}

\begin{thebibliography}{10}

\bibitem{yu2017renyi}
L.~Yu and V.~Y.~F. Tan.
\newblock R{\'e}nyi resolvability and its applications to the wiretap channel.
\newblock In {\em International Conference on Information Theoretic Security},
  pages 208--233. Springer, 2017.

\bibitem{Han}
T.~Han and S.~Verd{\'u}.
\newblock Approximation theory of output statistics.
\newblock {\em IEEE Trans. on Inform. Theory}, 39(3):752--772, 1993.

\bibitem{Hayashi06}
M.~Hayashi.
\newblock General nonasymptotic and asymptotic formulas in channel
  resolvability and identification capacity and their application to the
  wiretap channel.
\newblock {\em IEEE Trans. on Inform. Theory}, 52(4):1562--1575, 2006.

\bibitem{Hayashi11}
M.~Hayashi.
\newblock Exponential decreasing rate of leaked information in universal random
  privacy amplification.
\newblock {\em IEEE Trans. on Inform. Theory}, 57(6):3989--4001, 2011.

\bibitem{Liu}
J.~Liu, P.~Cuff, and S.~Verd{\'u}.
\newblock {$E_{\gamma}$}-resolvability.
\newblock {\em IEEE Trans. on Inform. Theory}, 63(5):2629--2658, 2017.

\bibitem{yu2018simulation}
L.~Yu and V.~Y.~F. Tan.
\newblock Simulation of random variables under {R\'enyi} divergence measures of
  all orders.
\newblock {\em arXiv preprint 1805.12451}, 2018.

\bibitem{yu2018asymptotic}
L.~Yu and V.~Y.~F. Tan.
\newblock Asymptotic coupling and its applications in information theory.
\newblock {\em IEEE Trans. on Inform. Theory}, 65, 2019.

\bibitem{Wyner}
A.~Wyner.
\newblock The common information of two dependent random variables.
\newblock {\em IEEE Trans. on Inform. Theory}, 21(2):163--179, 1975.

\bibitem{Cuff}
P.~Cuff.
\newblock Distributed channel synthesis.
\newblock {\em IEEE Trans. on Inform. Theory}, 59(11):7071--7096, 2013.

\bibitem{Bloch}
M.~R. Bloch and J.~N. Laneman.
\newblock Strong secrecy from channel resolvability.
\newblock {\em IEEE Trans. on Inform. Theory}, 59(12):8077--8098, 2013.

\bibitem{Han14}
T.~S. Han, H.~Endo, and M.~Sasaki.
\newblock Reliability and secrecy functions of the wiretap channel under cost
  constraint.
\newblock {\em IEEE Trans. on Inform. Theory}, 60(11):6819--6843, 2014.

\bibitem{Parizi}
M.~B. Parizi, E.~Telatar, and N.~Merhav.
\newblock Exact random coding secrecy exponents for the wiretap channel.
\newblock {\em IEEE Trans. on Inform. Theory}, 63(1):509--531, 2017.

\bibitem{hou2014effective}
J.~Hou and G.~Kramer.
\newblock Effective secrecy: Reliability, confusion and stealth.
\newblock In {\em Information Theory (ISIT), 2014 IEEE International Symposium
  on}, pages 601--605. IEEE, 2014.

\bibitem{Kumar}
G.~R. Kumar, C.~T. Li, and A.~El Gamal.
\newblock Exact common information.
\newblock In {\em IEEE International Symposium on Information Theory (ISIT)},
  pages 161--165. IEEE, 2014.

\bibitem{li2017distributed}
C.~T. Li and A.~El Gamal.
\newblock Distributed simulation of continuous random variables.
\newblock {\em IEEE Trans. on Inform. Theory}, 63(10):6329--6343, 2017.

\bibitem{yu2018on}
L.~Yu and V.~Y.~F. Tan.
\newblock On exact and $\infty$-{R\'enyi} common informations.
\newblock {\em arXiv preprint 1810.00295}, 2018.

\bibitem{iwamoto}
M.~Iwamoto and J.~Shikata.
\newblock Information theoretic security for encryption based on conditional
  {R\'enyi} entropies.
\newblock {\em Lecture Notes in Computer Science (Information Theoretic
  Security)}, 8317:103--121, 2014.

\bibitem{Shikata}
J.~Shikata.
\newblock Design and analysis of information-theoretically secure
  authentication codes with non-uniformly random keys.
\newblock {\em IACR Cryptology ePrint Archive}, 2015:250, 2015.

\bibitem{Bai2015}
S.~Bai, A.~Langlois, T.~Lepoint, D.~Stehl{\'e}, and R.~Steinfeld.
\newblock Improved security proofs in lattice-based cryptography: Using the
  {R{\'e}nyi} divergence rather than the statistical distance.
\newblock In {\em Advances in Cryptology--ASIACRYPT 2015}, pages 3--24, Berlin,
  Heidelberg, 2015. Springer Berlin Heidelberg.

\bibitem{yu2018wyner}
L.~Yu and V.~Y.~F. Tan.
\newblock Wyner's common information under {R{\'e}nyi} divergence measures.
\newblock {\em IEEE Trans. on Inform. Theory}, 64(5):3616--3632, 2018.

\bibitem{yu2018corrections}
L.~Yu and V.~Y.~F. Tan.
\newblock Corrections to ``{Wyner's} common information under {R\'enyi}
  divergence measures''.
\newblock {\em arXiv preprint arXiv:1810.02534}, 2018.

\bibitem{barron1986entropy}
A.~R. Barron.
\newblock Entropy and the central limit theorem.
\newblock {\em The Annals of Probability}, pages 336--342, 1986.

\bibitem{bobkov2016r}
S.~G. Bobkov, G.~P. Chistyakov, and F.~G{\"o}tze.
\newblock R\'enyi divergence and the central limit theorem.
\newblock {\em arXiv preprint arXiv:1608.01805}, 2016.

\bibitem{beigi2014quantum}
S.~Beigi and A.~Gohari.
\newblock Quantum achievability proof via collision relative entropy.
\newblock {\em IEEE Trans. on Inform. Theory}, 60(12):7980--7986, 2014.

\bibitem{dodis2013overcoming}
Y.~Dodis and Y.~Yu.
\newblock Overcoming weak expectations.
\newblock In {\em Theory of Cryptography}, pages 1--22. Springer, 2013.

\bibitem{Hayashi17}
M.~Hayashi and V.~Y.~F. Tan.
\newblock Equivocations, exponents, and second-order coding rates under various
  {R{\'e}nyi} information measures.
\newblock {\em IEEE Trans. on Inform. Theory}, 63(2):975--1005, 2017.

\bibitem{Tan}
V.~Y.~F. Tan and M.~Hayashi.
\newblock Analysis of remaining uncertainties and exponents under various
  conditional {R{\'e}nyi} entropies.
\newblock {\em IEEE Trans. on Inform. Theory}, 64(5), 2018.

\bibitem{chou2017coding}
R.~A. Chou, B.~N. Vellambi, M.~R. Bloch, and J.~Kliewer.
\newblock Coding schemes for achieving strong secrecy at negligible cost.
\newblock {\em IEEE Trans. on Inform. Theory}, 63(3):1858--1873, 2017.

\bibitem{Hayashi}
M.~Hayashi and R.~Matsumoto.
\newblock Secure multiplex coding with dependent and non-uniform multiple
  messages.
\newblock {\em IEEE Trans. on Inform. Theory}, 62(5):2355--2409, 2016.

\bibitem{Csiszar78}
I.~Csisz{\'a}r and J.~K{\"o}rner.
\newblock Broadcast channels with confidential messages.
\newblock {\em IEEE Trans. on Inform. Theory}, 24(3):339--348, 1978.

\bibitem{Csi97}
I.~Csisz\'{a}r and J.~{K\"{o}rner}.
\newblock {\em Information Theory: Coding Theorems for Discrete Memoryless
  Systems}.
\newblock Cambridge University Press, 2011.

\bibitem{vellambi2016sufficient}
B.~N. Vellambi and J.~Kliewer.
\newblock Sufficient conditions for the equality of exact and {Wyner} common
  information.
\newblock In {\em Communication, Control, and Computing (Allerton), 2016 54th
  Annual Allerton Conference on}, pages 370--377. IEEE, 2016.

\bibitem{vellambi2018new}
B.~N. Vellambi and J.~Kliewer.
\newblock New results on the equality of exact and {Wyner} common information
  rates.
\newblock In {\em 2018 IEEE International Symposium on Information Theory
  (ISIT)}, pages 151--155. IEEE, 2018.

\bibitem{yu2018exact}
L.~Yu and V.~Y.~F. Tan.
\newblock Exact channel synthesis.
\newblock {\em arXiv preprint arXiv:1810.13246}, 2018.

\bibitem{Erven}
T.~Van Erven and P.~Harremos.
\newblock {R{\'e}nyi} divergence and {Kullback-Leibler} divergence.
\newblock {\em IEEE Trans. on Inform. Theory}, 60(7):3797--3820, 2014.

\bibitem{Verdu}
S.~Verd{\'u}.
\newblock $\alpha$-mutual information.
\newblock In {\em Information Theory and Applications Workshop (ITA)}, pages
  1--6, 2015.

\bibitem{Fong}
S.~L. Fong and V.~Y.~F. Tan.
\newblock Strong converse theorems for classes of multimessage multicast
  networks: A {R{\'e}nyi} divergence approach.
\newblock {\em IEEE Trans. on Inform. Theory}, 62(9):4953--4967, 2016.

\bibitem{cachin1997entropy}
C.~Cachin.
\newblock {\em Entropy measures and unconditional security in cryptography}.
\newblock PhD thesis, ETH Zurich, 1997.

\bibitem{fehr2014conditional}
S.~Fehr and S.~Berens.
\newblock On the conditional {R{\'e}nyi} entropy.
\newblock {\em IEEE Trans. on Inform. Theory}, 60(11):6801--6810, 2014.

\bibitem{Wyner75}
A.~Wyner.
\newblock The wire-tap channel.
\newblock {\em Bell Labs Technical Journal}, 54(8):1355--1387, 1975.

\bibitem{CN04}
I.~Csisz\'ar and P.~Narayan.
\newblock Secrecy capacities for multiple terminals.
\newblock {\em IEEE Trans. on Inform. Theory}, 50(12):3047--3061, 2004.

\bibitem{watanabe2015optimal}
S.~Watanabe and Y.~Oohama.
\newblock The optimal use of rate-limited randomness in broadcast channels with
  confidential messages.
\newblock {\em IEEE Trans. on Inform. Theory}, 61(2):983--995, 2015.

\bibitem{xu2008broadcast}
J.~Xu and B.~Chen.
\newblock Broadcast confidential and public messages.
\newblock In {\em Information Sciences and Systems, 2008. CISS 2008. 42nd
  Annual Conference on}, pages 630--635. IEEE, 2008.

\bibitem{kobayashi2013secure}
D.~Kobayashi, H.~Yamamoto, and T.~Ogawa.
\newblock Secure multiplex coding attaining channel capacity in wiretap
  channels.
\newblock {\em IEEE Trans. on Inform. Theory}, 59(12):8131--8143, 2013.

\bibitem{Gohari}
A.~Gohari and V.~Anantharam.
\newblock Generating dependent random variables over networks.
\newblock In {\em Information Theory Workshop (ITW), 2011 IEEE}, pages
  698--702, 2011.

\bibitem{Goldfeld}
Z.~Goldfeld, P.~Cuff, and H.~H. Permuter.
\newblock Semantic-security capacity for wiretap channels of type {II}.
\newblock {\em IEEE Trans. on Inform. Theory}, 62(7):3863--3879, 2016.

\bibitem{Cuff10}
P.~Cuff, H.~Permuter, and T.~Cover.
\newblock Coordination capacity.
\newblock {\em IEEE Trans. on Inform. Theory}, 56(9):4181--4206, 2010.

\bibitem{Dembo}
A.~Dembo and O.~Zeitouni.
\newblock {\em Large Deviations Techniques and Applications}.
\newblock Springer-Verlag, 2nd edition, 1998.

\bibitem{Gallager}
R.~G. Gallager.
\newblock {\em Information Theory and Reliable Communication}, volume~2.
\newblock Springer, 1968.

\bibitem{tan2011large}
V.~Y.~F. Tan, A.~Anandkumar, L.~Tong, and A.~S. Willsky.
\newblock A large-deviation analysis of the maximum-likelihood learning of
  {Markov} tree structures.
\newblock {\em IEEE Trans. on Inform. Theory}, 57(3):1714--1735, 2011.

\bibitem{Sion}
M.~Sion.
\newblock On general minimax theorems.
\newblock {\em Pacific J. Math}, 8(1):171--176, 1958.

\bibitem{Cover}
T.~M. Cover and J.~A. Thomas.
\newblock {\em Elements of Information Theory}.
\newblock Wiley-Interscience, 2nd edition, 2006.

\bibitem{boucheron2013concentration}
S.~Boucheron, G.~Lugosi, and P.~Massart.
\newblock {\em Concentration inequalities: A nonasymptotic theory of
  independence}.
\newblock Oxford university press, 2013.

\bibitem{Mitzenmacher}
M.~Mitzenmacher and E.~Upfal.
\newblock {\em Probability and Computing: Randomized Algorithms and
  Probabilistic Analysis}.
\newblock Cambridge university press, 2005.

\bibitem{Gamal}
A.~El Gamal and Y.-H. Kim.
\newblock {\em Network Information Theory}.
\newblock Cambridge university press, 2011.

\bibitem{Bloch2011}
M.~Bloch and J.~Barros.
\newblock {\em Physical-layer Security: From Information Theory to Security
  Engineering}.
\newblock Cambridge University Press, 2011.

\bibitem{Yassaee}
M.~Yassaee, A.~Gohari, and M.~Aref.
\newblock Channel simulation via interactive communications.
\newblock {\em IEEE Trans. on Inform. Theory}, 61(6):2964--2982, 2015.

\end{thebibliography}

\begin{IEEEbiographynophoto}{Lei Yu} received the B.E. and Ph.D. degrees, both in electronic engineering, from University of Science and Technology of China (USTC) in 2010 and 2015, respectively. From 2015 to 2017, he was a postdoctoral researcher at the Department of Electronic Engineering and Information Science (EEIS), USTC. Currently, he is a research fellow at the Department of Electrical and Computer Engineering, National University of Singapore. His research interests include information theory, probability theory, and discrete mathematics. \end{IEEEbiographynophoto} 

\begin{IEEEbiographynophoto}{Vincent Y.\ F.\ Tan} (S'07-M'11-SM'15)  was born in Singapore in 1981. He is currently a Dean's Chair Associate Professor in the Department of Electrical and Computer Engineering  and the Department of Mathematics at the National University of Singapore (NUS). He received the B.A.\ and M.Eng.\ degrees in Electrical and Information Sciences from Cambridge University in 2005 and the Ph.D.\ degree in Electrical Engineering and Computer Science (EECS) from the Massachusetts Institute of Technology (MIT)  in 2011.  His research interests include information theory, machine learning, and statistical signal processing.

Dr.\ Tan received the MIT EECS Jin-Au Kong outstanding doctoral thesis prize in 2011, the NUS Young Investigator Award in 2014, the NUS Engineering Young Researcher Award in 2018, and the Singapore National Research Foundation (NRF) Fellowship (Class of 2018). He is also an IEEE Information Theory Society Distinguished Lecturer for 2018/9. He has authored a research monograph on {\em ``Asymptotic Estimates in Information Theory with Non-Vanishing Error Probabilities''} in the Foundations and Trends in Communications and Information Theory Series (NOW Publishers). He is currently serving as an Associate Editor of the IEEE Transactions on Signal Processing. \end{IEEEbiographynophoto}

\end{document}